\documentclass[prb,twocolumn,superscriptaddress]{revtex4}
\usepackage{graphicx}
\usepackage{amsfonts} 
\usepackage{amsmath}
\usepackage{amssymb}
\usepackage{bm}
\usepackage{hyperref}
\usepackage{slashed}
\usepackage{ulem}
\usepackage[latin1]{inputenc}

\begin{document}
\title{Chern-Simons superconductors and their instabilities}
\author{Rui Wang}
\affiliation{National Laboratory of Solid State Microstructures and Department of Physics, Nanjing University, Nanjing 210093, China}
\affiliation{Collaborative Innovation Center for Advanced Microstructures, Nanjing 210093, China}
\author{Baigeng Wang}
\email{bgwang@nju.edu.cn}
\affiliation{National Laboratory of Solid State Microstructures and Department of Physics, Nanjing University, Nanjing 210093, China}
\affiliation{Collaborative Innovation Center for Advanced Microstructures, Nanjing 210093, China}
\author{Tigran Sedrakyan}
\email{tsedrakyan@umass.edu}
\affiliation{Department of Physics, University of Massachusetts, Amherst, Massachusetts 01003, USA}


\date{\today }

\begin{abstract}

Two-dimensional quantum antiferromagnets host rich physics, including long-range ordering, high-$T_c$ superconductivity, quantum spin liquid behavior, topological ordering, a variety of other exotic phases, and quantum criticalities. Frustrating perturbations in antiferromagnets may give rise to strong quantum fluctuations, challenging the theoretical understanding of the many-body ground state.
Here we develop a method to describe the quantum antiferromagnets using fermionic degrees of freedom. The method is based on a formally exact mapping between spin exchange models and theories describing fermionic matter with the emergent $U(1)$ Chern-Simons gauge field. For the planar N\'{e}el state, this mapping self-consistently generates the Chern-Simons superconductor mean-field ground state of introduced spinless fermions. We systematically compare the Chern-Simons superconductor state with the planar N\'{e}el state at the level of collective modes as well as order parameters. We reveal qualitative and quantitative correspondences between these two states. We demonstrate that such a construction using the fractionalized excitations and Chern-Simons gauge field can not only describe the N\'{e}el order but can also be applied to study quantum spin liquids. Furthermore, we show that the confinement-deconfinement transitions from the N\'{e}el order to quantum spin liquids are signaled and characterized by the instabilities of Chern-Simons superconductors, driven by strong frustration. The results suggest observing and classifying the descendants of antiferromagnets, including other ordered states and unconventional superconductors, as well as emergent quantum spin liquids.

\end{abstract}

\maketitle

\section{introduction}
Antiferromagnetism originates from the correlation between electron spins, and it is a long-studied phenomenon in condensed matter physics \cite{Andersona}. The underlying physics is mainly captured by antiferromagnetic (AFM) Heisenberg exchange couplings between local spin-$S$ operators, $\hat{S}_{\bf r}$\cite{W. Marshall}. The ground state of an AFM in the three spatial dimensions (3D) usually has long-range orders described by spontaneous breaking of the spin rotational symmetry. The thermal fluctuations then attempt to restore the broken continuous symmetry, generating the Goldstone modes. These are the magnons representing the elementary magnetic excitations.

In 1D systems, the strong quantum fluctuations can melt the long-range order.
A nonlinear sigma model generally describes the AFM fluctuations up to a Wess-Zumino term \cite{efradkin}, which is further dependent on the parity of $2S$. This leads to the Haldanes' conjecture \cite{Haldanea,Haldaneb,Haldanec} that the 1D spin chains are disordered (critical) for $2S$ being even (odd). The conjecture was later validated and generalized to the notion of symmetry protected topological (SPT) phases \cite{mlevin,xchen,tsentila,anton} in $S=1$ Heisenberg chain with boundary modes. This line of studies stimulated the discovery of more 1D topological phases in the last decade \cite{Pollmann,amturner,Ying Ran}.

Zero-temperature antiferromagnetism is even more interesting in 2D, where there can be a strong competition between symmetry breaking and quantum fluctuations. As a result of the competition, the AFM N\'{e}el order in 2D behaves in some sense as a physically marginal platform where many effects become more manifested than in 1D and 3D. Drastic changes of ground state properties can be achieved through perturbation on top of an AFM order, e.g, enhancing the frustration \cite{RMoessner,Yildirim,DHLee,Sachdeva,HCJiang,ShouShuGong,Capriotti,LWang,Zhitomirsky,Singh,Capriottia,Takano,Doretto,VChubukov,GMZhang,WJHu,LingWangb,lingWangc,Poilblanc,Haghshenas,SMorita,cnvarney,Carrasquilla,Ciolo,zzhu,Plekhanov}, doping with carriers \cite{paleea,cgros,Paramekanti,Sorella,ctshih,srwhitea,tamaier,dsenechal,amstremblay,ctshihb,gbaskaran}, varying the lattice structure \cite{zhen weihong}, as well as introducing more correlation effects such as the Dzyaloshinskii-Moriya (DM) interaction \cite{gangchen,cmfong}.
Some of these perturbations can greatly enhance the effect of quantum fluctuations, driving the system into a completely different ground state as compared to the parent AFM order. The destabilization  of N\'{e}el AFM is the main focus of this work.

One of the most prominent examples of 2D systems where the drastic changes of the ground state are achieved is the cuprate high-$T_c$ superconductor whose main physics for Cooper pairing is believed to take place in 2D Cu-O plane. These changes of the ground state happen upon doping of an AFM Mott insulator \cite{paleea,cgros,Paramekanti,Sorella,ctshih,srwhitea,tamaier,dsenechal,amstremblay,ctshihb,gbaskaran,rmkonik,rblaughlinn,kyyangg,dsrokhsarr,gkotliarr,gabrierkot,schatterje}. Another major interest topic is the theoretical prediction of quantum spin liquids (QSL) stabilized in frustrated 2D quantum magnets \cite{pwandersonna,pwandersonnb,pmonthonxa,xxgwenn}. Both of the above phenomena have received significant attention and evoked fundamental developments, analytically and numerically, in strongly-correlated condensed matter systems.
Furthermore, much progress was recently achieved in the understanding of AFM topological materials \cite{runrundongli,ruiyuyu,kuroda,yshor}, such as the AFM topological insulator (TI) $\mathrm{MnTe_2Bi_4}$ \cite{dongqinzhang,yangong,shuatlee,jliyli,mmotrokovv,mmotrokov,jqyan,jlicwang,RCVidal}, whose topological surface states and the quantized anomalous Hall effect have been experimentally verified. Similar achievements also include AFM topological semimetals (TSMs) \cite{qiwangwang,dfliuliu,qiunanxu,jiaxinyin}. The AFM TIs and TSMs can be regarded as the product of the marriage of two kinds of physics, i.e., the AFM ordering of local moments and the spin-orbit couplings of itinerant electrons.

Our primary interest in this work is the study of  the effects destabilizing the AFM, yielding novel correlated states such as QSLs and, as a consequence, possibly high-$T_c$ superconductors. We remind that it has long been proposed that the QSLs can be essential ingredients in the mechanism of the high-$T_c$ superconductivity \cite{rmkonik,rblaughlinn,kyyangg,dsrokhsarr,gkotliarr,gabrierkot} and could be affecting the pseudogap regime of the underdoped cuprates \cite{paleea}. Thus, the key question we will address here is the attempt of understanding the nature of phase transitions from AFMs to QSLs, which are unconventional ones beyond the Landau's paradigm of symmetry breaking. In fact, envisaging the emergence of QSLs from 2D frustrated quantum magnets and understanding the underlying physical mechanisms of their formation is one of the critical challenges in quantum condensed matter physics.
To achieve this goal, we need to introduce a set of new methodologies. Thus, we present in this article a fundamental theoretical construction, with a particular focus on the quantum anti-ferromagnetism. We also briefly discuss the application of this construction to investigating the topological phase transitions. A detailed study on the this topic is presented by an independent work of ours, i.e., Ref.\cite{ruinew}, which introduces a Chern-Simons mean-field framework that greatly simplifies our understandings of certain topological phase transitions.

A hypothetical transition from the N\'{e}el state to the QSL can take place upon tuning a specific model parameter, $g$, representing a frustration parameter due to competing interactions, or the doping level of the system. As one approaches the transition point around $g=g_{crit}$, the rotational symmetry is restored. As a consequence, bosonic Goldstone modes of the N\'{e}el state are destroyed, and after crossing the transition point, the fractionalized excitations in QSLs are formed. Such drastic changes in the ground state are characterized by deconfinement of the fractional excitations, with an emergent gauge field that enriches the elementary excitations by bringing about the Abelian \cite{dixiao} or non-Abelian \cite{yizhuangyouu} geometric phase. The geometric phases, characterizing the QSLs, are of various kinds and are model-dependent, resulting in different types of emergent excitations in QSLs. We list and highlight the fractionalized excitations for typical QSLs in Table I. Correspondingly, there exist many unconventional phase transitions, characterized by the different natures of QSLs \cite{svisakov,TarunGrover,yanchengwangg,YangQiqi,cenkesubir,EunGookMoon}. The earliest specific examples of gapped spin liquids are the Kalmeyer-Laughlin chiral spin liquids (CSLs), the low-energy effective theory given by the Chern-Simons gauge theory \cite{Kalmeyer,Wena}. Other types of spin liquids with $\mathrm{Z_2}$ invariant gauge group (IGG) were originally proposed and classified in Ref. \cite{xiaogangwenw}, which were shown to be the ground states of some exactly solvable models such as Kitaev's toric code \cite{Kitaeva,Wenb,Kitaevb}.

\begin{table}[b]
\caption{\label{table1}
Emergent elementary excitations in typical types of QSLs.
}
\begin{ruledtabular}
\begin{tabular}{c|c|c}
CSL & \begin{tabular}{c}
$Z_2$ spin liquid\\
(toric code)
\end{tabular}  & \begin{tabular}{c}
U(1) Dirac\\
spin liquid
\end{tabular}   \\
\colrule
\begin{tabular}{c}
\textbf{semions}\\
fermionic spinons
\end{tabular}
&
\begin{tabular}{c}
\textbf{electric excitations(e)}\\
\textbf{magnetic excitations(m)}\\
fermions (f)
\end{tabular}
&\begin{tabular}{c}
\textbf{fermionic} \\
\textbf{spinons}
\end{tabular}
\end{tabular}
\end{ruledtabular}
\end{table}


An early example of unconventional transition proposed in the literature is the deconfinement quantum critical point (DQCP) from a N\'{e}el AFM order to a dimer order \cite{UMASS1,UMASS2,Senthila,levin,Vishwanath,Senthilb,Sandvik,Nahum}. The transitions from N\'{e}el AFM order to $Z_2$ spin liquids were later studied \cite{YangQiqi,Cenkea}. These are possible QSLs in proximity to the magnetically ordered N\'{e}el AFM state on a square lattice \cite{Chatterjee}. This lattice model later was suggested to have potential relevance to the cuprate superconductors \cite{Thomson}. The quantum fluctuations and the low-energy dynamics of the N\'{e}el AFM order in these studies can be described by a $\mathrm{O}(3)$ nonlinear sigma model (NL$\mathbf{\sigma}$M). In turn, the latter is equivalent to the $\mathbb{CP}^1$ field theory containing bosonic spinons field $z_{\alpha}$. The $\mathbb{CP}^1$ gauge field theory was used as a starting ingredient to formulate the quantum field theory describing the unconventional phase transitions, where the ordered phase is obtained via condensation of the elementary excitations. This elegant topological quantum field theory (TQFT) \cite{EdwardWitten,qingruiwang,pengyyee} treatment is, however, not without its downsides. Even though it captures the AFM quantum fluctuations around the quantum critical point (QCP), it misses the quantitative information about the energetics of the original spin Hamiltonian, making it difficult to obtain the whole phase diagram at quantitative level for a given microscopic model.

A more traditional theoretical approach is based on the slave-particle mean-field theories \cite{dyoshioka,fawangg,clkanee,yuanmingluu}, where one introduces parton representations, such as the Schwinger bosons, that only carry the spin degrees of freedom. In this framework, the magnetically ordered states are understood as condensates of Schwinger bosons with finite expectation value $\langle b\rangle\neq0$ of bosonic partons, $b$.
Whereas, the description of QSLs is more elusive because it relies on the prior knowledge of the property of ground states, which implies a particular mean-field ansatz. Nevertheless, one can find out all possible QSLs ansatz that preserves all the symmetries of original spin Hamiltonian based on the projected symmetry group (PSG) construction \cite{xiaogangwenw}, where the space group (SG) of the lattice model is the quotient between PSG and IGG, $\mathrm{SG}=\mathrm{PSG/IGG}$. It should be noted that, in these parton constructions, a single occupation constraint must be enforced on each site. This constraint implicitly generates strong gauge fluctuations that have to be fully taken into account in the theory.  Therefore, for any obtained mean-field ground state, the fluctuations around the mean-field saddle point, which manifest as the emergent gauge fields,  must be analyzed very carefully \cite{xiaogangwenw}. Since the fluctuations around the saddle point is highly dependent on the prior knowledge of the mean-field ansats, this powerful formalism cannot easily be applied to studying the general unconventional phase transitions in a systematic way.

This is also clear by observing that these parton constructions prefer different types of ``physical languages" for describing   the ordered state and the QSLs respectively, for example, the magnetically ordered states are described by the condensation $b$ field while QSLs are depicted by slave particles with emergent gauge degrees of freedom. Thus, it is very difficult, if not unlikely, to achieve a unified mean-field theory for any unconventional phase transitions within this framework.


\begin{figure}[htb]\label{fig1}
\includegraphics[width=\linewidth]{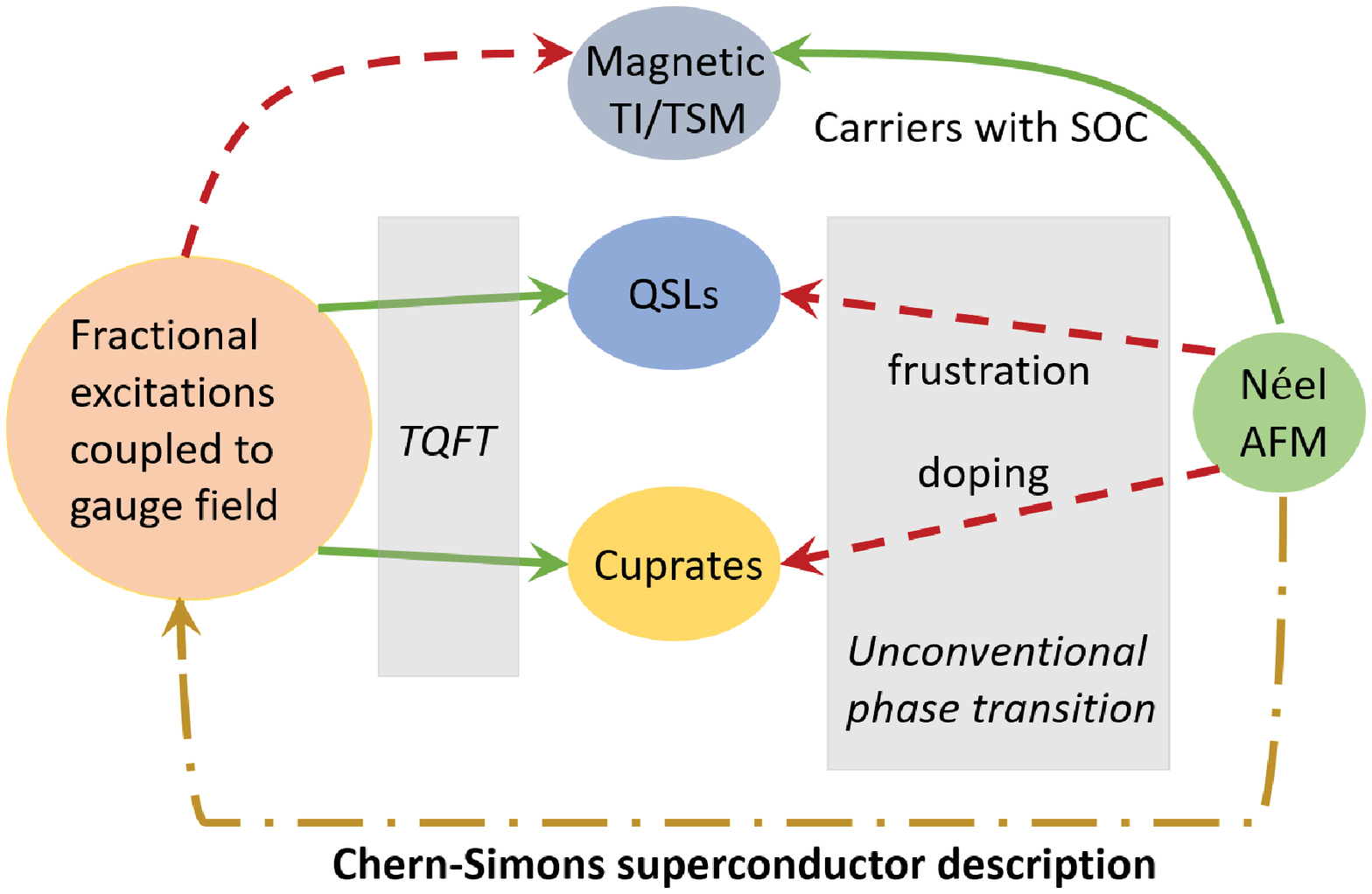}
\caption{(color online) The motivation for a theoretical construction to map the N\'{e}el AFM state to a model of fractional excitations coupled to gauge field. }
\end{figure}
The motivation of the present study is summarized in Fig.1. As shown by the red dashed arrows, it is known that the construction of the theory describing the unconventional phase transitions from the N\'{e}el AFM state to either QSLs or cuprate high-$T_c$ superconductivity is an open problem in modern condensed matter physics. Interestingly, since the gauge field theory of fractional excitations (matter fields) is a possible description of QSLs, and is related to cuprates (see the full green arrows on the left side), we then make a detour first to ask the following question: is there any way to describe the N\'{e}el AFM state also using fractional excitations and gauge field? If this is found, then we obtain a path connecting the two completely different types of phases, as indicated by the yellow dash-dot curve in Fig.1. Interestingly, this will then  naturally leads to a connected route from  the N\'{e}el AFM to QSLs or cuprates high-$T_c$ superconductors by firstly completing the yellow dash-dot path and then following the green full arrows in Fig.1.

This scheme, as long as achieved, can possibly lead to a ``global" mean-field-type theory to describe the evolution of the system from the ordered phase to the QSLs passing through the unconventional phase transitions. We note that this scheme also has potential applications in the field of magnetic topological materials such as the magnetic TIs \cite{dongqinzhang,yangong,shuatlee,jliyli,mmotrokovv,mmotrokov,jqyan,jlicwang,RCVidal} and magnetic TSMs \cite{qiwangwang,dfliuliu,qiunanxu,JianpengLiu,kkuroda,Muhammad,haoyang}. For an electronic system with significant spin-orbit-coupling and antiferromagnetism \cite{JianpengLiu,kkuroda,Muhammad,haoyang}, one can apply the scheme and map the model Hamiltonian to a gauge field theory coupled to the matter fields with two different flavors. One is the deconfined particles from  the local moments, and the other is the spin-orbit-coupled itinerant carriers intrinsic to the material.

In this paper, we firstly
construct a comprehensive theory of the N\'{e}el AFM using a Chern-Simons representation of spins and then discuss a scenario to investigate the unconventional phase transitions. We will briefly mention some examples of unconventional phase transitions \cite{Tigrana,ruia,ruinew,triangular,kagome,MS2,SGK2,SGK1}  that can be examined by the developed method. Specifically, the present work laid theoretical foundations on several aspects as following. (i) We introduce the Chern-Simons representation of spins. Unlike the conventional slave-particle representation, the $\mathrm{U}(1)$ Chern-Simons gauge field here  is explicit in representation itself. Moreover, because the Chern-Simons term gaps out the photons in Maxwell field theory. The photons belong to the high-energy sector compared to the matter fields, thus enables a rigorous treatment of the corresponding gauge fluctuations. (ii) We provide an alternative physical picture of N\'{e}el AFM state in the introduced representation, where the fractionalized excitations form Cooper pairs with $p\pm ip$ symmetry. Since the paired state is induced by the emergent Chern-Simons (CS) gauge field, we term the found superconducting state the CS superconductor \cite{Tigrana}. (iii) We perform detailed calculations about several physical observables, from which we demonstrate the physical correspondence of the N\'{e}el AFM state and the CS superconductor, and (iv) we present a possible advanced scenario to study the unconventional phase transitions induced by frustration.  We also discuss some obtained results on two specific models of frustrated quantum magnets. Interesting topological phase transitions into quantum spin liquids are found to be characterized by the instabilities of the CS superconductor.

\subsection{Summary of results}
We now highlight our results. We firstly introduce a CS representation of spin-half operators and show an exact mapping from the 2D XY models to a fractionalized fermionic theory coupled to a $\mathrm{U}(1)$ Chern-Simons gauge field. Then, we formulate a systematic approach to look for the mean-field ground state of the fermionized model for weak frustration. For generic XY spin models with planar N\'{e}el AFM ground state, either collinear or non-collinear, we show that the ground states are always captured by the superconducting state of the fractional excitations, whose pairing is induced by the $\mathrm{U}(1)$ CS gauge field.
Then, we study the detailed properties of the two states, namely the N\'{e}el AFM state and the CS superconductor. The results enable us to build up quantitative correspondences between the two states' low-energy Goldstone modes, the Higgs modes, and the spin orderings. These calculations validate that the CS superconductors are satisfactory descriptions of N\'{e}el AFMs, independent of whether the ordering is collinear or non-collinear, nor does it relies on the underlying lattice geometry. Last, we discuss the effects of frustration using the language of CS fermions. We show that the frustration is manifested by competing interactions induced by the CS gauge field, driving towards the instability of CS superconductors. The outline of the present work is illustrated by Fig.2.
\begin{figure}[htb]\label{fig2}
\includegraphics[width=\linewidth]{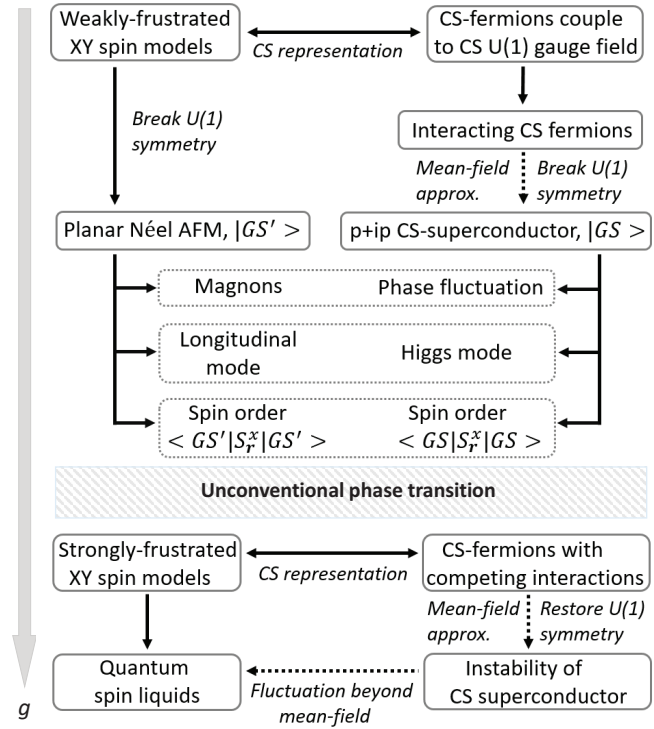}
\caption{Schematic outline of the present work: a weakly frustrated 2D XY spin model leads to the $p\pm ip$ CS superconductor following the method of Section II. Three different physical aspects of the N\'{e}el AFM states and the CS superconductors are studied and compared.
We also provide the outline for the strongly-frustrated cases with possible QSL ground states. The emergence of QSLs and the unconventional phase transition with tuning the frustration parameter $g$ can then be understood, in the mean-field level, as certain types of instabilities of CS superconductors that restore the $\mathrm{U}(1)$ symmetry. In the fermionic picture, the latter is driven by the competing interactions induced by the CS gauge field. With going beyond the mean-field theory, the instabilities can capture the corresponding QSLs state.  }
\end{figure}

The starting point of this work is the spin-half XY models on different 2D lattices, which can lead to planar N\'{e}el AFM ground state for weak frustration. We show results for square, honeycomb, and triangular lattices, while the method can be readily applied to other 2D models with different lattice symmetries. We firstly consider the non-frustrated or weakly-frustrated case where the ground states can be precisely obtained using numerical methods such as density matrix renormalization group (DMRG). As motivated by Fig.1, we aim to find an alternative physical description of planar N\'{e}el antiferromagnetism. This is achieved by introducing CS representation, where a spin-half operator is exactly described by a spinless fermion (termed the CS fermion) coupled to a nonlocal string operator dependent on the fermion density throughout the whole lattice. After the mapping, the string operators can generate a lattice $\mathrm{U}(1)$ gauge field. Because the XY model has $\mathrm{U}(1)$ symmetry with the conservation of total $S^z$, the total number of CS fermions, which is proportional to total $S^z$, is conserved. Moreover, for planar N\'{e}el order, $\langle S^z\rangle=0$ at each site, implying the half-filling condition of CS fermions at each site. As will be shown below, this leads to specific constraints of the CS flux. Given the flux condition and the lattice geometry, one can arrive at enlarged unit cells enclosing different sublattices. This will be discussed in detail below using the square and triangular lattices as examples.

Although we deal with purely 2D systems, there is a similarity with fermionization techniques developed in 1D. It is well known that a 1D transverse Ising model can be transformed into a Kitaev's 1D $p$-wave superconductor model for specific parameters via the Jordan-Wigner transformation \cite{E. Lieb,S. Katsura}. Here, we suggest that the 2D XY planar N\'{e}el AFM's, after the CS fermionization, can be described as stable mean-field ground states where the CS fermions form Cooper pairs with $p+ip$ symmetry, as indicated in Fig.2. This is achieved by firstly integrating out the $\mathrm{U(1)}$ CS gauge field to obtain a low-energy effective fermionic field theory with nonlocal interaction between CS fermions \cite{Tigrana,ruia,triangular}.  The resultant nonlocal interaction generally yields a vertex with $p$-wave symmetry, independent of the underlying lattice symmetries. Within the self-consistent mean-field calculation, we show that the $p$-wave vertex favors chiral $p+ip$ wave pairing order parameters, spontaneously breaking the time-reversal symmetry, in accordance with the antiferromagnetism. The $p+ip$ CS-superconductor belongs to the $\mathrm{DIII}$ class of the 10-fold Altland-Zirnbauer classification, displaying the chiral Majorana edge state at the boundary of the 2D lattices. Similar to the 1D transverse Ising model, the boundary/edge mode found here reflects the bulk topology of the CS superconductors and the $p+ip$ nature of the order parameter. Whereas, after the transformation back to the spin language, the latter becomes a nonlocal one that renders the bulk topology not explicitly observable.

Although the mean-field CS superconductor solution leads to the spin rotational symmetry broken state, the extent to which it  describes the planar N\'{e}el AFM state needs to be answered.
Here we investigate and compare the essential physical properties of the two states and find qualitative and quantitative correspondences. Because both the N\'{e}el AFM and the CS superconductor spontaneously break the continuous $\mathrm{U(1)}$ symmetry, one expects the occurrence of the Goldstone modes as well as the Higgs mode as collective excitations for both states. The Goldstone modes in the N\'{e}el AFM are physically manifested as the magnons, which should be compared with the CS superconductor phase fluctuations. On the other hand, the Higgs mode of the superconductor state, akin to the Higgs particle in high energy physics, originates from the amplitude fluctuations of the order parameter \cite{Varma}. This needs to be compared with the longitudinal mode of the planar N\' {e}el AFM order; the latter is identified as the magnitude fluctuations of the spin order parameter \cite{Barankov}. The longitudinal mode is well-defined and immune to dissipation into the Goldstone modes in XY antiferromagnets because of the presence of particle-hole symmetry, which in turn results in an effective Ginzberg-Landau field theory with Lorentz symmetry \cite{David Pekker} that forbids the mixing of transverse and longitudinal modes.
As shown in Fig.3, we compare the collective modes of the two states on different lattices. Remarkably good quantitative agreements are obtained, suggesting the physical similarities of the two states, as indicated by the dashed boxes in Fig.2.

Let us denote mean-field ground state of the CS superconductor  by $|GS\rangle$ while the planar N\'{e}el AFM state by $|GS^{\prime}\rangle$ (see Fig.2). To compare these two states, one can evaluate the spin expectation value $\langle GS|\hat{S}^x_{\mathbf{r}}|GS\rangle$ for the CS superconductor. Direct correspondence with the N\'{e}el AFM can be revealed if $\langle GS|\hat{S}^x_{\mathbf{r}}|GS\rangle$ exhibits alternating finite spin polarization for different sublattices within in a unit cell.
 To this end, we show that the boundary condition of the spin model plays a key role in the fermionic language, which, in the thermodynamic  limit, brings about the doubly-degenerate Bogoliubov vacuum states with even and odd fermion parity (FP) respectively. This inevitably makes the direct calculation of $\langle GS|\hat{S}^x_{\mathbf{r}}|GS\rangle$ unlikely because the ground state in thermodynamic limit can be an arbitrary superposition of the doubly-degenerate states. Therefore, we study the local response of the CS superconductor to an infinitesimal local magnetic field, $B$. We show that the CS superconductor has a divergent magnetic susceptibility at $B\rightarrow0$ in thermodynamic limit, as long as the external perturbation field is asymmetric on different sublattices. The two-fold degeneracy of the ground state is thus broken by infinitesimal $B$, generating the alternating spin polarization with respect to the CS superconductor description, strongly suggesting their physical correspondences to the planar N\'{e}el AFM orders.

\begin{figure}[htb]\label{fig3}
\includegraphics[width=\linewidth]{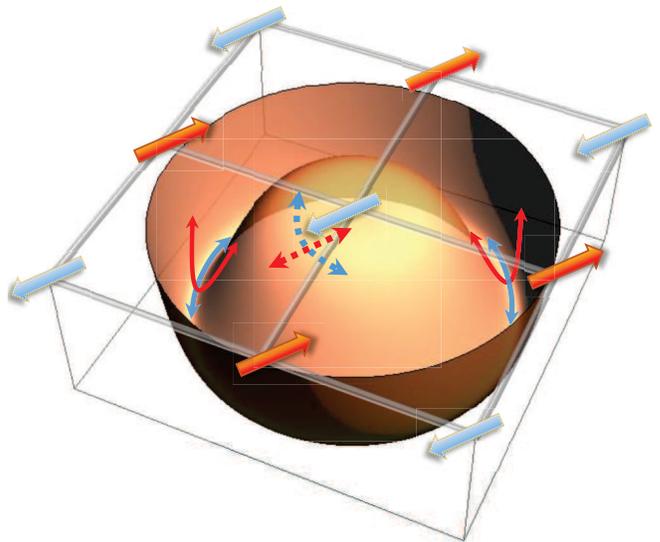}
\caption{(color online) The collective modes in CS superconductors and the planar N\'{e}el AFM state. A lattice is shown with N\'{e}el spin order on A and B sublattice. The magnons and longitudinal modes are displayed by the light (blue), and dark (red) dashed arrows. The ``Mexican hat" represents the free energy potential of the symmetry broken CS superconductors. The phase fluctuation and the Higgs mode are indicated by the full light (blue) and dark (red) arrows, respectively.
}
\end{figure}

The above-mentioned results complete the first part of the paper outlined in Fig.2, i.e., the description of the stable ground state of the weakly frustrated XY models. Then, we consider the effect of frustration by including further neighboring exchange couplings on the XY models. In the fermionic language, the frustration introduces further neighbor hoppings of CS fermions as well as a mediating CS $\mathrm{U}(1)$ gauge field. Following the same approach used to study the weakly-frustrated cases, the $\mathrm{U}(1)$ gauge field now induces additional interactions except for the one that is responsible for the CS superconductivity \cite{ruia}. The advantage of the formalism then becomes obvious. The spin model with tunable frustration $g$ is now transformed into a unified form described by interacting CS fermions, allowing us to perform analysis by the many-body techniques for fermions. We show that, for the strongly-frustrated cases, the instabilities of the CS superconductors are very typical phenomena due to the competition between ordering and fluctuation.
Moreover, we point out that certain types of instabilities can restore the $\mathrm{U}(1)$ without breaking further symmetries. It is natural to expect them to serve as the mean-field signals for the unconventional phase transitions towards QSLs. The nature of the resultant state can be further understood in a controlled way by going beyond the mean-field theory and restoring the fluctuations of order parameters.

The above procedure realizes the two-step scheme illustrated by Fig.1, i.e., firstly, building the connection between the N\'{e}el AFM and the fractional excitations with the gauge field via the CS superconductor description, and secondly, interpreting the formation of QSLs via the topological quantum field theories of fractionalized excitations.

Here we would to highlight an independent recent  work of ours,  i.e., Ref. \onlinecite{ruinew}, where, based on the fermionization approach here, we  work out a Chern-Simons mean-field theory for analyzing and predicting topological phase transitions in frustrated quantum magnets. By using the proposed method, we are above to predict an interesting topological phase transition from
the N\'{e}el AFM state to a non-uniform chiral spin liquid on a honeycomb XY model. Remarkably, such a novel transition, which is widely known to beyond the Landau's paradigm of symmetry breaking, is unambiguously shown to still enjoy an unprecedented  mean-field description. Another related example is the deconfinement transition from the $120^\circ$ state  on triangular lattice to a new helical spin liquid state, proposed by Ref. \onlinecite{triangular}.

The remaining part of this paper is organized as follows. In Sec IIA, we introduce the CS fermionization of a generic 2D XY model. Then, we formulate a systematic framework to determine the CS superconductor mean-field ground state of the fermionized model in Sec IIB. It is applied to three different lattice models, giving rise to corresponding mean-field theories discussed in Sec.IIC. In Sec.III, we move to the topic of collective modes of the CS superconductors. In Sec.IIIA, we present the study of the Higgs mode in detail by considering the leading Feynman diagram that restores the broken $\mathrm{U}(1)$ symmetry beyond Hartree-Fock mean-field level. The dispersion of the Higgs mode of CS superconductors is obtained from the Bethe-Salpeter equations, which we solve analytically.  In Sec.IIIB, the longitudinal mode from the planar N\'{e}el AFM is calculated following Feynman's conjecture, which was originally proposed to study the excitations in $\mathrm{He}^3$ superfluid \cite{rpfeyman}. The longitudinal mode is reduced to the evaluation of two spin-spin correlation functions, which we precisely obtained using DMRG. Remarkably well correspondences of the collective modes are obtained at a quantitative level. The calculation of the spin ordering in Sec.IV is  further divided into two subsections. Sec.IVA focuses on the generalization of the fermionization method to the case with periodic boundary conditions on a torus. This is a necessary step as we show that the boundaries lead to the doubly degenerate CS superconductor ground state with even and odd FP in the thermodynamics limit. In Sec.IVB, we then study the spin order from the CS superconductors. In Sec.V, we investigate the strongly-frustrated XY models by extending the CS superconductor mean-field theory to account for frustration.  A detailed discussion of  unconventional phase transition is presented based on two specific models. In Sec.VI, we give a summary and provide more discussions, with the emphasize on further applications of the proposed theory to related topics including the QSLs, doped AFM and superconductivity, exactly solvable models of topologically ordered states, as well as the finite temperature formalism where the Kosterliz-Thouless transition can play an important role.


\section{Chern-Simons superconductor description of a 2D planar N\'{e}el order}

\subsection{CS fermionization}
We begin with the fermionization of the 2D XY spin exchange model on a bipartite lattice and consider the weak frustration case with a planar N\'{e}el AFM ground state. The Hamiltonian under study has the following form,
\begin{equation}\label{eq1}
 H_{XY}=\sum_{\mathbf{r},\mathbf{r}^{\prime}}J_{\mathbf{r},\mathbf{r}^{\prime}}(S^x_{\mathbf{r}}S^x_{\mathbf{r}^{\prime}}+S^y_{\mathbf{r}}S^y_{\mathbf{r}^{\prime}}),
\end{equation}
where $J_{\mathbf{r},\mathbf{r}^{\prime}}>0$ is the local antiferromagnetic exchange interaction between neighboring sites. Generalization of the results to more complicated spin models also including the Ising term are possible as discussed in the closing section. We however use the Hamiltonian Eq.\eqref{eq1} as the typical example supporting a planar N\'{e}el AFM state to demonstrate a fermion description of the latter. Such a description will help one to obtain more insights into the physics of unconventional phase transitions. In the first part of this work, we consider the lattice to be free from geometric frustration, and consider only the nearest exchange couplings $J_{\mathbf{r},\mathbf{r}^{\prime}}=J\delta_{\mathbf{r},\mathbf{r}+\mathbf{e}_j}$ with $\mathbf{e}_j$ the nearest vector bonds on the lattice.

We now introduce the CS fermionization of spin-$1/2$ operators,
\begin{equation}\label{eq2}
 S^{\pm}_{\mathbf{r}}=f^{\pm}_{\mathbf{r}}U^{\pm}_{\mathbf{r}},
\end{equation}
where the spinless CS fermions $f^{\pm}_{\mathbf{r}}$ are attached to a string operator defined as
\begin{equation}\label{eq3}
 U^{\pm}_{\mathbf{r}}=e^{\pm ie\sum_{\mathbf{r}^{\prime}\neq \mathbf{r}}\mathrm{arg}(\mathbf{r}-\mathbf{r}^{\prime})n_{\mathbf{r}^{\prime}}}.
\end{equation}
Here $e$ is the CS charge. It can take odd integer values to guarantee that the $\mathrm{SU}(2)$ algebra of the spin-half operators is preserved. The string operator is nonlocal in the sense that it includes a sum of particle number operators $n_{\mathbf{r}}=f^{\dagger}_{\mathbf{r}}f_{\mathbf{r}}=S^z_{\mathbf{r}}+1/2$ throughout the whole lattice. Its non-locality potentially facilitates the study of topologically ordered states, characterized by long-range quantum entanglement. Compared to the Schwinger particle representation, the above representation does not artificially enlarge the local Hilbert space and therefore is free from additional constraints.
Inserting Eq.\eqref{eq2},\eqref{eq3} into Eq.\eqref{eq1}, we obtain
\begin{equation}\label{eq4}
 H=\sum_{\mathbf{r},\mathbf{r}^{\prime}}J_{\mathbf{r},\mathbf{r}^{\prime}}(f^{\dagger}_{\mathbf{r}}e^{ieA_{\mathbf{r},\mathbf{r}^{\prime}}}f_{\mathbf{r}^{\prime}}+h.c.),
\end{equation}
where a factor $1/2$ has been absorbed into $J_{\mathbf{r},\mathbf{r}^{\prime}}$. $A_{\mathbf{r},\mathbf{r}^{\prime}}$ is the $\mathrm{U}(1)$ gauge field generated by the string operators, i.e., $A_{\mathbf{r},\mathbf{r}^{\prime}}=U^+_{\mathbf{r}}U^-_{\mathbf{r}^{\prime}}=\sum_{\tilde{\mathbf{r}}\neq\mathbf{r}}\mathrm{arg}(\mathbf{r}-\tilde{\mathbf{r}})n_{\tilde{\mathbf{r}}}
-\sum_{\tilde{\mathbf{r}}\neq\mathbf{r}^{\prime}}\mathrm{arg}(\mathbf{r}^{\prime}-\tilde{\mathbf{r}})n_{\tilde{\mathbf{r}}}$. Here the CS charge $e$ appears in front of $A_{\mathbf{r},\mathbf{r}^{\prime}}$, which can take all odd integers.

To obtain more intuition about the $\mathrm{U}(1)$ gauge field, here we transform it into the continuum form. Considering $A_{\mathbf{r},\mathbf{r}^{\prime}}$ defined on bonds connecting two nearest neighbor sites, we define its continuous counterpart as following. Due to translation invariance, $A_{\mathbf{r},\mathbf{r}^{\prime}}=A_{\mathbf{r}-\mathbf{r}^{\prime}}$, and the Taylor expansion near $\mathbf{r}$ gives $A_{\mathbf{r},\mathbf{r}^{\prime}}=\mathbf{A}_\mathbf{r}\cdot(\mathbf{r}-\mathbf{r}^{\prime})$, with the argument approaching $\mathbf{r}-\mathbf{r}^{\prime}\rightarrow0$. Therefore, the $\mathrm{U}(1)$ lattice gauge field is cast in the continuous form into a local vector potential $\mathbf{A}_\mathbf{r}$. By taking the derivative with respect to $A_{\mathbf{r},\mathbf{r}^{\prime}}$, one obtains that
\begin{equation}\label{eq5}
 \mathbf{A}_{\mathbf{r}}=\sum_{\mathbf{r}^{\prime}\neq\mathbf{r}}\frac{\mathbf{e}_z\times(\mathbf{r}-\mathbf{r}^{\prime})}{|\mathbf{r}-\mathbf{r}^{\prime}|^2}n_{\mathbf{r}^{\prime}}.
\end{equation}
In analogy with the vector potential of electrodynamics, $\mathbf{A}_{\mathbf{r}}$ from the quantum magnet generates a gauge flux in a closed contour centered at a generic site, say $\mathbf{r}_0$. Namely, $B_{\mathbf{r}_0}=\oint_{\mathbf{r}_0}\mathbf{A}_{\mathbf{r}}\cdot d\mathbf{r}=\sum_{\mathbf{r}^{\prime}\neq\mathbf{r}}n_{\mathbf{r}^{\prime}}\oint_{\mathbf{r}_0} \frac{\mathbf{e}_z\times(\mathbf{r}-\mathbf{r}^{\prime})}{|\mathbf{r}-\mathbf{r}^{\prime}|^2}d\mathbf{r}$. After introducing the complex coordinates $z=x+iy$ and $z^{\prime}=x^{\prime}+iy^{\prime}$ to represent $\mathbf{r}=(x,y)$ and $\mathbf{r}^{\prime}=(x^{\prime},y^{\prime})$ respectively, the complex integral is easily calculated by counting the residuals enclosed by the contour in the complex plane, leading to the fact that $-i\oint_{z_0} dz\frac{(z-z^{\prime})^{\star}}{|z-z^{\prime}|^2}$ equals to $2\pi$ for $z^{\prime}$ enclosed by the contour and equals to $0$ otherwise. Therefore, the flux reads as $B_{\mathbf{r}_0}=2\pi n_{\mathbf{r}_0}$, where $n_{\mathbf{r}_0}$ is the number operator of f-fermions enclosed by the contour centered at $\mathbf{r}_0$.
From above, we see that the Gauss law  $B_{\mathbf{r}}=2\pi n_{\mathbf{r}}$ is  an essential requirement in the fermionization approach.

To enforce the flux rule $B_{\mathbf{r}}=2\pi n_{\mathbf{r}}$, we introduce in the functional representation the term, $\int d\mathbf{r}A^0_{\mathbf{r}}(\frac{B_{\mathbf{r}}}{2\pi}-n_{\mathbf{r}})$, where $A^0_{\mathbf{r}}$ is the Lagrangian multiplier field and it enters into the functional integral measure of the partition function. We see from above that, the flux rule leads to an effective chemical potential for the f-fermions $\int d\mathbf{r}A^0_{\mathbf{r}}n_{\mathbf{r}}$, as well as another term $\frac{1}{2\pi}\int d\mathbf{r}A^0_{\mathbf{r}}B_{\mathbf{r}}=\frac{1}{4\pi}\int d\mathbf{r}\epsilon_{ij} A^0_{\mathbf{r}}\partial_j A^i_{\mathbf{r}}$, topologically equivalent to a CS action up to boundary term. Therefore, once the fermionization is performed to represent the spin
one-half operators, the CS fermions derived from the 2D XY
models are automatically coupled to a CS $\mathrm{U}(1)$ gauge field, whose action is obtained as
\begin{equation}\label{eq6}
\begin{split}
 S_{XY}&=\int dt\sum_{\mathbf{r}}\hat{f}^{\dagger}_{\mathbf{r}}(i\partial_t-A^0_{\mathbf{r}})\hat{f}_{\mathbf{r}}-\sum_{\mathbf{r},\mathbf{r}^{\prime}}t(\hat{f}^{\dagger}_{\mathbf{r}}e^{ieA_{\mathbf{r},\mathbf{r}^{\prime}}}\hat{f}_{\mathbf{r}^{\prime}}+h.c.)\\
 &-\frac{1}{2\pi}\int dt\sum_{\mathbf{r}}B_{\mathbf{r}}A^0_{\mathbf{r}},\
\end{split}
\end{equation}
This is the fermionized action describing the XY model Eq.\eqref{eq1}. The mapping via CS fermionization is exact to this step. We note in passing that, the CS $\mathrm{U}(1)$ gauge field and CS action derived here has an intricate connection with the topological Hopf term of the $\mathbb{CP}^1$ theory \cite{Dzyaloshinskii,fradkinstone} describing the quantum fluctuations of N\'{e}el AFM order. The study of the connection between the two independent theories should be an interesting topic worth further investigations.

\subsection{Reformulation as a theory of interacting CS fermions}
To understand the ground state properties of the system described by the fermionized action Eq.\eqref{eq6},  we propose the following key steps.
\begin{itemize}
 \item Setting free the CS fermions by ``turning off" the gauge field, and obtaining the locations of the energy minima of the free model in momentum space, $\mathbf{Q}_i$, $i=1,...,N$. $N$ counts the degeneracy of the energy minima.
 \item Attaching an non-fluctuating $\mathrm{U}(1)$ gauge field $\overline{A}_{\mathbf{r},\mathbf{r}^{\prime}}$ satisfying $\overline{A}_{\mathbf{r}_1,\mathbf{r}_2}+\overline{A}_{\mathbf{r}_2,\mathbf{r}_3}+...\overline{A}_{\mathbf{r}_N,\mathbf{r}_1}=2\pi \langle n_{lp}\rangle$, where $\langle n_{lp}\rangle$ is the number of fermions enclosed by a generic closed loop on the lattice. For loop enclosing the plaquette with $N_{pl}$ lattice sites, this implies that $\overline{A}_{\mathbf{r}_1,\mathbf{r}_2}+\overline{A}_{\mathbf{r}_2,\mathbf{r}_3}+...\overline{A}_{\mathbf{r}_N,\mathbf{r}_1}=B=2\pi\nu N_{pl}=\pi N_{pl}$ for half-filling $\nu=1/2$. Note that the total $S_z$ should be zero for the ground state of XY models, therefore half-filling is required in the fermion picture.
 \item Making expansion of the fermionic theory near the energy minima with taking into account the non-fluctuating $\mathrm{U}(1)$ gauge field.
 \item Restoring the fluctuation of the CS gauge field and integrating out the gauge field fluctuation to obtain a low-energy effective field theory with interacting CS fermions.
\end{itemize}
We will demonstrate these steps in the following sections.
\subsubsection{Identification of energy minima.}

As the first step, we intentionally turn off the gauge field $A_{\mathbf{r},\mathbf{r}^{\prime}}=0$,  then Eq.\eqref{eq6} can be readily diagonalized leading to the single-particle dispersion of the CS fermions. One can then find out the energy minima of the CS fermion spectrum in the first Brillouin zone (BZ). We denote the lattice momentum of the minima as $\mathbf{Q}_i$ and $i=1,...,N$. Some essential information of the ground state can be found from $\mathbf{Q}_i$. With $A_{\mathbf{r},\mathbf{r}^{\prime}}=0$, the dispersion obtained from Eq.\eqref{eq4} is the exactly the same as the single-particle spectrum under the hardcore boson representation. Therefore, the energy minima suggest the $\mathbf{k}$-point where the bosons intend to condense, indicating the nesting vector of the magnetic orders. Besides, the spin structure factor, $S(\mathbf{q})=\int d\mathbf{r}d\mathbf{r}^{\prime}\langle \hat{S}^a(\mathbf{r})\hat{S}^a(\mathbf{r}^{\prime})\rangle e^{i\mathbf{q}\cdot(\mathbf{r}-\mathbf{r}^{\prime})}$ should display peak at these points. For the case with $N>1$, the hardcore bosons condensation will spontaneously take place at only one of the degenerate points $\mathbf{Q}_i$. The degeneracy of $\mathbf{Q}_i$ is essentially the result of the underlying lattice space group. Therefore, the condensation at one of the minima indicates the spontaneous breaking of certain symmetry element of the space group, in addition to the $\mathrm{U}(1)$ symmetry of the XY model.

\subsubsection{Attaching a non-fluctuating $\mathrm{U(1)}$ gauge field.}

The free CS fermionic spectrum obtained in the previous step does not represent the system's correct excitations because the $\mathrm{SU(2)}$ algebra of the spin-half operators is lost upon disregarding the CS gauge field. To restore the $\mathrm{U(1)}$ gauge field, we firstly decompose the lattice $\mathrm{U(1)}$ phase $A_{\mathbf{r},\mathbf{r}^{\prime}}$ into a sum of non-fluctuating and fluctuating parts: $A_{\mathbf{r},\mathbf{r}^{\prime}}=\overline{A}_{\mathbf{r},\mathbf{r}^{\prime}}+\delta A_{\mathbf{r},\mathbf{r}^{\prime}}$.
As we have demonstrated before, the CS fermionization requires the Gauss' law $B_{\mathbf{r}_0}=\oint_{\mathbf{r}_0}\mathbf{A}_{\mathbf{r}}\cdot d\mathbf{r}=2\pi n_{\mathbf{r}_0}$, which on a lattice, reads as $A_{\mathbf{r}_1,\mathbf{r}_2}+A_{\mathbf{r}_2,\mathbf{r}_3}+...A_{\mathbf{r}_N,\mathbf{r}_1}=2\pi n_{lp}$.
At the mean-field level, one then requires that \cite{footnotes1}
\begin{equation}\label{eq7}
 \overline{A}_{\mathbf{r}_1,\mathbf{r}_2}+\overline{A}_{\mathbf{r}_2,\mathbf{r}_3}+...\overline{A}_{\mathbf{r}_N,\mathbf{r}_1}=2\pi \langle n_{lp}\rangle,
\end{equation}
where $\langle n_{lp}\rangle$ is the ground state expectation of $n_{lp}$. Note that $n_{lp}$ denotes the total number of fermions shared by all the sites enclosed by the loop \cite{footnotes2}.

Equivalent to Eq.\eqref{eq7}, we in fact require
\begin{equation}\label{eq8}
 \langle\delta A_{\mathbf{r}_1,\mathbf{r}_2}+\delta A_{\mathbf{r}_2,\mathbf{r}_3}+...\delta A_{\mathbf{r}_N,\mathbf{r}_1}\rangle=0.
\end{equation}
In other words, the gauge field's fluctuation does not change the half-filling of CS fermions, which is in accordance with a planar magnetic order.

 Eq.\eqref{eq4} is then cast into
\begin{equation}\label{eq9}
\overline{H}_{XY}=t\sum_{\mathbf{r},\mathbf{r}^{\prime}}[\hat{f}^{\dagger}_{\mathbf{r}}e^{ie\overline{A}_{\mathbf{r},\mathbf{r}^{\prime}}}e^{ie\delta A_{\mathbf{r},\mathbf{r}^{\prime}}}f_{\mathbf{r}^{\prime}}+h.c.].
\end{equation}
With neglecting the fluctuating field, $\delta A_{\mathbf{r},\mathbf{r}^{\prime}}=0$, we arrive at an approximated Hamiltonian $\overline{H}_{XY}$ describing the CS fermions decorated by $\mathrm{U(1)}$ phases. This is, of course, a better approximation than the free CS fermions in the last step.
To enforce the flux rule in Eq.\eqref{eq7}, we note that there are in general three different cases depending on the underlying lattice geometry, more specifically, on the number of lattice sites $N_{pl}$ in a unit plaquette: (a) $N_{pl}=1$ (e.g., the square lattice, Fig.4(a)), (b) $N_{pl}>1$ (e.g., the unit hexagon contains two lattice sites on the honeycomb lattice, Fig.4(b)), and (c) $N_{pl}<1$ (e.g. the unit triangular consists of $1/2$ site on the triangular lattice, Fig.4(c)).
In case (a), one arrives at the $\pi$-flux state under half-filling, according to Eq.\eqref{eq7}. Namely, the flux through each plaquette equals to either $\pi$ or $-\pi$. For lattices belonging to case (b) with even number of $N_{lp}$, the flux can be found to be $B=0$ module $2\pi$ for all plaquettes, therefore $\overline{A}_{\mathbf{r},\mathbf{r}^{\prime}}$ can be gauged out. For case (c), the $\pi$-fluxes must be distributed in an area that encloses $1/N_{pl}$ plaquettes. We will show for the triangular lattice case that, if there exists several distributions that are energetically degenerate, the system will spontaneously break the degeneracies by taking a generic flux configuration. In the hardcore boson representation, this corresponds to the condensation of bosons at a generic $\mathbf{Q}_i$, as discussed above.
In cases (a) and (c), the generated flux patterns will inevitably enlarge the unit cell compared to that of the lattice. This is consistent with the planar N\'{e}el AFM ground state where the formation of alternating spin order enlarges the lattice unit cell. We show in Fig.4(a)-(c) the gauge fluxes on the square, honeycomb, and triangular lattice as typical representatives for the above three cases.

\begin{figure}[thb]\label{fig4}
\includegraphics[width=\linewidth]{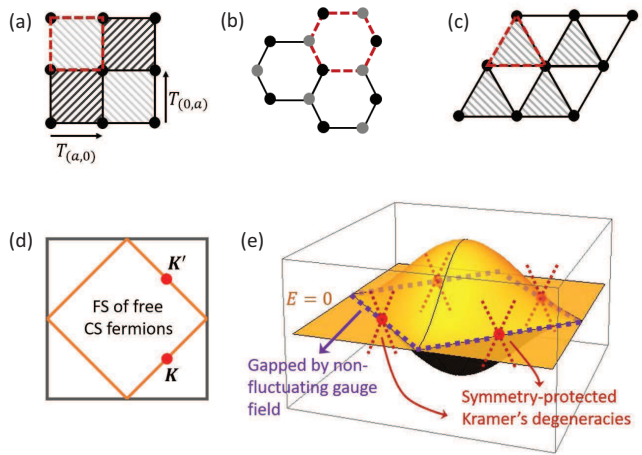}
\caption{(color online) The non-fluctuating flux distribution is shown on (a) square, (b) honeycomb, and (c) triangular lattices, respectively. The red dashed boundaries denote the unit plaquette for the three lattices. The two regions with opposite shadings represent the $\pi$ and $-\pi$ fluxes. The unshaded area represents the zero-flux region. On the triangular lattice, two flux states related by $Z_2$-symmetry are degenerate in energy, corresponding to two different types of hardcore boson condensation at $\mathbf{Q}_1$ and $\mathbf{Q}_2$, respectively\cite{triangular}. We show one of the flux states in (c), while the other is obtained by an interchange of the shaded and unshaded area. (d) The Fermi surface of free CS fermions on a square lattice, which becomes a loop degeneracy after the BZ-folding for infinitesimal $\overline{A}_{\mathbf{r},\mathbf{r}^{\prime}}$, i.e., the purple dashed lines in (e). The degenerate loop is gapped out due to the finite $\overline{A}_{\mathbf{r},\mathbf{r}^{\prime}}$, except for four Kramer's degenerate nodes (two inequivalent ones, $\mathbf{K}$ and $\mathbf{K}^{\prime}$ in (d)), as indicated by the red points, which are enforced by remaining symmetries of $\overline{H}_{XY}$. }
\end{figure}

\subsubsection{Low-energy effective theory in the presence of non-fluctuating gauge field.}
In the last step, we have obtained an approximated model $\overline{H}_{XY}$, i.e., Eq.\eqref{eq9} with $\delta A_{\mathbf{r},\mathbf{r}^{\prime}}=0$. To restore the fluctuation of gauge field, we have to consider nonzero $\delta A_{\mathbf{r},\mathbf{r}^{\prime}}$. The  CS fermions energy spectrum ( $\delta A_{\mathbf{r},\mathbf{r}^{\prime}}=0$) can be readily obtained from $\overline{H}_{XY}$. After insertion of the non-fluctuating flux state, as illustrated in Fig.4(a)-(c), the sublattice degrees of freedom emerge and the CS fermions $\hat{f}_{\mathbf{k},\alpha}$ ($\hat{f}^{\dagger}_{\mathbf{k},\alpha}$) carry the sublattice index  $\alpha$. Then, one can formally obtain the Hamiltonian in the diagonalized basis as,
\begin{equation}\label{eq10}
\overline{H}_{XY}=\sum_{\mathbf{k},\alpha,\beta}\hat{f}^{\dagger}_{\mathbf{k},a}\epsilon_{\alpha,\beta}(\mathbf{k})\hat{f}_{\mathbf{k},b}=\sum_{\mathbf{k},n}\hat{f}^{\dagger}_{\mathbf{k},n}\epsilon_{n}(\mathbf{k})\hat{f}_{\mathbf{k},n},
\end{equation}
where $\epsilon_{\alpha,\beta}$ denotes the single-particle Hamiltonian of CS fermions in the sublattice space, $n$ represents the band index, and $\epsilon_n(\mathbf{k})$ is the energy spectrum of CS fermions for different bands, $n$. The non-fluctuating CS gauge field $\overline{A}_{\mathbf{r},\mathbf{r}^{\prime}}$ and the caused fluxes modulate the energy spectrum and whose effects are implicit in $\epsilon_n(\mathbf{k})$. It should be noted that the CS charge $e$ in front of $\overline{A}_{\mathbf{r},\mathbf{r}^{\prime}}$ is also implicit in $\epsilon_n(\mathbf{k})$, and can take any odd integer values due to the compactness of gauge field theory on the lattice.

In specific models, as we will show below, the gapless Dirac nodes will generally occur. Before proceeding, we illustrate the fact that the Dirac nodes are enforced by symmetries of the $\overline{H}_{XY}$ using the square lattice model as an example. We assume that one starts from the free CS fermion model and gradually turns on the non-fluctuating gauge field $\overline{A}_{\mathbf{r},\mathbf{r}^{\prime}}$, namely, the fluxes in each square plaquette is gradually increased from zero in Fig.4(a). For $\overline{A}_{\mathbf{r},\mathbf{r}^{\prime}}=0$, the free fermion model enjoys a square Fermi surface (FS) at half-filling indicated by Fig.4(d).
 For an infinitesimal phase, $\overline{A}_{\mathbf{r},\mathbf{r}^{\prime}}=0^+$, as indicated by the alternating shaded regions in Fig.4(a), the system already develops different sublattices, leading to BZ folding. This generates a degenerate energy contour along the FS, as shown by the dashed blue lines in Fig.4(e). With further increasing $\overline{A}_{\mathbf{r},\mathbf{r}^{\prime}}$, the larger gauge field introduces stronger couplings between CS fermions on different sublattices, reshaping the energy spectrum of the CS fermions and gapping out the degenerate contour as indicated by Fig.4(e). Finally, $\overline{A}_{\mathbf{r},\mathbf{r}^{\prime}}$ is increased to the value such that the $\pi$-flux rule is satisfied, i.e., the flux in each of the square plaquette becomes $\pm\pi$. The question is: will the degenerate contour be completely gapped out or some gapless nodes remain at certain $\mathbf{k}$-points?

The answer is only dependent on the symmetries of Eq.\eqref{eq10}.  On the square lattice, with staggered $\pm\pi$ flux as shown in Fig.4(a), we can construct the united symmetric operations using the TRS operator $\Theta$ and the translation operators, $T_{(a,0)}$ and $T_{(0,a)}$, i.e., $U_1=\Theta T_{(a,0)}$ and $U_2=\Theta T_{(0,a)}$. It is clear that the Bloch wave function $u(\mathbf{k})$ satisfies $U^2_{1/2}u(\mathbf{k})=e^{2ik_{x/y}}u(\mathbf{k})$, such that $U^2_{1/2}$ is anti-unitary for $\mathbf{k}=(\pm\pi/2,k_y)$ or $\mathbf{k}=(k_x,\pm\pi/2)$. Then, Kramers degeneracies can be identified at $\mathbf{k}=(\pm\pi/2,\pm\pi/2)$, which are the only TRS invariant points along the lines $\mathbf{k}=(\pm\pi/2,k_y)$ or $\mathbf{k}=(k_x,\pm\pi/2)$, as indicated by the red dots in Fig.4(e). This example on the square lattice indicates the general existence of Dirac nodes from $\overline{H}_{XY}$, enforced by its symmetry. On the other lattices with a certain distribution of fluxes, one can construct corresponding united operators $U_{i}=\Theta T_{k_i\mathbf{a}_i}$, where $k_i$ with $i=1,2$ are integers depending on the fluxes distribution and the underlying lattice symmetry. Therefore, the similar symmetry analysis would suggest robust gapless touchings between conduction and valence CS fermions. We denote the location of these Dirac nodes as $\mathbf{K}_i$, $i=1,2...$, in the following.

At half-filling, as in the XY antiferromagnets, the FS of CS fermions exactly passes through the Dirac nodes. One can therefore safely make expansion of $\overline{H}_{XY}$ with respect to the lattice momentum, arriving at the low-energy effective theory near $\mathbf{K}_i$, i.e.,
\begin{equation}\label{eq11}
  \overline{H}_{XY}=\sum_{\mathbf{r}}v_F\hat{f}^{\dagger}_{\mathbf{r},\alpha}(\mathbf{K}_1)\sigma^i_{\alpha\beta}(-i\boldsymbol{\nabla}^i)\hat{f}_{\mathbf{r},\beta}(\mathbf{K}_1)+...
\end{equation}
where $\mathbf{k}^i=-i\boldsymbol{\nabla}^i$  is measured from the Dirac point $\mathbf{K}_1$, the ellipsis denotes the  terms expanded near other nodes. $\sigma$ is the Pauli matrix denoting the sublattice degrees of freedom.  $v_F$ is the Fermi velocity derived near the Dirac nodes.
It is proportional to exchange coupling $ J $ of the spin model Eq.\eqref{eq1}, but can generally be anisotropic and have different values associating with different Dirac nodes.

\subsubsection{Fluctuating gauge field.}
Now we are ready to restore the fluctuating gauge field $\delta A_{\mathbf{r},\mathbf{r}^{\prime}}$. In the low-energy effective theory Eq.\eqref{eq11}, $\delta A_{\mathbf{r},\mathbf{r}^{\prime}}$ is minimally coupled to the CS fermions, leading to
\begin{equation}\label{eq12}
  H_{XY}=\sum_{\mathbf{r}}v^i_F\hat{f}^{\dagger}_{\mathbf{r},\alpha}(\mathbf{K}_1)\sigma^i_{\alpha\beta}(-i\boldsymbol{\nabla}^i+e\delta A_{\mathbf{r}})\hat{f}_{\mathbf{r},\beta}(\mathbf{K}_1)+...
\end{equation}
where, according to Sec.IIA, one has introduced the continuum form, i.e., for $\mathbf{r}-\mathbf{r}^{\prime}\rightarrow0$, $\delta A_{\mathbf{r}-\mathbf{r}^{\prime}}\equiv\delta A_{\mathbf{r}}\cdot(\mathbf{r}-\mathbf{r}^{\prime})$. Taking into account the CS term and the Lagrangian multiplier in Eq.\eqref{eq6}, we obtain the action of the low-energy effective gauge field theory, that is a good approximation to capture the ground state of the spin exchange model, i.e.,
\begin{equation}\label{eq13}
\begin{split}
  S_{XY}&=\int d\mathbf{r}dt\hat{f}^{\dagger}_{\mathbf{r},\alpha}(\mathbf{K}_i)\sigma^0_{\alpha\beta}(i\partial_0-A^0_{\mathbf{r}})\hat{f}_{\mathbf{r},\beta}(\mathbf{K}_i)\\
  &-\int d\mathbf{r}dt\hat{f}^{\dagger}_{\mathbf{r},\alpha}(\mathbf{K}_i)\sigma^i_{\alpha\beta}(-i\partial_i+eA_{\mathbf{r}})\hat{f}_{\mathbf{r},\beta}(\mathbf{K}_i)+S_{CS},
\end{split}
\end{equation}
where we used the notation $A_{\mathbf{r}}$ instead of $\delta A_{\mathbf{r}}$ for brevity. The CS term $S_{CS}$ originates from the flux rule, inherited from the last term in Eq.\eqref{eq6}. The derivation of Eq.\eqref{eq13} from the XY Hamiltonian Eq.\eqref{eq1} is exact in low-energy, because we only  made a long-wave approximation to derive long-wave physics near the emergent symmetry-enforced Dirac nodes. $S_{XY}$ in Eq.~\eqref{eq13} is a quite general result, suggesting that one can understand the 2D antiferromagnetism from a 2+1D quantum-electrodynamics-type theory but with a CS rather than the Maxwell term.

In Eq.\eqref{eq13}, the matter field is the gapless Dirac CS fermions. This naturally suggests us to integrate out the degrees of freedom, $A^{\mu}_{\mathbf{r}}$, which shows up in a bilinear form in $S_{CS}$, giving rise to a general theory describing interacting Dirac CS fermions living in sublattice space with multiple valleys:
\begin{equation}\label{eq14}
\begin{split}
 S_{f}&=\int dt[\int d\mathbf{r} \hat{f}^{\dagger}_{\mathbf{r},\alpha}(\mathbf{K}_i) \sigma^{\mu}_{\alpha\beta}i\partial_{\mu}\hat{f}_{\mathbf{r},\beta}(\mathbf{K}_i)+\int d\mathbf{r}d\mathbf{r}^{\prime} \\
 & V^{\alpha,\beta,\rho,\sigma}_{\mathbf{r}-\mathbf{r}^{\prime}}(\mathbf{K}_i,\mathbf{K}_j)\hat{f}^{\dagger}_{\mathbf{r},\alpha}(\mathbf{K}_i)\hat{f}_{\mathbf{r},\beta}(\mathbf{K}_i)\hat{f}^{\dagger}_{\mathbf{r}^{\prime},\rho}(\mathbf{K}_j)\hat{f}_{\mathbf{r}^{\prime},\sigma}(\mathbf{K}_j)].
\end{split}
\end{equation}
Formally, the Dirac CS fermions interact via a nonlocal vertex $V^{\alpha,\beta,\rho,\sigma}_{\mathbf{r}-\mathbf{r}^{\prime}}$ which is proportional to the CS charge $e$. Since the XY spin model with planar N\'{e}el order is mapped to interacting CS Dirac fermions, we expect the physical nature of the long-range N\'{e}el state should be captured by the spontaneous symmetry breaking of the Dirac fermions and the conventional spin-wave theory.

We note that the gauge field theory, Eq.\eqref{eq12}, is no longer compact after making the long-wave expansion near the Dirac nodes and the coupling of the CS fermions to the gauge field is proportional to the CS charge $e$. This leads to a $e$-dependent many-body theory $S_f$, Eq.\eqref{eq14}. The $e$-dependence, which is absent in the original lattice field theory, is an inevitable theoretical artifact brought by the long-wavelength approximation. We are going to show below that $e$-dependent physical quantities generated by the theory $S_f$ are all proportional to $e\Lambda v_F$, which is the characteristic energy scale of interacting Dirac fermions. On the other hand, the characteristic energy scale of the spin model is the exchange coupling $J$. To make a quantitative comparison between the two theories, one needs to require that $e\Lambda v_F$ being comparable to $J$. Hence, the lower energy $v_F\Lambda$ we are focusing on, the larger $e$ is implicit in the interacting fermionic theory. Since the long-wave approximation we made is accurate in the long-wavelength limit $v_F\Lambda\rightarrow0$, we expect to obtain an accurate description showing a very weak $e$-dependence for $e\gg 1$.

\subsection{CS superconductor mean-field theories on different lattices}
After mapping from the XY spin model to the interacting Dirac CS fermions, we are now in a position to investigate the mean-field ground state of Eq.\eqref{eq14}, using typical lattices as examples.  We discuss the triangular lattice in more details as it is more complicated case that leads to a non-collinear N\'{e}el AFM order. The results for the honeycomb and square lattice are also provided, in order to facilitate the study in the next sections.

\subsubsection{Non-collinear N\'{e}el order and CS superconductor on the triangular lattice}
Starting from Eq.\eqref{eq9} on the triangular lattice, two degenerate energy minima with $\mathbf{Q}_1=(-2\pi/3a,-2\pi/\sqrt{3}a)$, $\mathbf{Q}_2=(2\pi/3a,-2\pi/\sqrt{3}a)$ can be identified from the single-particle CS fermion spectrum after turning off both $\overline{A}_{\mathbf{r},\mathbf{r}^{\prime}}$ and $\delta A_{\mathbf{r},\mathbf{r}^{\prime}}$. In the hardcore boson picture, the bosons will condense in one of the two degenerate $\mathbf{Q}$ points, spontaneously breaking  the $Z_2$ symmetry. From the nesting vector $\mathbf{Q}_i$, $i=1,2$, one can determine the configuration of the spin order, which is the 120 degree planar N\'{e}el state, as shown in Fig.5(a).

Then we turn on the non-fluctuating gauge field $\overline{A}_{\mathbf{r},\mathbf{r}^{\prime}}$, which generates the flux in Fig.4(c). The phase $\overline{A}_{\mathbf{r},\mathbf{r}^{\prime}}$ that satisfies Eq.\eqref{eq7} can be determined up to the gauge redundancy. The obtained $\overline{A}_{\mathbf{r},\mathbf{r}^{\prime}}$ enlarges the unit cell by six times and decreases the BZ to one sixth of that for the original lattice. Diagonalizing the tight-binding CS fermion model $\overline{H}_{XY}$ with the non-fluctuating gauge field, two inequivalent Dirac nodes located at $\mathbf{K}=(\pi/6a,-\pi/2\sqrt{3}a)$ and $\overline{\mathbf{K}}=(-\pi/6a,\pi/2\sqrt{3}a)$ can be obtained in the first BZ. After expansion, the low-energy effective Hamiltonian around each of the nodes can be derived. Because of the BZ folding, the Dirac spinor is of six-dimension. We introduced two sets of indices $\rho=1,2,3$ and $\alpha=1,2$ to decompose the six-dimensional Dirac spinor into three copies of two-dimensional Dirac spinors in the sublattice space. This leads to the effective Dirac Hamiltonian, $H_0=H_{\mathbf{K}}(\mathbf{k})+H_{\overline{\mathbf{K}}}(\mathbf{k})$, where $H_{\mathbf{K}}(\mathbf{k})$ reads as,
\begin{equation}\label{eq16}
 H_{\mathbf{K}}(\mathbf{k})=v_F\sum_{\mathbf{k},\alpha,\beta,\rho} \hat{f}^{\dagger}_{\mathbf{k},\alpha,\rho}\boldsymbol{\sigma}_{\alpha\beta}\cdot\mathbf{k}\hat{f}_{\mathbf{k},\beta,\rho}.
\end{equation}
The three copies of Dirac spinors in the above Hamiltonian, $\rho=1,2,3$, are in accordance with the three emergent sublattices of the 120-degree N\'{e}el state from the spin XY model. $p_i=\mathbf{k}\cdot\mathbf{e}_i$ and $\mathbf{e}_i$, $i=1,2,3$, are the unit vectors of the NN bond in triangular lattice, which are of the length $a$ as shown by Fig.5(a). The Hamiltonian describing the other Dirac cone state at $\overline{\mathbf{K}}$ is obtained from the time-reversal transformation applied to Eq.\eqref{eq16}.

Following the step proposed in the last section, we restore the fluctuating gauge field $\delta A_{\mathbf{r},\mathbf{r}^{\prime}}$ to the low-energy Dirac fermions $H_0$, and then integrate out the gauge field fluctuation.
A nonlocal interaction between Dirac CS fermions is obtained as,
\begin{equation}\label{eq17}
 H_{int}=\sum_{\mathbf{k},\mathbf{k}^{\prime},\mathbf{q},\rho}V^{\alpha\alpha^{\prime}\beta\beta^{\prime}}_{\mathbf{q}}\hat{f}^{(a)\dagger}_{\mathbf{k},\alpha,\rho}
\hat{f}^{(b)\dagger}_{\mathbf{k}^{\prime}+\mathbf{q},\alpha^{\prime},\rho}\hat{f}^{(b)}_{\mathbf{k}^{\prime},\beta,\rho}\hat{f}^{(a)}_{\mathbf{k}+\mathbf{q},\beta^{\prime},\rho},
\end{equation}
where the derived interaction vertex is of the following form as
\begin{equation}\label{eq18}
 V^{\alpha\alpha^{\prime}\beta\beta^{\prime}}_{\mathbf{q}}=2\pi ie[\sigma^i_{\alpha\beta}\delta_{\alpha^{\prime},\beta^{\prime}}+\delta_{\alpha\beta}(\sigma^i_{\alpha^{\prime},\beta^{\prime}})^{\mathrm{T}}]\epsilon_{l,m}A^m_{\mathbf{k}}e^l_i,
\end{equation}
where $A^m_{\mathbf{k}}=k^m/k^2$, $m=1,2$ and $\epsilon_{l,m}$ the antisymmetric Levi-Civita tensor. $a,b=\mathbf{K},\overline{\mathbf{K}}$  denote the two different Dirac cones in the first BZ, as shown by the left figure in the bracket ``interacting Dirac fermions" in Fig.5. Both intra- and inter-valley interactions are mediated by the fluctuating gauge field. Here, since the ground state of the spin model is known to be a N\'{e}el AFM state, which is a condensate with momentum $\mathbf{Q}_i$, we only look for the mean-field theory that can describe the same physics in the CS fermion picture. Let us first consider intra-valley interaction.
Assuming that a mean-field order is stabilized, any bilinear mean-field orders from CS fermions then will enjoy the total momentum either as $\mathbf{K}_{tot}=2\mathbf{K}=(\pi/3a,-\pi/\sqrt{3}a)$ or $\mathbf{K}_{tot}=2\overline{\mathbf{K}}=(-\pi/3a,\pi/\sqrt{3}a)$. Since $\mathbf{K}_{tot}\neq \mathbf{Q}_{1,2}$ (mod $\mathbf{G}\equiv l_1\mathbf{b}_1+l_2\mathbf{b}_2$, with $\mathbf{b}_{1,2}$ the reciprocal vector in Fig.5), therefore no mean-field theory from the intra-valley interaction is able to describe the N\'{e}el AFM state.
On the other hand, the mean-field orders from inter-valley interaction always carry the total momentum $\mathbf{K}_{tot}=0$, which is equal to $\mathbf{Q}_i$ up to the reciprocal vector, consistent with 120 degree N\'{e}el state that corresponds to condensation at $\mathbf{Q}_i$.
Therefore, we show that by examining the total momentum of the possible mean-field orders, one can determine whether the inter-valley or the intra-valley interaction plays the key role. This can efficiently simplify Eq.\eqref{eq18} and facilitate the mean-field study of the possible ground state.

It is straightforward to construct a mean-field theory for inter-valley interaction.
For a given sublattice $\rho$, two types of bosonic mean-field orders can be introduced via Hubbard-Stratonovich decomposition, i.e.,
\begin{eqnarray}
 \Delta^{\alpha\alpha^{\prime}}_{\mathbf{k},\rho} &=& -2i\pi e\sum_{\beta\beta^{\prime}}V^{\alpha\alpha^{\prime}\beta\beta^{\prime}}_{\mathbf{k}-\mathbf{k}^{\prime}}\langle \hat{f}_{-\mathbf{k}^{\prime},\beta,\rho}\hat{f}_{\mathbf{k}^{\prime}\beta^{\prime},\rho}\rangle, \\
 \chi^{\alpha\alpha^{\prime}}_{\mathbf{k},\rho} &=& -2i\pi e\sum_{\beta\beta^{\prime}}V^{\alpha\alpha^{\prime}\beta\beta^{\prime}}_{\mathbf{k}-\mathbf{k}^{\prime}}\langle \hat{f}^{\dagger}_{\mathbf{k}^{\prime},\beta,\rho}\hat{f}_{\mathbf{k}^{\prime}\beta^{\prime},\rho}\rangle.
\end{eqnarray}
The former is usually stabilized for weakly frustrated XY models with nearest-neighbor interaction. At the same time, we find that the latter could only arise with stronger frustration \cite{ruinew}. Thus, a superconductor state of Dirac CS fermions becomes the most stable mean-field ground state with weak frustration. Besides, Eq. \eqref{eq18} clearly indicates that all the nonzero components of the vertex $V^{\alpha\alpha^{\prime}\beta\beta^{\prime}}_{\mathbf{q}}$ are proportional to $\mathcal{A}^x_{\mathbf{k}}-i\mathcal{A}^y_{\mathbf{k}}$ where $\mathcal{A}^i_{\mathbf{k}}=\epsilon_{l,m}A^m_{\mathbf{k}}e^l_i$. The $p+ip$ interaction vertex therefore, energetically favors a $p+ip$-wave rather than a normal $s$-wave pairing state. We term the $p+ip$ paired state of Dirac CS fermions emergent from XY spin models the CS superconductors.

\subsubsection{Collinear N\'{e}el order and CS superconductivity on honeycomb and square lattices.}

For the honeycomb and square lattices, similar derivation leads to the mean-field theory of the CS superconductors. As shown by the outlined mechanism in Fig.5, weakly-frustrated quantum XY models lattice enjoy collinear N\'{e}el AFM order on both square and honeycomb, as a result of spontaneous symmetry breaking of the $\mathrm{U}(1)$ invariant spin exchange model.
After CS fermionization and following the same procedure as before, CS superconductor states can be found on square and honeycomb lattice as well.
Since the derivation is similar to that of the triangular case,  we do not show the details  but straightforwardly present the self-consistent mean-field  equations and their solutions.

For the honeycomb lattice, despite the CS Dirac fermions, the gauge-field induced an inter-valley interaction which reads as,
\begin{equation}\label{eq20}
H_{int}= \sum_{\mathbf{k},\mathbf{k}^{\prime},\mathbf{q}}V^{\alpha\alpha^{\prime}\beta^{\prime}\beta}_{\mathbf{q}}\hat{f}^{\dagger}_{\mathbf{k}\alpha}\hat{\overline{f}}^{\dagger}_{\mathbf{k^{\prime}+q}\alpha^{\prime}}\hat{\overline{f}}_{\mathbf{k^{\prime}},\beta^{\prime}}\hat{f}_{\mathbf{k+q}\beta}
\end{equation}
with the interaction vertex
\begin{equation}\label{eq21}
 V^{\alpha\alpha^{\prime}\beta^{\prime}\beta}_{\mathbf{q}}=-2\pi i e v_F\epsilon^{ij}(\sigma^i_{\alpha\beta}\delta_{\alpha^{\prime}\beta^{\prime}}+\delta_{\alpha\beta}\sigma^{i T}_{\alpha^{\prime}\beta^{\prime}})A^j_{\mathbf{q}},
\end{equation}
where we used $\hat{f}$ and $\hat{\overline{f}}$ to distinguish the CS fermions from the two different Dirac nodes on the honeycomb lattice. $\alpha$, $\beta$ represent for the sublattice degrees of freedom on honeycomb lattice. $A^j_{\mathbf{q}}=q^j/|q|^2$ such that the $p+ip$ wave nature is implicit in the interaction vertex in Eq.\eqref{eq21}.

In the basis $\Psi_{\mathbf{k}}=[\hat{f}_{\mathbf{k},A},\hat{f}_{\mathbf{k},B},\hat{\overline{f}}^{\dagger}_{-\mathbf{k},A},\hat{\overline{f}}^{\dagger}_{-\mathbf{k},B}]^T$, the mean-field Hamiltonian describing the CS superconductor on the honeycomb lattice can be obtained via the Hubbard-Stratonovich decomposition, which is cast into a simple form as,
\begin{equation}\label{eq20a}
 H_{MF}=v_F\mathbf{k}\cdot\boldsymbol{\sigma}\tau^3+\hat{\Delta}_{\mathbf{k}}\tau^++\hat{\Delta}^{\dagger}_{\mathbf{k}}\tau^-,
\end{equation}
where the pairing potential above is a 2 by 2 matrix lying in the sublattice space as
\begin{equation}\label{eq21a}
 \hat{\Delta}_{\mathbf{k}}= \Delta_{3k}\sigma^0+i(\Delta_{0k,x}\sigma^y-\Delta_{0k,y}\sigma^x),
\end{equation}
where $\Delta_{0k,x}=\Delta_{0k}k_x/k$, $\Delta_{0k,y}=\Delta_{0k}k_y/k$, and $\Delta_{0k}$, $\Delta_{3k}$ are the two  superconductor  order parameters that characterize the mean-field state. Minimizing the mean-field ground state energy, and after integration over the polar angle of momentum $\mathbf{k}$, the self-consistent equations of the order parameters are obtained as,
\begin{equation}\label{eq22}
  \Delta_{0k}=\frac{ev_F}{2}\sum_{a=\pm}\int^k_0dk^{\prime}\frac{k^{\prime}\Delta_{3k^{\prime}}}{kE^{(a)}_{k^{\prime}}},
\end{equation}
and
\begin{equation}\label{eq23}
  \Delta_{3k}=\frac{ev_F}{2}\sum_{a=\pm}\int^{\Lambda}_kdk^{\prime}\frac{\Delta_{0k^{\prime}}+av_Fk^{\prime}}{kE^{(a)}_{k^{\prime}}},
\end{equation}
where $E^{(a)}_{k}=\sqrt{|av_k\mathbf{k}+\Delta_{0\mathbf{k}}|^2+|\Delta_{3\mathbf{k}}|^2}$.  As we have discussed before, in the long-wave length limit where the effective theory becomes a accurate description, one expects a larger CS charge $e$ in order to make the theory to be of the same characteristic energy as the original spin XY model. We are therefore interested in the large $e$ case. For $e\gg1$, it is found that nontrivial solutions of the order parameters always exists. Meanwhile, the mean-field equation, Eq.\eqref{eq22} and Eq.\eqref{eq23}, are reduced to the following form,
\begin{equation}\label{eq24}
  \Delta_{0k}=ev_F\int^k_0dk^{\prime}\frac{k^{\prime}\Delta_{3k^{\prime}}}{kE_{k^{\prime}}}
\end{equation}
and
\begin{equation}\label{eq25}
  \Delta_{3k}=ev_F\int^{\Lambda}_kdk^{\prime}\frac{\Delta_{0k^{\prime}}}{E_{k^{\prime}}}
\end{equation}
where $E_{k^{\prime}}=\sqrt{\Delta^2_{0\mathbf{k}}+\Delta^2_{3\mathbf{k}}}$. In long-wave length limit, by making Taylor expansion in terms of $k$, the solutions can be found as $\Delta_{3k}=0.445e\Lambda v_F$ and $\Delta_{0k}=ev_Fk/2$.

Similarly, the effective Hamiltonian of a CS superconductor for the square lattice is obtained as
\begin{equation}\label{eq29}
 H_{MF}=v_F\boldsymbol{\sigma}\cdot\mathbf{k}\frac{\tau^3+1}{2}+v_F\boldsymbol{\sigma}^{T}\cdot\mathbf{k}\frac{1-\tau^3}{2}+\tau^+\hat{\Delta}_{\mathbf{k}}+\tau^-\hat{\Delta}^{\dagger}_{\mathbf{k}},
\end{equation}
and pairing potential lies in the sublattice space as
\begin{equation}\label{eq30}
 \hat{\Delta}_{\mathbf{k}}=\Delta_{0k,x}\sigma^0-i\Delta_{0k,y}\sigma^3+i\Delta_{3k}\sigma^2.
\end{equation}
where $\Delta_{0k,x}=\Delta_{0k}\frac{k_x}{k}$ and $\Delta_{0k,y}=\Delta_{0k}\frac{k_y}{k}$. The mean-field self-consistent equations enjoy similar form as Eq.\eqref{eq22} and Eq.\eqref{eq23}, and are not written explicitly here for brevity.

Let us summarize the results of this section. We have obtained self-consistent mean-field ground states on three typical lattices. We note that that the CS superconductor description is very general. With a straightforward generalization, it can be applied to study all weakly-frustrated 2D XY spin models supporting the N\'{e}el AFM ground state, either collinear or non-collinear. The physical mechanism accounting for the formation of CS superconductors is concisely demonstrated in Fig.5.

\begin{figure}[h]\label{fig5}
\includegraphics[width=\linewidth]{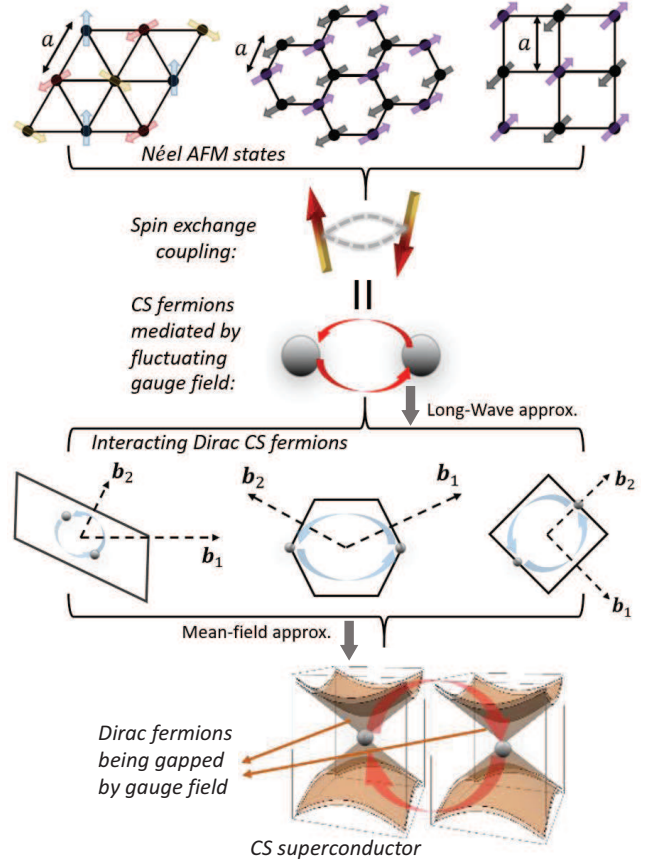}
\caption{(Color online) Shown is the mechanism for the formation of CS superconductors on triangular, honeycomb, and square lattice. The non-collinear or collinear N\'{e}el AFM state can be stabilized due to the symmetry breaking of the spin exchange model. On the other hand, the spin exchange interaction is translated to CS fermions mediated by fluctuating gauge field through an exact mapping. In long-wavelength approximation, one arrives at Dirac CS fermions at symmetry-enforced $\mathbf{k}$-points, with nonlocal interaction induced by the fluctuating gauge field. The grey dots indicate the Dirac nodes, $\mathbf{K}$, and $\overline{\mathbf{K}}$, on triangular, honeycomb, and square lattice in the first BZs. The noninteracting part of the low-energy effective Hamiltonian, $\overline{H}_{XY}$, violates the translation symmetry of the original lattice, enlarging the unit cell and introducing BZ folding. The reciprocal vectors of each the lattice are indicated by $\mathbf{b}_1$ and $\mathbf{b}_2$, e.g., for the triangular lattice, $\mathbf{b}_1=(4\pi/3,0)$ and $\mathbf{b}_2=(-2\pi/3a,-2\pi/\sqrt{3}a)$. Stable mean-field ground states where the Dirac CS fermions are paired are found, as a general result for the three typical lattices and can be proved to exist on all weakly-frustrated lattices that support planar N\'{e}el AFM ground states. The Dirac CS fermions from $\overline{H}_{XY}$ are gapped by the Cooper pairing with $p\pm ip$ symmetry.
}
\end{figure}


\section{Collective modes of a CS superconductor}
As suggested in Fig.5, the XY spin exchange model, which generates the N\'{e}el AFM state, is mapped in low-energy to a CS superconductor mean-field ground state. One would then naturally expect physical correspondences between the two phases and ask if the CS superconductor is a good description of the N\'{e}el AFM state in the CS representation.  A more careful investigation is therefore needed to compare the physical quantities on two sides. In this section, we discuss the comparison of the collective excitation modes. For demonstration, we use the quantum XY model on honeycomb lattice as an example. The generalization to other cases such as the square and triangular lattice are straightforward.

\subsection{Higgs mode from a CS superconductor}
The CS superconductor, a pair condensate that breaks the $\mathrm{U(1)}$ symmetry, should possess collective modes at zero temperature. These include the low-energy Goldstone mode and the gapped Higgs mode, whose physical origins are the phase and the amplitude fluctuation of the pairing order parameter, respectively, as indicated in Fig.3.
In this subsection, we first present a detailed study on the Higgs mode of a CS superconductor and then make comparison with the longitudinal mode of the N\'{e}el AFM order.

To calculate the Higgs mode of CS superconductors, we should consider the effect of the gauge-field-induced interaction $H_{int}$, Eq.\eqref{eq20} with going beyond the mean-field level. To facilitate the study, we use the Nambu formulation and make the sublattices explicit, where the creation and annihilation operators for the CS fermionic fields are written as two copies of Nambu spinors as, $\Psi_{\mathbf{k}\alpha}=[\hat{f}_{\mathbf{k},\alpha},\hat{\overline{f}}^{\dagger}_{-\mathbf{k},\alpha}]$, $\alpha=A,B$.  The interaction $H_{int}$ Eq.\eqref{eq20} is then rewritten in this basis as
\begin{equation}\label{eq31}
  H_{int}=-\sum_{\mathbf{k},\mathbf{k}^{\prime},\mathbf{q}}V^{\alpha\alpha^{\prime}\beta\beta^{\prime}}_{\mathbf{q}}\Psi^{\dagger}_{\mathbf{k}-\mathbf{q},\alpha}\overline{\tau}^+\Psi_{\mathbf{k},\beta^{\prime}}\Psi^{\dagger}_{\mathbf{k}^{\prime}+\mathbf{q},\beta}\overline{\tau}^-\Psi_{\mathbf{k}^{\prime},\alpha^{\prime}},
\end{equation}
where we have defined the Pauli matrix $\overline{\tau}^{\pm}=(\tau^0\pm\tau^z)/2$ in Nambu space.
Then, we are in a position to study $H_{int}$ in Eq.\eqref{eq20} beyond the mean-field theory. In the mean-field theory, the self-energy, i.e., the renormalization to the non-interacting CS Dirac fermions is obtained at the Hartree-Fock level, which is an approximation that breaks gauge $\mathrm{U}(1)$ symmetry. Both the Higgs mode and the Goldstone mode are originated from fluctuations that attempt to restore the broken symmetry.  The renormalization of the interaction vertex that restores the  $\mathrm{U}(1)$ symmetry generates the Bethe-Salpeter equations of the paired state.

Our focus here is to extract the Higgs mode of the CS superconductor, which originates from the fluctuation of the superconducting order parameter magnitude. Therefore, rather than solving Bethe-Salpeter equations, here we only need to consider the renormalization of the order parameter. Following Sec.IIC, the bare order parameter is obtained in the mean-field level by contraction of two Nambu spinors in the interaction $H_{int}$, Eq.\eqref{eq31}, leaving the interaction vertex two external legs, as represented by the first diagram (denoted by $\gamma$) on the right-hand-side of Fig.6.
In the following we will refer to this contracted vertex as the order parameter. Similarly, the renormalized superconductor order parameter is represented by the left-hand-side diagram in Fig.6, denoted by $\hat{\Gamma}_{\beta^{\prime}\beta}(\mathbf{k}+\mathbf{q},\mathbf{k})$, with sublattice indices $\beta^{\prime}$, $\beta$. Here, $\mathbf{q}$ is the transfer of momentum during the interaction process. The external legs denote the propagators in Nambu space. Thus, for given $\beta^{\prime}$ and $\beta$, $\hat{\Gamma}_{\beta^{\prime}\beta}(\mathbf{k}+\mathbf{q},\mathbf{k})$ can be understood as a vector residing in the 2 by 2 Nambu space.
Moreover, the superconductor order parameters are off-diagonal in Nambu space, therefore $\hat{\Gamma}_{\beta^{\prime}\beta}(\mathbf{k}+\mathbf{q},\mathbf{k})$ always resides in $\tau^x$-$\tau^y$ plane. The Feynman diagram in Fig.6 then, in fact, indicates a set of equations that self-consistently determines the vector  $\hat{\Gamma}_{\beta^{\prime}\beta}$ in the $\tau^x$-$\tau^y$ plane.
\begin{figure}[thb]\label{fig6}
\includegraphics[width=\linewidth]{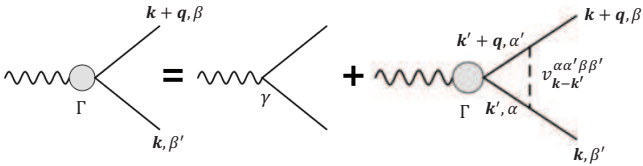}
\caption{(color online) Renormalization of the CS superconducting order parameter due to the interaction Eq.\eqref{eq31}. }
\end{figure}

To make simplifications, we recall that the bare order parameter of the CS superconductor $\hat{\Delta}_{\mathbf{k}}$ in Eq.\eqref{eq21a}  (shown as vertex $\gamma$ in Fig.6) can be rewritten in the Nambu formulation with explicit sublattice indices as $\hat{\Delta}_{\mathbf{k},\alpha\beta}$, which has the following form,
\begin{equation}\label{eq34}
   \hat{\Delta}_{\mathbf{k},\alpha\beta}= \Delta_{3k}\sigma^0_{\alpha\beta}\tau^x+i(\Delta_{0kx}\sigma^y_{\alpha\beta}-\Delta_{0ky}\sigma^x_{\alpha\beta})\tau^y.
\end{equation}
Hence, for given $\alpha$ and $\beta$, the bare order parameter $\hat{\Delta}_{\mathbf{k},\alpha,\beta}$ is also a vector in the $\tau^x$-$\tau^y$ plane of the Nambu space. Besides, we know from Eq.\eqref{eq34} that the diagonal terms (in sublattice space), $\Delta_{\mathbf{k},11}$ and $\Delta_{\mathbf{k},22}$ $(\Delta_{\mathbf{k},11}=\Delta_{\mathbf{k},22})$, point toward $\tau^x$-direction, whereas, the off-diagonal terms lie along the $i\tau^y$-direction. In a more compact form they can be rewritten as $\hat{\Delta}_{\mathbf{k},12}=-\hat{\Delta}^{\star}_{\mathbf{k},21}=i\tau^y\Delta_{0k}e^{-i\theta}$ with $\theta$ being the angle of $\mathbf{k}$. As discussed in the last section, the directions of $\Delta_{\mathbf{k},11/22}$ and $\Delta_{\mathbf{k},12/21}$ in Nambu space are determined by the symmetry of the interaction vertex  $V^{\alpha\alpha^{\prime}\beta^{\prime}\beta}_{\mathbf{q}}$, which is clear from the mean-field Hamiltonian of the CS superconductor. Thereby, for stable mean-field order parameters, their symmetries should not be altered by the perturbation around the mean-field solutions. One thus can expect that the renormalized order parameter $\hat{\Gamma}_{\alpha\beta}(\mathbf{k}+\mathbf{q},\mathbf{k})$ inherits the symmetries, such that its diagonal terms $\hat{\Gamma}_{11/22}(\mathbf{k}+\mathbf{q},\mathbf{k})$ and the off diagonal terms $\hat{\Gamma}_{12/21}(\mathbf{k}+\mathbf{q},\mathbf{k})$ must be in parallel with $\tau^x$- and $i\tau^y$-direction, respectively. More specifically, we then express the renormalized order parameter by the components along the $\tau^x$, $i\tau^y$ directions, and require that $\hat{\Gamma}_{11}=\hat{\Gamma}_{22}\equiv\Gamma_{11}\tau^x$, $\hat{\Gamma}_{12}=-\hat{\Gamma}^{\star}_{21}\equiv i\tau^y\Gamma_{12}$, where the momentums are implicit for brevity.  The components $\Gamma_{\alpha\beta}$ satisfy $\Gamma_{12}=-\Gamma^{\star}_{21}$, $\Gamma_{11}=\Gamma_{22}$. Moreover, in the long-wave length limit $|\mathbf{q}|, |\mathbf{k}|\ll\Lambda$, we know from the last section that $\Gamma_{11}$ is approximately a constant independent of momentums (with the leading order being  quadratic) and $\Gamma_{12}\propto e^{-i\theta}$ which is a requirement by the $p+ip$ feature of  $V^{\alpha\alpha^{\prime}\beta^{\prime}\beta}_{\mathbf{q}}$.

With the above analysis, the self-consistent relation corresponding to Fig.6 yields a Bethe-Salpeter-type equation. This is a nontrivial generalization of the normal $s$-wave superconductor case because of the complication by the sublattice degrees of freedom and the $p+ip$ symmetry. It describes the fluctuations of the magnitude of mean-field order parameters characterizing the CS superconductor. The Higgs mode can be established by solving  the equations \cite{Varma}. After a lengthy calculation whose details are included in the Appendix A, we find that the Higgs mode enjoys the following dispersion in the long-wave and large $e$ limit as
\begin{equation}\label{eq42}
\begin{split}
  \nu&=\sqrt{4\Delta^2_{3k^{\prime}}+2\Delta^2_{0q}}\simeq2\Delta_{3k^{\prime}}+\Delta^2_{0q}/2\Delta_{3k^{\prime}}\\
  &=0.89e\Lambda v_F+0.281e\Lambda v_F \overline{q}^2.
\end{split}
\end{equation}
Here we inserted in the last step the mean-field solutions for order parameters $\Delta_{3k}=0.445e\Lambda v_F$ and $\Delta_{0k}=ev_Fk/2$. We have also introduced a normalization of the wave vector by defining $\overline{q}=q/\Lambda$. From Eq.\eqref{eq42}, we know that the dispersion of Higgs mode of the CS superconductor has an energy gap, $2\Delta_{3k^{\prime}}=0.89e\Lambda v_F$, thus $e\Lambda v_F$ is regarded as the characteristic energy scale of CS superconductors. Although the gap is energy scale dependent, we can extract from Eq.\eqref{eq42} an energy scale-independent quantity that captures the feature of Higgs mode dispersion, i.e., the ratio between the gap and the coefficient in front of the dispersion $\overline{q}^2$. The ratio $0.89e\Lambda v_F/0.281e\Lambda v_F=3.167$ is an inherent physical quantity that characterizes the collective mode of the CS superconductor state. This quantity should be further compared with that evaluated from the planar N\'{e}el AFM state.

\subsection{Longitudinal fluctuation mode in a  N\'{e}el AFM state}
Having studied the Higgs mode of the CS superconductor, let us investigate its counterpart in a N\'{e}el AFM state, i.e., the longitudinal mode. In the general  Ginzberg-Landau theory with a complex order parameter field $\Psi(\mathbf{r},t)$, the stability of the longitudinal fluctuations in the condensed matter systems is more subtle than that of the Higgs particles in high-energy physics. This is because, unlike the particle physics which respects the Lorentz symmetry, there is no insurance of the Lorentz symmetry in condensed matter systems, such that there allows a decay channel from the amplitude mode to the phase modes \cite{David Pekker}. Only a few condensed matter systems have been proposed to support the well-defined amplitude fluctuations as an analog of the Higgs particles. The superconductors at low temperatures attracted the most attention \cite{Shimanoryo,Shermand}. Superconductors at low temperatures ($T\ll T_c$) enjoy perfect particle-hole symmetry near the Fermi surface. Therefore, the dynamical term of its corresponding Ginzberg-Landau theory respects the Lorentz invariance. This is the reason why we can obtain in the last subsection a well-defined amplitude mode from the CS superconductor at zero temperature in the long-wavelength limit. Another impressive condensed matter system is the antiferromagnets \cite{taohong,normandr}. For AFM states stabilized in a Heisenberg spin model, one usually does not expect the well-defined amplitude mode because the ground state, which breaks the $\mathrm{SU(2)}$ symmetry, is in general particle-hole asymmetric, therefore allows the decay into phase fluctuations. However, the XY antiferromagnetic, e.g., the N\'{e}el AFM ground state emergent from the XY spin model studied in this work, enjoys the particle-hole symmetry strictly. The corresponding coarse-grained field theory, being Lorentz invariant, stabilizes a well-defined amplitude mode in the long-wavelength limit, consistent with the CS superconductor. In the following, in order to be clear in terms of terminologies, we term the amplitude mode in the CS superconductor and the one in the N\'{e}el AFM state the Higgs mode and the longitudinal mode, respectively.

Previous literatures mainly study the longitudinal modes in magnetically ordered states starting from the field theoretical formalism \cite{dPodolsky}, because it is more convenient to evaluate the collective modes in a coarse-grained description than a microscopic picture. In this way, for example, the longitudinal mode from an AFM Heisenberg model can then be evaluated in the effective $\mathrm{O}(3)$ NL$\sigma$M \cite{dPodolsky}. Here, the CS superconductor state is derived from the microscopic spin model. In order to compare the two states with each other precisely, it is desirable to investigate the longitudinal fluctuation from the microscopic spin model rather than from the coarse-grained field theory. The former scheme is more advantageous as it directly compares at the quantitative level the collective modes of the two states.

Now we consider the oscillation of the magnetic orders on top of a planar N\'{e}el AFM state. We still use the honeycomb lattice as an example, while the following formulations can be generalized to other lattices without any technical difficulties. We start with a planar N\'{e}el ground state where opposite magnetization emerges on the two sublattices. Without losing generality, one can align the magnetization along x-direction by rotating the reference coordinates. That way, the spin operator $\hat{s}^x_a$ (with $a=A,B$ the sublattice index) takes opposite expectation values at different sublattices with $\langle s^x_A\rangle=-\langle s^x_B\rangle$.
Fluctuation of the order parameter $\langle s^x_a\rangle$ leads to the well known magnons, which
describe the spin-flip excitations on the lattice. The corresponding quasi-particle operators are bosons associated to the ``rotated" spin-raising and -lowing operators $\hat{\tilde{S}}^{\pm}=\hat{S}^z\mp i\hat{S}^y$ as
\begin{equation}\label{eq43}
 \hat{\tilde{S}}^+_{\mathbf{r},A}=\sqrt{1-\hat{a}^{\dagger}_{\mathbf{r}}\hat{a}_{\mathbf{r}}}\hat{a}_{\mathbf{r}}\simeq\hat{a}_{\mathbf{r}},
\end{equation}
and
\begin{equation}\label{eq44}
 \hat{\tilde{S}}^-_{\mathbf{r},A}=\hat{a}^{\dagger}_{\mathbf{r}}\sqrt{1-\hat{a}^{\dagger}_{\mathbf{r}}\hat{a}_{\mathbf{r}}}\simeq\hat{a}^{\dagger}_{\mathbf{r}},
\end{equation}
where the approximation is made with the assumption of low magnon density for a stable N\'{e}el ordering, i.e., $\hat{a}^{\dagger}_{\mathbf{r}}\hat{a}_{\mathbf{r}}\sim\langle \hat{a}^{\dagger}_{\mathbf{r}}\hat{a}_{\mathbf{r}}\rangle\ll1$. The magnons from the B sublattice can be introduced similarly as above.

Magnons defined in Eq.\eqref{eq43} and Eq.\eqref{eq44} are the spin-flip fluctuation of ground state, i.e., the transverse collective mode. The longitudinal mode then corresponds to the fluctuation of the magnitude of $\langle \hat{S}^x_a\rangle$. Since $\hat{S}^x_{\mathbf{r},A}=1/2-\hat{a}^{\dagger}_{\mathbf{r}}\hat{a}_{\mathbf{r}}$, this physically corresponds to the fluctuation of the magnon density with respect to the ground state. Therefore, the longitudinal excitations here are similar to those in the helium superfluid, which are collective excitations of boson density on top of the superfluid ground state, as firstly studied by Feynman. Following the seminar paper by Feynman \cite{rpfeyman}, such collective mode perturbs the vacuum ground state in a way such that the resulting wave function becomes a plane-wave superposition of local boson densities of the ground state. Following this spirit, in the case of the planar N\'{e}el AFM state, we can write down the ansatz of the wave function describing the longitudinal excitation as
\begin{equation}\label{eq45}
 |\Psi^e_{\mathbf{q}}\rangle=\frac{1}{\sqrt{N}}\sum_{\mathbf{r}}e^{i\mathbf{q}\cdot\mathbf{r}}\hat{S}^x_{\mathbf{r}}|0\rangle,
\end{equation}
where the vacuum state $|0\rangle$ denotes the N\'{e}el AFM ground state. As discussed above, we know that $|\Psi^e_{\mathbf{q}}\rangle$ describes the magnon density wave with momentum $\mathbf{q}$ on top of the vacuum state.

The energy of the excitation state $|\Psi^e_{\mathbf{q}}\rangle$ can be calculated via $E_{\mathbf{q}}=\langle\Psi^e_{\mathbf{q}}|H |\Psi^e_{\mathbf{q}}\rangle/ \langle\Psi^e_{\mathbf{q}}|\Psi^e_{\mathbf{q}}\rangle-\langle0|H|0\rangle$, which is measured from the energy of the vacuum ground state. Introducing the excitation operator $X_{\mathbf{q}}|0\rangle= |\Psi^e_{\mathbf{q}}\rangle$, i.e., $X_{\mathbf{q}}=\sum_{\mathbf{r}}e^{i\mathbf{q}\cdot\mathbf{r}}\hat{S}^x_{\mathbf{r}}/\sqrt{N}$, inserting which into $E_{\mathbf{q}}$, one obtains
\begin{equation}\label{eq45add}
 E_{\mathbf{q}}=\frac{\langle X^{\dagger}_{\mathbf{q}}H X_{\mathbf{q}}\rangle}{\langle\Psi^e_{\mathbf{q}}|\Psi^e_{\mathbf{q}}\rangle}-\langle H\rangle=\frac{\langle X^{\dagger}_{\mathbf{q}}[H,X_{\mathbf{q}}] \rangle}{\langle\Psi^e_{\mathbf{q}}|\Psi^e_{\mathbf{q}}\rangle},
\end{equation}
here and in the following, we use $\langle...\rangle$ to represent for the expectation with respect to the vacuum ground state $\langle0|...|0\rangle$. Using the condition that $X^{\dagger}_{\mathbf{q}}=X_{-\mathbf{q}}$ (due to the fact that $\hat{S}^x_{\mathbf{r}}$ is a Hermitian operator), the numerator of Eq.\eqref{eq45add}, $N_{\mathbf{q}}$, can be derived as $N_{\mathbf{q}}=\langle[X_{-\mathbf{q}},[H,X_{\mathbf{q}}]]\rangle/2$, while the denominator, after expansion, is the spin structure factor of the lattice model defined as $S_{\mathbf{q}}=\sum_{\mathbf{r},\mathbf{r}^{\prime}}e^{i\mathbf{q}\cdot(\mathbf{r}-\mathbf{r}^{\prime})}\langle\hat{S}^x_{\mathbf{r}}\hat{S}^x_{\mathbf{r}^{\prime}}\rangle/N$.  Therefore, the energy of the fluctuation of magnon density with momentum $\mathbf{q}$ is obtained as,
\begin{equation}\label{eq46}
 E_{\mathbf{q}}=\frac{N_{\mathbf{q}}}{S_{\mathbf{q}}}.
\end{equation}
It approximately produces the dispersion of the longitudinal model of the XY N\'{e}el order, according to the analysis above. The Feynman's ansatz above has been proposed by Ref.\cite{yxian} to study the longitudinal mode in Heisenberg spin models. Here, we apply the method to the XY antiferromagnets whose longitudinal mode has no ambiguity because the underlying Lorentz invariance that forbids the decay into a pair of Goldstone modes as discussed above.

Insertion of $X_{\mathbf{q}}=\frac{1}{\sqrt{N}}\sum_{\mathbf{r}}e^{i\mathbf{q}\cdot\mathbf{r}}\hat{S}^x_{\mathbf{r}}$ into $N_{\mathbf{q}}$ generates a correlation function. In the long-wave limit $|\mathbf{q}|\rightarrow0$ \cite{footnote1}, it reads as,
\begin{equation}\label{eq47}
  N_{\mathbf{q}}=\frac{J}{4N}\sum_{\mathbf{r},j}\langle -\hat{S}^y_{\mathbf{r}}\hat{S}^y_{\mathbf{r}+\mathbf{e}_j}+\hat{S}^z_{\mathbf{r}}\hat{S}^z_{\mathbf{r}+\mathbf{e}_j}\rangle,
\end{equation}
where $\mathbf{e}_j$, $j=1,2,3$ denotes the three nearest neighbor bond vectors of a given site on the honeycomb lattice.
It is clear that in this way, the evaluation of the longitudinal mode is simplified to the calculation of certain sets of correlation functions with respect to the N\'{e}el AFM state. In order to obtain more precise results, we apply DMRG to calculate the correlation functions and then obtain $S_{\mathbf{q}}$ and $N_{\mathbf{q}}$ on the honeycomb lattice. The calculation is performed with cylindrical geometry where periodical boundary condition is taken along $y$-direction, and the zigzag boundary is taken at $x=0$ and $x=N_x$. The calculated $E_{\mathbf{\overline{q}}}$ along the $\Gamma-X$ direction in the BZ is shown in Fig.7, where $\overline{\mathbf{q}}$ is the wave vector normalized $\mathbf{q}$ by the magnitude of the wave vector at the BZ boundary, $\Lambda_0=2\pi/3a$.
\begin{figure}[thb]\label{fig7}
\includegraphics[width=\linewidth]{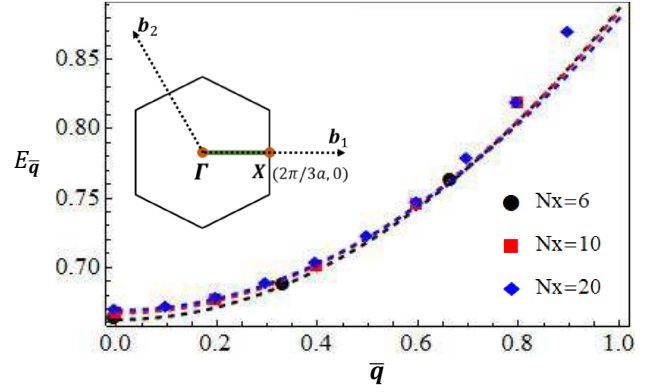}
\caption{(Color online) DMRG calculation of the longitudinal modes' dispersion on a cylindrical geometry with the zigzag boundary and the circumference $N_y=6$. Data for finite-size $N_x=6$, $N_x=10$ and $N_x=20$ are shown. The dispersion is plotted along the $\Gamma-X$ direction in the BZ, as shown in the inset. $\overline{\mathbf{q}}$ is the wave vector $\mathbf{q}$ normalized by the BZ boundary. In the calculation, we keep 2000 states for the finite DMRG in the form of matrix product state, which can reach the truncation error less than $10^{-9}$ for the nearest neighbor XY model on the honeycomb lattice.}
\end{figure}
The black sphere, red square and blue rhombus data curve in Fig.7 show the dispersion with increasing system size of $N_x=6$, $N_x=10$, and $N_x=20$ respectively. The larger $N_x$, the more data is collected in the discrete reciprocal space. As shown clearly, The longitudinal mode dispersion is weakly dependent on $N_x$ for $N_x\geq6$. Moreover, the data from DMRG can be well fitted by quadratic dispersion with
 \begin{equation}\label{eq50}
   E_{\overline{\mathbf{q}}}=c_1+c_2\overline{q}^2
 \end{equation}
where $c_1$ and $c_2$ are the fitting constant parameters. The normalized $\overline{q}$ is dimensionless, therefore both $c_1$ and $c_2$ in Eq.\eqref{eq50} are of the same dimension as energy.
\begin{figure}[thb]\label{fig8}
\includegraphics[width=\linewidth]{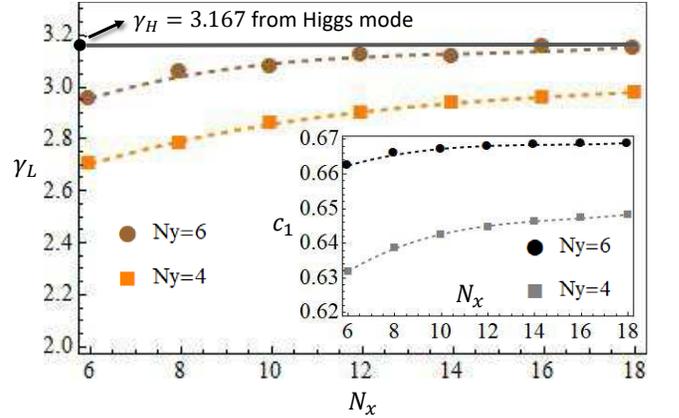}
\caption{(color online) The ratio $\gamma_{L}=c_1/c_2$ with $c_1$ and $c_2$ being extracted by fitting the DMRG data in Fig.7 to Eq.\eqref{eq50} in the regime $0<\overline{q}\lesssim1/5$. The black line $\gamma_H=3.167$ is the ratio of the Higgs mode evaluated from the CS superconductor on a honeycomb lattice. The inset shows calculated energy gap $c_1$ of the longitudinal mode for different $N_y$.}
\end{figure}

Recall that in the low-energy description of a CS superconductor, we obtain the Higgs mode dispersion as Eq.\eqref{eq42}. By comparing Eq.\eqref{eq42} and Eq.\eqref{eq50}, we found that the longitudinal mode of the N\'{e}el AFM state agrees very well, in the algebraic form, with the Higgs mode of a CS superconductor. Both display an energy gap for $\overline{q}=0$ and a leading quadratic $\overline{q}$ dispersion. It should be noted that although the Higgs mode is derived from a low-energy effective description of the CS superconductor while the longitudinal mode is evaluated numerically from the lattice spin model, quantitative comparisons between the two modes still makes sense and matters in long-wavelength regime.

Let us now compare the two modes quantitatively. As discussed in the last section, the Higgs mode enjoys a dimensionless quantity that characterizes its dispersion, i.e., the ratio between the energy gap and the quadratic dispersion efficient, $\gamma_{H}$,. We find that $\gamma_{H}=0.890e\Lambda v_F/0.281e\Lambda v_F=3.167$, as indicated by the horizontal black line in Fig.8. On the other hand, we obtain $c_1$ and $c_2$ by fitting the DRMG results to Eq.\eqref{eq50} in the long-wavelength regime, as shown by the dashed curves in Fig.7. This generates the ratio from the longitudinal mode, $\gamma_{L}=c_1/c_2$, as shown for different lattice sizes in Fig.8. It is found that with increasing $N_y$, $\gamma_{L}$ is gradually enlarged. For $N_y=6$ and with increasing $N_x$, $\gamma_{L}$ display a gradual and perfect saturation to the predicted value of $\gamma_{H}=3.167$ from the Higgs mode of the CS superconductor. The obtained excellent quantitative consistency strongly suggests a precise correspondence between the Higgs mode of a CS superconductor and the longitudinal mode of the N\'{e}el AFM state.

In addition to the magnitude fluctuation, there is a phase mode associated with the ground state of a CS superconductor.
We have studied and compared the phase fluctuation mode in the CS superconductor with the spin wave mode of the planar N\'{e}e AFM in our previous study, i.e., Ref. \cite{ruia}. Remarkably good quantitative consistence is found between the two modes especially for $e\gg 1$. This further supports our previous observation that CS superconductor becomes a more accurate low-energy description for lager $e$.  To summarize, we have established quantitative correspondence between the collective modes of the CS superconductor and N\'{e}el AFM state, namely, the consistency between the magnons and the phase fluctuations, and the excellent match between the longitudinal mode and the Higgs mode, as indicated previously by  Fig.2.

Last, we would like to discuss the stability of CS superconductors in the large $e$ limit. As shown above, when evaluated in the units of $ev_F\Lambda$, the velocity of the phase fluctuation mode display very weak $e$-dependence and saturates to the predicted value calculated from the spin-wave picture, as can be found in Ref.\cite{ruia}. Moreover, in the units of $ev_F\Lambda$, the Higgs mode obtained at large $e$ is also $e$-independent, as is clear from Eq.\eqref{eq42}. According to the discussion above, the $e$-independence of the physical quantities in the large $e$ limit justify the long-wave approximation of the lattice gauge theory. Therefore, the CS superconductor states should serve as accurate descriptions of the planar N\'{e}el AFMs in low-energy.

\section{The spin ordering from CS superconductivity}
In this section, we will investigate a more direct correspondence of the CS superconductor and the N\'{e}el AFM order, i.e, the spin orderings.  To proceed, we need some additional preparations and make generalization of the Chern-Simons fermionization to the lattice with periodic boundary conditions.

\subsection{Fermion parity-dependent boundary condition}
We firstly would like to draw the readers' attention to the intuitive similarity between the  CS superconductor with $p\pm ip$ pairing symmetry and Kitaev's 1D spinless $p$-wave superconductor. The latter can be exactly mapped from a 1D transverse Ising model, while the former is mapped from the 2D XY spin model (with additional mean-field approximation). The 1D spinless $p$-wave and the 2D $p\pm ip$ CS superconductors are topological states in the sense that they have a robust bulk topology and enjoy the Majorana boundary modes. To simplify the calculations, we consider the periodic boundary condition on a 2D lattice in the following section. Recalling that the boundary condition plays an important role and has connection with the ground state wave function of the 1D transverse Ising model, we are motivated to firstly study the physical consequences of taking a periodic boundary condition and generalize the CS fermionization, Eq.\eqref{eq2} and Eq.\eqref{eq3}, to the case where the Hamiltonian is defined on a compact torus.

Following the detailed analysis, which is included in the Appendix B, we show that special attention needs to be paid for the exchange coupling terms crossing the boundaries, as shown by the dashed curves in Fig.9(a), where we use the square lattice as an example for demonstration.  In the fermion language, these terms are cast into the hopping crossing the boundaries, as indicated by the red dashed curves in Fig.9(b). Due to the presence of boundary, the CS fermions receive an additional $Z_2$ factor $(-1)^{N_e-1}$ once they hop across the boundary under periodical boundary condition, where $N_e$ is the total number of CS fermions. Thus, one obtains a FP-dependent boundary condition for the CS fermions, which are summarized in the following as,
\begin{itemize}
  \item For $N_e$ being odd, one has $f(1,j)=f(N_{x}+1,j)$, a periodic boundary condition for CS fermions, such that $k_x=\frac{2\pi n_x}{N_x}$, $k_y=\frac{2\pi n_y}{N_y}$ with $n_{x,y}$ the integer taking the values $-N_{x,y}/2$, $-N_{x,y}/2+1$,..., $N_{x,y}/2$. Here, we use $f(x,y)$ to denote the fermion operator defined on 2D lattice coordinate $(x,y)$.
  \item For $N_e$ being even, one has $f(1,j)=-f(N_{x}+1,j)$, an anti-periodic boundary condition (APBC) for CS fermions. The APBC then generates a shift of the $\mathbf{k}$-lattice with $k_x=\frac{\pi(2n_{x}+1)}{N_x}$, $k_x=\frac{\pi(2n_y+1)}{N_y}$, with $n_{x,y}$ the integer taking the values $-N_{x,y}/2$, $-N_{x,y}/2+1$,..., $N_{x,y}/2-1$.
\end{itemize}
The FP-dependent boundary condition and the corresponding shift of the momentum lattice are schematically plot in Fig.9.
\begin{figure}[htbp]\label{fig9}
\includegraphics[width=\linewidth]{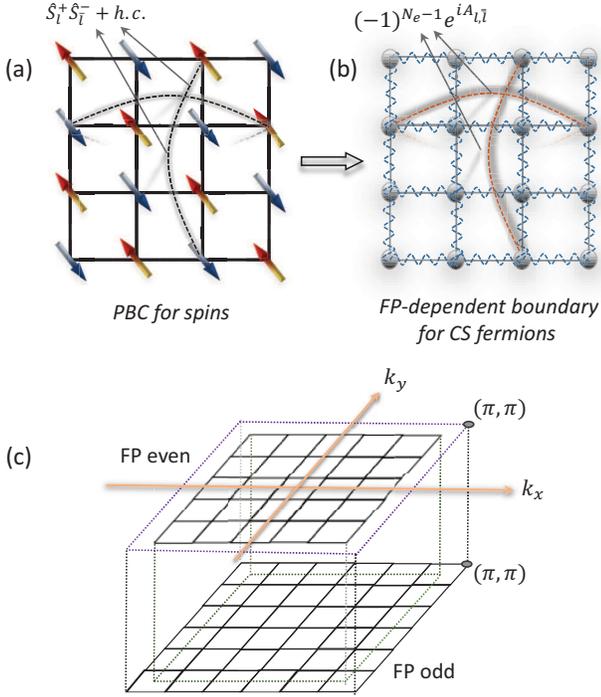}
\caption{(color online) (a) The XY spin model with a periodic boundary condition. (b) The fermion parity-dependent boundary condition for the CS fermions coupled to the gauge field (wavy lines). (c) The odd and even fermion parities leading to shifted $\mathbf{k}$-lattice in the BZ. }
\end{figure}

\subsection{Fermion parity-dependent ground state}
To evaluate the spin ordering with respect to the ground state, we need to firstly study the ground state wave function of CS superconductors with considering the above FP-dependent boundary condition. As demonstrated in detail in the Appendix D, a Bogoliubov transformation can be made to obtain the ground state wave function of the CS superconductor as
\begin{equation}\label{eq73}
  |GS\rangle_e=-\prod_{\mathbf{k}}U_{\mathbf{k}}e^{G_{ab}f^{\dagger}_{-\mathbf{k},a}f^{\dagger}_{\mathbf{k},b}}|0\rangle,
\end{equation}
where $G_{ab}$ is the matrix in sublattice space,  $U_{\mathbf{k}}$ is an overall function. Both $G_{ab}$ and $U_{\mathbf{k}}$ are related to a transformation matrix $\hat{R}$ , as shown explicitly in the Appendix D. From Eq.\eqref{eq73} it is seen that the ground state is derived as a coherent state of Cooper pairs of CS fermions from both inter- and intra- sublattice, as one can expect by making an analogy with the BCS theory.

The wave function $|GS\rangle_e$ is the Bogoliubov vacuum in the sense that any annihilation operators of Bogoliubov particles will annihilate $|GS\rangle_e$. Then, $|GS\rangle_e$ describes the state where Cooper pairs are created on top of the fermionic vacuum, so that $|GS\rangle_e$ has even FP with even $N_e$. Except for $|GS\rangle_e$,  there is also another degenerate Bogoliubov vacuum with odd FP.

To clearly show this, we firstly regularize the CS superconductor onto a lattice. As shown by Appendix C, it is found that the momentum $\mathbf{Q}=(\pi,\pi)$ is a particular $\mathbf{k}$-point, where the spinless CS fermion evades forming pair with its time-reversal partner, i.e., the CS fermions with momentum $\mathbf{Q}$ and $\mathbf{-Q}$ are unpaired. Moreover, we have derived in Sec.IVA that for the even FP, we must enforce the anti-periodic boundary condition of fermions, such that $k_x=\pi(2n_x+1)/N_x$ and $k_y=\pi(2n_y+1)/N_y$. The discrete momentum space is shifted, as shown by Fig.9(c), and there are no CS fermions that enjoy the exact lattice momentum $\mathbf{k}=\mathbf{Q}=(\pi,\pi)$. Therefore, for even FP,  all CS fermions form pairs, generating the ground state wave function $|GS\rangle_e$ above, with the subscript $e$ representing even parity.

On the other hand, for the odd parity sector of the Bogoliubov vacuum, one has to, by definition, add a fermion to the state $|GS\rangle_e$. We recall that for odd FP, we derived in Sec.IVA that instead of the antiperiodic boundary condition, a periodic boundary condition must be satisfied by the CS fermions, resulting in the discrete $\mathbf{k}$-space indicated by the lower plane in Fig.9(c). Compared to the $\mathbf{k}$-space for even FP, the key difference here for the odd parity is that the $\mathbf{k}=\mathbf{Q}=(\pi,\pi)$ point is now a physical state occupied by a CS fermion. As shown by the Appendix C, the CS fermion operator occupying $\mathbf{k}=\mathbf{Q}$ is a superposition of CS fermions on different A and B lattices, i.e, $\tilde{\hat{f}}_{\mathbf{Q}}=\frac{1}{\sqrt{2}}(\hat{f}_{\mathbf{Q},A}-\hat{f}_{\mathbf{Q},B})$. Then, the ground state wave function for the odd FP sector is given by
\begin{equation}\label{eq76}
  |GS\rangle_o=-\tilde{\hat{f}}^{\dagger}_{\mathbf{Q}} \prod_{\mathbf{k}}U_{\mathbf{k}}e^{G_{ab}f^{\dagger}_{-\mathbf{k},a}f^{\dagger}_{\mathbf{k},b}}|0\rangle.
\end{equation}
One can readily check that $|GS\rangle_o$ is also a Bogoliubov vacuum. Therefore, we have analytically extracted two Bogoliubov vacuum wave functions $|GS\rangle_e$, $|GS\rangle_o$, for the even and odd FP case, respectively. $|GS\rangle_o$ differs from $|GS\rangle_e$ by the creation of an addition CS fermions. In the thermodynamic limit, the reciprocal lattices for the even and odd FP approach to each other, resulting in the doubly degenerate ground state, $|GS\rangle_e$ and $|GS\rangle_o$.


\subsection{Measurement of the N\'{e}el  spin order parameter from a CS superconductor}
With all the above preparations, we are now able to study the spin ordering  of a CS superconductor.
We are interested in the thermodynamic limit where the two Bogoliubov vacuum states are degenerate, as shown by the degenerate energy levels with different ground states in Fig.10. Because of the degeneracy, it is difficult to obtain useful physical information by directly considering the expectation value of a spin operator, e.g.,
$\hat{S}^x_{\mathbf{r}}=(\hat{S}^++\hat{S}^-)/2$ or $\hat{S}^y_{\mathbf{r}}=(\hat{S}^+_{\mathbf{r}}+\hat{S}^-_{\mathbf{r}})/(2i)$,
because it seems that the true ground state can be a generic superposition of $|GS\rangle_o$ and $|GS\rangle_e$. Formally, if we evaluate the spin $\hat{S}^x$ operator with either one of the two Bogoliubov vacuum states, we can write down
\begin{equation}\label{eq77}
  \langle \hat{S}^x_{\mathbf{r}}\rangle= _{o,e}\langle GS|\hat{S}^x_{\mathbf{r}}|GS\rangle_{{o,e}},
\end{equation}
\begin{equation}\label{eq78}
  \hat{S}^x_{\mathbf{r}}=\frac{1}{2}(\hat{f}^{\dagger}_{\mathbf{r}}e^{i\alpha_{\mathbf{r}}}+\hat{f}_{\mathbf{r}}e^{-i\alpha_{\mathbf{r}}}),
\end{equation}
where $\alpha_{\mathbf{r}}$ is a string of operators defined as $e^{\pm i\alpha_{\mathbf{r}}}=U^{\pm}_{\mathbf{r}}$ from Eq.\eqref{eq3}.  Since $\alpha_{\mathbf{r}}$ consists of billinear combinations of CS fermion operators, $\hat{S}^x_{\mathbf{r}}$ contains odd number of fermionic operators. $\hat{S}^x_{\mathbf{r}}$ therefore changes the FP of the ground state, leading to $\langle \hat{S}^x_{\mathbf{r}}\rangle=0$ for both $|GS\rangle_o$ or $|GS\rangle_e$, which seems to be in contradictory with the planar N\'{e}el order.
\begin{figure}[bp]\label{fig10}
\includegraphics[width=\linewidth]{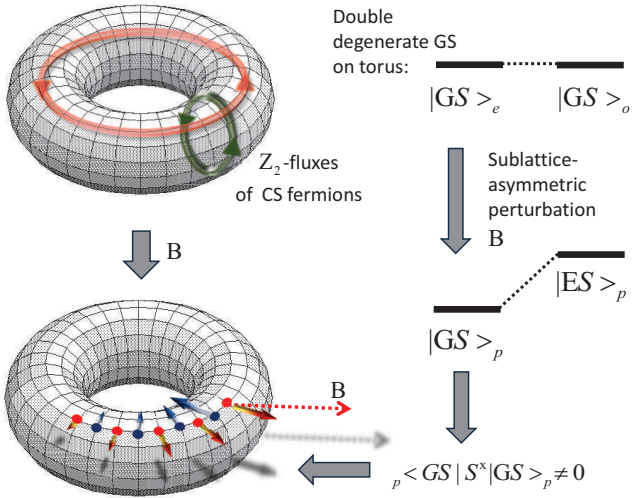}
\caption{(color online) The schematic plot for the emergence of the staggered spin susceptibility. The CS superconductor on a torus enjoys an additional $Z_2$-flux and the double degenerate ground state (GS) in the thermodynamic limit. With applying an infinitesimal sublattice-asymmetric perturbation $B$, the double degeneracy of the ground state is lifted, resulting in a finite magnetization $_{p}\langle GS|\hat{S}^x|GS\rangle_{p}\neq0$. The field-induced magnetization is staggered with respect to different sublattices. The blue and red dots and the opposite arrows schematically represent for the induced spin order along the $x$-direction on the torus. }
\end{figure}

It should be noted that one expects the spontaneous symmetry breaking only in the thermodynamic limit, however Eq.\eqref{eq77}, Eq.\eqref{eq78} has ambiguity when applied in the thermodynamic limit where the ground state can be a superposition of the two degenerate Bogoliubov vacuum states. Therefore, instead of calculating the spin order directly, one should resort to other approaches. One way is to
calculate of spin-spin correlation function $\langle \hat{S}^i_{\mathbf{r}}\hat{S}^j_{\mathbf{r}^{\prime}}\rangle$ instead of the expectation value of spins. However, once transforming to CS fermions, the spin-spin correlation function acquires complicated combinations of string operators whose analytic derivation is complicated. Here, we are only interested in a qualitative physical property of the CS superconductor ground state. Therefore, we adopt an alternative method that is commonly used to capture the spontaneous symmetry breaking of a system with degenerate ground states in thermodynamic limit. Namely, rather than calculating the spin order directly, we focus in the following on the ``susceptibility" of the system under the application of an infinitesimal local external field.

Theoretically, to probe the spin order of the system, we apply an infinitesimal perturbation, i.e., a local magnetic field to the CS superconductor. Note that the CS superconductors, physically different from normal superconductors, are not bothered by the Meissner effect, and a local magnetic field at $\mathbf{r}_0$ with strength $B$ is coupled to the CS fermions in the following way:
\begin{equation}\label{eq79}
  H^{\prime}=-B\hat{S}^x_{\mathbf{r}_0}=-\frac{B}{2}[\hat{f}^{\dagger}_{\mathbf{r}_0,a}e^{i\alpha_{\mathbf{r}_0}}+h.c.].
\end{equation}
Here, $a$ is the sublattice index, $a=A$ or $B$, depending on to which sublattice the local field is applied. $H^{\prime}$ acts as a perturbation to the ground state. Therefore, given a CS superconductor on a torus and in the thermodynamic limit, we can solve the problem by using a degenerate perturbation theory in the two-dimensional Hilbert space expanded by $|GS\rangle_o$ and $|GS\rangle_e$. The perturbation matrix in this space reads as:
\begin{equation}\label{eq80}
  H^{\prime}=\left(
               \begin{array}{cc}
                 _{o}\langle GS| H^{\prime}|GS\rangle_o & _{o}\langle GS| H^{\prime}|GS\rangle_e \\
                 _{e}\langle GS| H^{\prime}|GS\rangle_o & _{e}\langle GS| H^{\prime}|GS\rangle_e \\
               \end{array}
             \right)
\end{equation}
It is easy to see that the diagonal terms, $_{o}\langle GS| H^{\prime}|GS\rangle_o= ~_{e}\langle GS| H^{\prime}|GS\rangle_e=0$ because $H^{\prime}$ changes the FP. The off-diagonal term is then cast into:
\begin{equation}\label{eq81}
\begin{split}
  &_{o}\langle GS| H^{\prime}|GS\rangle_e=\langle\tilde{\hat{f}}_{\mathbf{Q}} H^{\prime}\rangle_e\\
  &=\langle(\hat{f}_{\mathbf{Q},A}-\hat{f}_{\mathbf{Q},B})[(-\frac{B}{2\sqrt{2}})(\sum_{\mathbf{k}}\hat{f}^{\dagger}_{\mathbf{k},a})e^{i\alpha_{0}}+h.c.]\rangle_e,
\end{split}
\end{equation}
where we have set $\mathbf{r}_0=0$ without losing any generality and $\langle...\rangle_{o,e}$  is short for $_{e,o}\langle GS|...|GS\rangle_{e,o}$.  The remaining task is then to calculate the off-diagonal matrix. After a length derivation included in the Appendix E, we find that the two off-diagonal terms are finite constants, as long as the perturbation is sublattice-asymmetric. Thus, under such infinitesimal perturbation, the doubly degenerate ground state is mixed, forming a perturbed ground state, $|GS\rangle_p$, as indicated by Fig.10. We then evaluate the spin order with respect to the perturbed ground state and calculate the spin susceptibility defined by $\chi_{\alpha}=\lim_{B\rightarrow0}\frac{_p\langle GS|\hat{S}^x_{\mathbf{r},\alpha}|GS\rangle_p}{B}$, with $\alpha=A,B$. Then, we find out the A and B sublattice enjoy the oppositely divergent susceptibility  $\chi_B=-\chi_A=\infty$ as an inherent feature of the CS superconductors in the thermodynamic limit. It indicates that CS superconductors have a natural tendency to lift the double degeneracy. This observation directly shows the physical correspondence of the CS superconductor and a planar N\'{e}el AFM state with respect to the spin ordering, accomplishing the third correspondence indicated by the dashed box in Fig.2. The above calculations and main results are summarized in Fig.10.

To summarize this section, we note that the CS superconductor description, followed from the CS fermionization, indeed bares physical correspondence with the N\'{e}el AFM state. The two states display collective modes in remarkable consistency with each other at the quantitative level. We also revealed that the CS superconductors, derived from the 2D spin-half XY modes, have a doubly degenerate ground state on a torus in the thermodynamic limit, which leads to a strong and intrinsic susceptibility to staggered magnetization for different sublattices. These results clearly demonstrate that the CS superconductor is a physical description of N\'{e}el AFM state, which is entirely different in formalism from the spin-wave theory.


\section{Unconventional phase transition as an instability of the Chern-Simons superconductor}
\subsection{General scheme for CS fermionization with frustration}
This section further discusses the remaining generalization of the above theory, which makes it applicable to frustrated spin models where an unconventional phase transitions may occur. Now let us follow Fig.2 and increase some parameter $g$
describing the frustration of the system. The frustration can be introduced in many cases by considering further neighboring interactions; therefore, we can consider variations of the exchange couplings in a given spin model. We restrict ourselves to a theoretical situation where the model can cross an unconventional phase transition starting from the N\'{e}el AFM order to a QSLs with tuning the couplings (see, for example, Ref.~\onlinecite{triangular}). Our focus in this section is to provide a systematic framework for studying such transitions using the proposed methods.

Let us firstly consider Eq.\eqref{eq1} with the most general couplings $J_{\mathbf{r},\mathbf{r}^{\prime}}$. We regard the coupling as $J_{\mathbf{r},\mathbf{r}^{\prime}}(g)$ as tunable in the Hamiltonian $H_{XY}[J_{\mathbf{r},\mathbf{r}^{\prime}}(g)]$, where $g\in[0,1]$ is an introduced parameter that characterizes the change of $J_{\mathbf{r},\mathbf{r}^{\prime}}$. We assume that for $g=0$, the theory starts from a simple situation, e.g., only the nearest neighbor coupling $J_{\mathbf{r},\mathbf{r}^{\prime}}=J_{\langle\mathbf{r},\mathbf{r}^{\prime}\rangle}$, which leads to a N\'{e}el AFM state. With gradually increasing $g$, the coupling $J_{\mathbf{r},\mathbf{r}^{\prime}}$ finds itself being tuned, which finally generates a complicated spin model with strong frustration at $g\sim 1$. The traditional spin wave theory can only be well applied to the starting point of the theory with small $g$, while it can hardly capture the ground state in the strong frustration regime with large $g$, nor can it describe the whole transition process.

Following the procedure in Sec.II and Sec.III, for a generic Hamiltonian $H_{XY}[J_{\mathbf{r},\mathbf{r}^{\prime}}(g)]$ with a general coupling $J_{\mathbf{r},\mathbf{r}^{\prime}}$, we can always fermionize the spin model exactly to the theory of CS fermions coupled to lattice gauge field, as in Eq.\eqref{eq4} but with $J_{\mathbf{r},\mathbf{r}^{\prime}}$ a tuning parameter,
\begin{equation}\label{eq95a}
 H(g)=\sum_{\mathbf{r},\mathbf{r}^{\prime}}J_{\mathbf{r},\mathbf{r}^{\prime}}(g)(f^{\dagger}_{\mathbf{r}}e^{iA_{\mathbf{r},\mathbf{r}^{\prime}}}f_{\mathbf{r}^{\prime}}+h.c.).
\end{equation}
Here and in what follows, we omitted the CS charge $e$ for brevity.  With tuning $g$, the variation of $J_{\mathbf{r},\mathbf{r}^{\prime}}$ is mapped to the tuning of the hopping coefficients of the CS fermions in Eq.\eqref{eq95a}. Moreover, the change of the ground state with tuning $g$ can generally lead to the changing of CS fermion density $n_{\mathbf{r}}$ which in turn reshapes the gauge field $A_{\mathbf{r},\mathbf{r}^{\prime}}$, which is a string operator consists of $n_{\mathbf{r}}$ operators.
Therefore, both the gauge field and the ground state of $f$-fermions are dependent on $J_{\mathbf{r},\mathbf{r}^{\prime}}(g)$ and transform with increasing $g$. The unconventional phase transition at $g=g_c$ then must be accompanied by a qualitative change of the behavior of the gauge field $A_{\mathbf{r},\mathbf{r}^{\prime}}$. As has been studied in previous sections, for $g\ll g_c$, the CS superconductor ground state suggests that the gauge field behaves as the glue that sticks two CS fermions together with $p\pm ip$ wave pairing symmetry. The condensation of Cooper pairs of CS fermions generates a mass gap for the gauge field and ``Higgs" its $\mathrm{U(1)}$ gauge symmetry to $\mathrm{Z}_2$ (which is further broken by infinitesimal field in thermodynamics limit as discussed in the last section). With increasing $g$, one can expect that the gauge field will start to lose its viscosity for $g\sim g_c$ in order to generate a deconfined phase of CS fermions.

It is difficult to obtain the evolution of the ground state and the gauge field directly from Eq.\eqref{eq95a}. It should be noted that the CS fermionization is, in essence, a nonlocal representation of spin operators by fractionalized particles. Thus, it is possible to describe the QSL state for $g\sim 1$ using the same fractionalized particles and Chern-Simons gauge field, akin to the conventional parton theories. In slave-particle theories, one usually obtains the mean-field solutions of possible disordered states and then achieves the nature of the QSL by going beyond the mean-field, which leads to the gauge field fluctuation. Here, we illustrate in the following how to construct a similar mean-field theory to describe the QSLs at $g\sim 1$ using CS fermions. Moreover, one great advantage of our method is that it could provide a ``global" Landau-type mean-field theory, which can not only describe  the QSL at $g\sim 1$, but also the  CS superconductors at $g=0$.

At $g=0$, we have established a stable CS superconductor mean-field theory. We can take $g=0$ as the starting reference point and consider gradually increasing $g$ from zero. For weak $g$, despite the gauge field, the low-energy theory is the Dirac CS fermions Eq.\eqref{eq11}, which are the symmetry-protected Kramer's degeneracies as shown by Fig.4(e). We consider the case where the tuning of $g$ does not break the corresponding symmetry and keep the gapless nodes intact, as is the case when introducing further neighboring interactions on honeycomb or square lattice. Then, we generally arrive at the following low-energy effective Hamiltonian after switching on the gauge field, i.e.,
\begin{equation}\label{eq96}
  H=\sum^N_{i=1}\sum_{\mathbf{r},\alpha,\beta}f^{\dagger}_{\mathbf{r},i,\alpha}\epsilon_{i,\alpha\beta}(-i\boldsymbol{\nabla}+\mathbf{A}_{\mathbf{r}})f_{\mathbf{r},i,\beta},
\end{equation}
where $\alpha$, $\beta$ are the subscripts for sublattices, $i=0,1,...,N$ denotes the $i$-th gapless nodes in the first BZ. $\epsilon_{i,\alpha\beta}$ is the low-energy Hamiltonian around the $i$-th gapless node with the momentum $\mathbf{k}=-i\boldsymbol{\nabla}$ measured from the corresponding node. For $g=0$, the effective single-particle Hamiltonian $\epsilon_{i,\alpha\beta}$ is reduced to the Dirac Hamiltonian in Eq.\eqref{eq11}, while with tuning $g$, correction terms take place and are included in the $\epsilon_{i,\alpha\beta}$ term,  leading to deviation from the linear dispersion which is less manifested for larger wave vectors away from the gapless nodes. Taking into the correction terms, we can make Taylor expansion of $\epsilon_{i,\alpha\beta}$ in terms of $\mathbf{k}$, leading to
\begin{equation}\label{eq96a}
  \epsilon_{i,\alpha\beta}(-i\boldsymbol{\nabla}+\mathbf{A}_{\mathbf{r}})=\epsilon^{(1)}_{i,\alpha\beta}(-i\boldsymbol{\nabla}+\mathbf{A}_{\mathbf{r}})+\epsilon^{(2)}_{i,\alpha\beta}(-i\boldsymbol{\nabla}+\mathbf{A}_{\mathbf{r}})+...
\end{equation}
where $-i\nabla+\mathbf{A}_{\mathbf{r}}$ appear in the low-energy window as the argument of function $\epsilon_{i,\alpha,\beta}$. The first and the second term in Eq.\eqref{eq96a} represent for the linear and quadratic expansion, respectively, and the ellipse denotes for the higher order corrections.

The low-energy effective Hamiltonian Eq.\eqref{eq96} can be further separated into a non-interacting pure fermionic model,
\begin{equation}\label{eq97}
  H_0=\sum^N_{i=1}\sum_{\mathbf{r},\alpha,\beta}f^{\dagger}_{\mathbf{r},i,\alpha}\epsilon^{(1)}_{i,\alpha,\beta}(-i\boldsymbol{\nabla})f_{\mathbf{r},i,\beta},
\end{equation}
where the quadratic and higher order kinetic terms are irrelevant in long-wave length regime, and a gauge field term
\begin{equation}\label{eq98}
  H_g=\sum^N_{i=1}\sum_{\mathbf{r},\alpha,\beta}f^{\dagger}_{\mathbf{r},i,\alpha}\tilde{\epsilon}_{i,\alpha,\beta}(\mathbf{A}_{\mathbf{r}})f_{\mathbf{r},i,\beta},
\end{equation}
where  $\tilde{\epsilon}_{i,\alpha,\beta}$ is the expansion of the gauge field from Eq.\eqref{eq96a}. Namely
\begin{equation}\label{eq98a}
  \tilde{\epsilon}_{i,\alpha,\beta}(\mathbf{A}_{\mathbf{r}})=\epsilon^{(1)}_{i,\alpha\beta}(\mathbf{A}_{\mathbf{r}})+\epsilon^{(2)}_{i,\alpha\beta}(i\boldsymbol{\nabla}\cdot\mathbf{A}_{\mathbf{r}})
  +\epsilon^{(2)}_{i,\alpha\beta}(\mathbf{A}^2_{\mathbf{r}})+...
\end{equation}
As has been introduced in Sec.IIB, we take into account the CS action and integrate out the gauge field in Eq.\eqref{eq98}, leading to $H_I=H^{(1)}_I+H^{(2)}_{I}+...$, where the two interactions formally can be written as
\begin{equation}\label{eq99}
  H^{(1)}_I=\sum_{ij}\sum_{\mathbf{r},\alpha,\beta,\rho,\sigma}V^{(1)\alpha,\beta,\rho,\sigma}_{i,\mathbf{r}-\mathbf{r}^{\prime}}f^{\dagger}_{\mathbf{r},i,\alpha}f_{\mathbf{r},i,\beta}f^{\dagger}_{\mathbf{r}^{\prime},j,\rho}f_{\mathbf{r}^{\prime},j,\sigma},
\end{equation}
and
\begin{equation}\label{eq100}
  H^{(2)}_I[g]=\sum_{ij}\sum_{\mathbf{r},\alpha,\beta,\rho,\sigma}V^{(2)\alpha,\beta,\rho,\sigma}_{i,\mathbf{r}-\mathbf{r}^{\prime}}f^{\dagger}_{\mathbf{r},i,\alpha}f_{\mathbf{r},i,\beta}f^{\dagger}_{\mathbf{r}^{\prime},j,\rho}f_{\mathbf{r}^{\prime},j,\sigma}.
\end{equation}
$H^{(1)}_I$ is the gauge-field induced interaction originated from $\epsilon^{(1)}_{i,\alpha\beta}(\mathbf{A}_{\mathbf{r}})$ in Eq.\eqref{eq98a}. $H^{(2)}_I$ is the newly generated interaction by frustration $g$, which is originated from $\epsilon^{(2)}_{i,\alpha\beta}(i\boldsymbol{\nabla}\cdot\mathbf{A}_{\mathbf{r}})$ in Eq.\eqref{eq98a}. It is model-dependent and therefore is not written explicit here for the general analysis.
Note that more higher-order interactions can emerge  from the higher order expansion of $A_{\mathbf{r}}$, which are denoted by the ellipsis in $H_I$.

Above, we have formally mapped the frustrated spin exchange model to a CS fermion model with competing interaction, $H=H_0+H^{(1)}_I+H^{(2)}_I+...$. With $g=0$, the Hamiltonian $H$ is reduced to $H_0+H^{(1)}_I$,  where $H_0$ describes the Dirac CS fermions and $H^{(1)}_I$, e.g., reads as Eq.\eqref{eq20} on the honeycomb lattice, thereby leading to the CS superconductors as studied before. Therefore, after the CS fermionization, the effect of frustration is mapped to more competing interactions between CS fermions, $H^{(2)}_I[g]$.
This fermionic picture provides a systematic way to investigate the unconventional quantum phase transitions. Since the mapping is mathematically exact in the long-wavelength limit and the orders of expanded interaction are controllable in a perturbative sense, we expect to have a  mean-field theory by studying the competition of gauge-field-induced interactions on CS fermions. Here, the CS superconductor is stabilized for $g\ll g_c$ is destabilized by $H^{(2)}_I[g]$ and the higher-order terms.

To observe the effect of the frustration, we can first study the Hamiltonian $H_0+H^{(2)}_I[g]$ with large $g$ (and higher-order correction term if necessary), with neglecting the interaction $H^{(1)}_I$ whose effect is to stabilize the CS superconductors. The traditional many-body theories, such as the perturbation renormalization group, can be applied to determine the most favorable mean-field orders from $H_0+H^{(2)}_I[g\sim 1]$. Hubbard-Stratonovich transformation can then be formulated by introducing bosonic orders consist of a bilinear combination of CS fermions, which leads to a mean-field description of the possible ground state with neglecting the fluctuation of the bosonic orders. This is in analogy with the slave-particle mean-field description of QSLs, where mean-field orders are ground state expectation values of bilinear terms formed by slave particles \cite{xiaogangwenw}. Therefore, the self-consistent solution stabilized by $H_0+H^{(2)}_I[g\sim 1]$ can capture the deconfined phase at with strong frustration at the mean-field level in the CS fermion language. After obtaining the mean-field orders at large $g$, we can study the total Hamiltonian $H_0+H^{(1)}_I+H^{(2)}_I[g]$ by mean-field treatment to both $H^{(1)}_I$ and $H^{(2)}_I$ at the same time. To this end, one can introduce simultaneously the mean-field order parameter for both the CS superconductor, $\Delta_{\alpha\beta}$, and the deconfined phase, say $\chi_{\alpha\beta}$, and search for a self-consistent solution of $\Delta_{\alpha\beta} (g)$ and $\chi_{\alpha\beta}(g)$ with tuning $g$.

Not all types of instabilities of CS superconductors suggest unconventional phase transitions, as indicated by the dashed arrow at the bottom of Fig.2.  One has to further explore the physical nature of the resultant CS mean-field state by going beyond the mean-field theory. By considering the fluctuations of the mean-field order parameters and integrating out the fermionic fields, we will arrive at a low-energy effective theory that implies whether any topological excitations exist in the state predicted by the CS mean-field theory. On the other hand, we can explicitly examine whether the CS mean-field state breaks any symmetries of the original spin Hamiltonian or not. If a completely disordered mean-field state is observed with emergence of topological excitations, then we can safely draw the conclusion that the system ends up into a QSL. For example, if the CS mean-field state breaks TRS and parity symmetry, and meanwhile enjoys a low-energy effective theory with a CS term whose coefficient is $K=2$, then a chiral spin liquid is found, as the result of the instability of the CS superconductors \cite{ruinew}, as illustrated by Fig.2.

\subsection{Discussion of the applications to  specific models}
After presenting the general scheme above, we discuss the application of the method to study the unconventional phase transitions from AFM to QSLs. Our discussion will be based on two specific models, namely, the $J_1$-$J_2$ XY model on the triangular lattice and the honeycomb lattice, respectively.

With the application of the general scheme in the last section to specific models, we can generally map the frustrated quantum spin models to Dirac fermions with multiple interactions, as in Eq.\eqref{eq99} and Eq.\eqref{eq100}. For example, for both triangular and honeycomb $J_1$-$J_2$ XY model, the emergent spinless CS Dirac fermions are subject to inter-and intra-valley interactions. However, the two models are different from each other in the following two aspects. First, the emergent interactions enjoy different forms as a result of the distinct lattice symmetries.  Second, there is a threefold degeneracy of Dirac nodes at each of the two valleys for the triangular lattice, making the total fermion flavor 6 in the first BZ. In comparison, there are no additional degeneracies at the Dirac valleys for the honeycomb model. Therefore the total fermion flavor is only 2.

The total fermion flavor is an important factor that affects the ground state, as it has been proved in the context of parton mean-field theory that gapless Dirac spin liquid is stable against gauge fluctuations under the large N limit \cite{Hermele}. This indicates that, on the triangular lattice, the frustration can favor the gapless Dirac QSL after the CS superconductor is destabilized, leaving the CS Dirac fermions intact. However, the CS Dirac fermions are more fragile against gauge fluctuations on the honeycomb lattice, generating gapped phases. Using the scheme proposed in the last section, we verified the above expectation via detailed mean-field calculations.  Here, we generally outline the new findings while the detailed calculations and results are presented in Ref. \cite{ruinew} and Ref.\cite{triangular}

For the $J_1$-$J_2$ XY model on the honeycomb lattice, we find that the CS superconductor phase is stable for $J_2/J_1\lesssim0.22$, corresponding to a planar N\'{e}el AFM, as indicated by Fig.11. For $J_2/J_1\gtrsim0.22$ and after applying a small TRS-breaking perturbation, the newly generated fermion-fermion interaction $H^{(2)}_I[g]$ destabilizes the CS superconductors and drives the system into a topological excitonic insulator phase, where the CS fermions and their hole excitations are paired, as indicated by Fig.11(b). The exciton order parameter gaps out the Dirac nodes and gives rise to the Chern number $1/2$ for each of the two Dirac valleys. Therefore, the topological exciton insulator is characterized by total Chern number $C=1$, and consequently, exhibits the chiral edge state along the boundaries.  Moreover, this state is coupled to the Chern-Simons gauge field with level $k=2$, implying the existence of semionic excitations, a fingerprint signature of CSL. Therefore, our methods, after application to the perturbed $J_1$-$J_2$ XY model on the honeycomb lattice, predict an interesting, unconventional phase transition from the planar N\'{e}el AFM to the chiral spin liquid. Tensor network calculations also support this \cite{ruinew}, which reveals numerical signatures for chiral spin liquid for $J_2/J_1\gtrsim0.22$, e.g., the entanglement spectrum consistent with the $\mathrm{SU}(2)_1$ Wess-Zumino-Novikov-Witten conformal field theory.
\begin{figure}[tbp]\label{fig11}
\includegraphics[width=\linewidth]{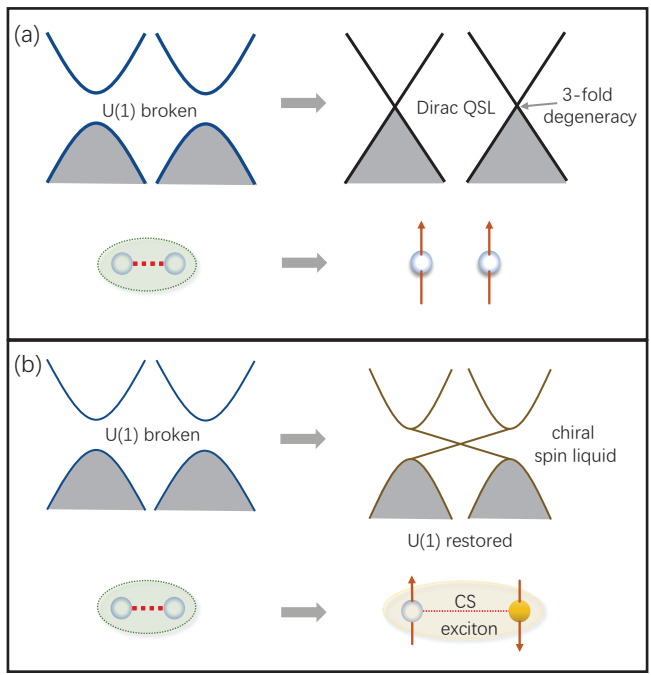}
\caption{(color online) Illustration of two different types of the instabilities of CS superconductors, both of which lead to QSLs. (a) The transition from the N\'{e}el AFM to the gapless Dirac QSL is predicted for the $J_1$-$J_2$ XY model on the triangular lattice. In the fermion picture, the transition is cast into the destabilization of the $\mathrm{U}(1)$ broken CS superconductor. Under strong frustration, the CS fermions are set free from the Cooper pairs and then behave as itinerant and deconfined fermions coupled to gauge field. The linear Dirac dispersion is maintained. (b) The transition from N\'{e}el AFM to the chiral spin liquid is predicted for the perturbed $J_1$-$J_2$ XY model on honeycomb lattice. In the fermion picture, this topological phase transition is cast in mean-field level as the transition from the CS superconductor to a topological CS exciton insulator. The latter phase is coupled to CS gauge fluctuations with level $k=2$. }
\end{figure}

For the $J_1$-$J_2$ XY model on the triangular lattice \cite{triangular}, instead of the gapped QSL predicted on the honeycomb lattice, a gapless helical Dirac spin liquid can be found using our proposed scheme. In this model, the CS fermions show robustness against gauge fluctuation because of the large number of fermion flavors. Consequently, the newly generated fermion-fermion interaction $H^{(2)}_I[g]$ cannot easily gap out the Dirac nodes. Thus, for the strong frustration regime, the ground state is gapless Dirac CS fermions after the CS superconductor is destabilized. This is pictorially illustrated by Fig.11(a).

Based on the above results, now we can summarize what the proposed methods have taught us about topological phase transitions. It is known from above that there are two different types of instabilities of CS superconductors, although they both generate QSLs with fractionalized excitations and gauge fluctuations. First, gapless Dirac spin liquids are possible if the fermion flavors are large enough, as with the $J_1$-$J_2$ XY model on the triangular lattice. The Cooper pairs of CS fermions become unstable for strong frustration, and the CS fermions are no longer paired, setting free the CS Dirac fermions. The latter behave as the itinerant deconfined particles coupled to the gauge field. This type of instability of CS superconductors generates gapless QSLs, whose underlying mechanism is schematically plotted in Fig.11(a).  Second, the CS fermions from the $\mathrm{U}(1)$-broken Cooper pairs can also form other types of orders that respect the $\mathrm{U}(1)$ symmetry, gapping out the Dirac nodes, as we discussed on the honeycomb lattice. In this case, the fluctuation of the order parameter can be nontrivial as long as the gapped fermionic mean-field state enjoys nonzero Chern numbers. This leads to emergent CS terms coupled to the fermions and alters their statistics, generating the anyonic excitations of QSLs. An example of this kind (chiral spin liquid) is pictorially illustrated in Fig.11(b).

Importantly, it should be noted that the proposed method implies a vital message, namely, specific topological phase transitions can still be captured within the familiar mean-field framework, as long as the CS representation is adequately adopted. This construction should endow the mean-field theories with new applications in previously inaccessible fields. We note that some other related topics, as well as examples, can be found in Refs.~\onlinecite{Tigrana,ruia,triangular,kagome,MS2,SGK2,SGK1}.


%

\section{Conclusion and Discussion}
The QSLs and the high $T_c$ superconductors, which are descendants states from long-range antiferromagnetic order, have generated enormous theoretical and experimental interest for the last decades. On the experimental front, on one hand, new candidate materials for QSLs and their properties at low temperatures have been discovered and reported in recent years \cite{Jackeli,Chaloupka,ysingh,Gretarsson,hschun,yjyu}. On the other hand, new findings in cuprate superconductors have motivated more in-depth investigations of the pairing mechanisms of high-$T_c$ materials \cite{Grissonnanche,bmichon}. Theoretically, much progress has been made in both fields. Based on the development of large-scale numerical techniques such as quantum Monte-Carlo, DMRG, and tensor network states, there have been increasing accumulations of numerical evidence of the QSL ground states in realistic models related to materials. Due to the possible connections between the QSLs and the high $T_c$ superconductors, the thorough understanding of the former (especially on square lattices) could provide much insights to the pseudogap regime \cite{cmvarmma,tvallaa}, the bad metal \cite{vjemery,seana} and the non-Fermi liquid \cite{Stewart,Hiroshii}, which are still open questions in condensed matter physics. Besides, concerning the formation of QSLs, a long-standing and crucial question is to obtain comprehensive understandings of the unconventional phase transitions from a long-range magnetically ordered state, e.g., the N\'{e}el AFM state, to the QSLs.

Both QSLs and high-$T_c$ superconductors can be generated, in many physical situations, from perturbation of a N\'{e}el AFM order.
In this work, we propose a different theoretical view of the N\'{e}el AFM state, which could be useful and shed light on the unconventional phase transitions, and possibly the relation between QSLs and high $T_c$ superconductors.

Specifically, we systematically study the 2D XY spin models whose ground states are N\'{e}el AFM states, using the CS representation. Effective superconducting states are obtained, where the CS fermions form pairs due to the effect of the gauge field. To verify the physical correspondence between the CS superconductor and the N\'{e}el AFM state, we show:

(i) The collective modes evaluated in each of the two theories are equivalent to each other. These include equivalence of (a) the phase fluctuation mode of the CS superconductor with magnons of the N\'{e}el state
and  (b) the Higgs mode of the CS superconductor with the longitudinal mode of the N\'{e}el state.

(ii) The magnetic susceptibility evaluated from the ground state of the CS superconductor strongly suggests the formation of antiferromagnetic N\'{e}el ordering. It shows a strong response to weak perturbations, thus the staggered magnetization is generated, consistent with the N\'{e}el AFM state.

These results convincingly support the proposal that the CS superconductor can be an alternative description that captures the major physics of the planar N\'{e}el AFM long-range order, based on the language of fractionalized excitations as well as gauge field.

 The usefulness of this new construction becomes clear when the nontrivial perturbation is applied onto the N\'{e}el AFM, resulting in possible QSLs for strong frustration or high-$T_c$ superconductors with proper doping. To demonstrate this point, we make further generalization of the theory of CS superconductors, and propose a general scenario  to study an unconventional phase transition from the N\'{e}el AFM order to QSLs. Some results obtained from specific frustrated spin models are also discussed.

In this scenario, the key advantage of the introduced CS superconductor description becomes fully manifested: it describes the N\'{e}el state using the fractionalized excitations and the gauge field, which are the most crucial degrees of freedom that characterize QSLs, as illustrated by Table.I. Therefore, after understanding the N\'{e}el AFM as an effective superconducting state in this language, it becomes possible to construct a global self-consistent mean-field theory to account for many unconventional phase transitions that are originally beyond the Landau's paradigm. Therefore, the proposed CS mean-field theory is advantageous as compared to earlier proposals as it treats both QSL and ordered states using the same sets of degrees of freedom.

We now conclude by discussing possible future directions. First, it is interesting to apply the theory to more concrete frustrated XY models, which can host QSLs. The technique,  together with some support from large-scale numerical calculations, can also be used as a systematic way to predict novel phases and determine the phase diagrams.

Second, there is a need to generalize the CS superconductor description from the planar XY N\'{e}el order to describe the N\'{e}el AFM state stabilized in models with full $\mathrm{SU}(2)$ symmetry (Heisenberg models). The Ising term of the Heisenberg model brings about an additional interaction between CS fermions. The Hubbard-Stratonovich decoupling of this interaction would lead to additional mean-field orders whose fluctuation in general leads to an emergent $\mathrm{SU}(2)$ gauge field rather than the $\mathrm{U}(1)$ gauge field studied in this work. It is interesting to study how the system can stabilize the N\'{e}el AFM order by breaking the spin $\mathrm{SU}(2)$ symmetry in the language of CS fermions. Once this generalization is developed, the CS fermionic field theory can be applied to more spin models relevant to realistic materials.

Third, a more exciting application would be the study of the doping effect of CS superconductors. With additional doped carriers, one would expect the coupling between the CS fermions and the carriers can play a vital role that drives the instability of CS superconductors. Important questions along this line include: whether there is a proximity effect of the effective Cooper pairs onto the doped carriers, and if yes, what is the pairing symmetry of the induced superconductivity?

Fourth, it is straightforward to generalize the current zero temperature theory based on CS fermions to finite temperatures. It is interesting to consider thermal effects in the regime where a QSL is stabilized. Besides, at finite temperatures, the Kosterliz-Thouless transition, accompanied by the proliferation of vortices and anti-vortices, will automatically take place in the XY models with the N\'{e}el AFM state ground state. It would be very natural to look for signatures of vortex solutions in the CS fermion picture.

Fifth, the proposed fermionization scheme can also find promising applications in impurity problems in frustrated magnets. Our recent study on the non-magnetic impurity problem in the flux phases of spin liquids is a typical example \cite{ruiwangkondo}, where Kondo behavior was found to take place as a result of the gauge fluctuations after using the Chern-Simons fermionization. The emergent Kondo phenomena can serve as the finger-print experimental feature to identify the deconfined phase with gauge fluctuations.

 Last, since the CS fermionic representation contains string-type nonlocal operators, it is worthwhile to apply this transformation to some exactly-solvable models that host a topologically ordered ground state with long-range quantum entanglement. The nontrivial topology of the exactly known ground state can find its physical manifestation in combinations of string operators and CS fermions. This would provide physical insights towards the long-range quantum entanglements as well as their intimate connections with original model Hamiltonian, which will in turn bring about a richer understanding of the topologically ordered states.

 \begin{acknowledgments}
This work was supported by the Youth Program of National Natural Science Foundation of China (No. 11904225), the National Key  R\&D Program of China (Grant No. 2017YFA0303200), and by NSFC under Grants No. 11574217 and No. 60825402.
T.A.S. acknowledges startup funds from UMass Amherst.
\end{acknowledgments}

\appendix


\section{Technical details in derivation of the Higgs mode of CS superconductor}
In this section, we present the details of derivation of the Higgs mode of a CS superconductor on honeycomb lattice. The single-particle Green's function of the CS superconductor can be read off from its mean-field theory $H_{MF}$ in the main text. We keep the sublattice index explicit and write the Green's function as a set of matrices in the Nambu space, i.e., $G^0_{\alpha\beta}$. Specifically, they are given by
\begin{equation}\label{eq32}
  G^0_{11}=-\frac{1}{U}\left(
                         \begin{array}{cc}
                           a_1 & a_2 \\
                           a_2 & a_1 \\
                         \end{array}
                       \right),
 ~ G^0_{12}=-\frac{1}{U}\left(
                         \begin{array}{cc}
                           c_1 & c_2 \\
                           -c_2 & -c_1 \\
                         \end{array}
                       \right),
\end{equation}
\begin{equation}\label{eq33}
  G^0_{21}=-\frac{1}{U}\left(
                         \begin{array}{cc}
                           c^{\star}_1 & -c^{\star}_2 \\
                           c^{\star}_2 & -c^{\star}_1 \\
                         \end{array}
                       \right),
 ~ G^0_{22}=-\frac{1}{U}\left(
                         \begin{array}{cc}
                           b_1 & b_2 \\
                           b_2 & b_1 \\
                         \end{array}
                       \right),
\end{equation}
where $U$, $a_1$, $a_2$, $b_1$, $b_2$, $c_1$, and $c_2$ are functions of $\mathbf{k}$ and Matsubara frequency $i\omega_n$ as we utilize the imaginary-time formalism of propagators, which are of the following forms:
\begin{eqnarray}
\begin{split}
  U &= (\Delta^2_{3k}+|\mathbf{\Delta}_{0k}-v_F\mathbf{k}|^2+\omega^2)\\
  &\times (\Delta^2_{3k}+|\mathbf{\Delta_{0k}}+v_F\mathbf{k}|^2+\omega^2), \\
  a_1 &= i\omega(\mathbf{\Delta}^2_{0k}+\Delta^2_{3k} +v^2_Fk^2+\omega^2)+2\Delta_{3k}v_F\mathbf{\Delta}_{0k}\cdot\mathbf{k}, \\
  a_2 &= \Delta_{3k}(\Delta^2_{0k}+\Delta^2_{3k}+\omega^2+k^2v^2_F)+i2v_F\omega\mathbf{\Delta}_{0k}\cdot\mathbf{k}, \\
  b_1 &= i\omega(\mathbf{\Delta}^2_{0k}+\Delta^2_{3k}+v^2_Fk^2+\omega^2)-2\Delta_{3k}v_F\mathbf{\Delta}_{0k}\cdot\mathbf{k}, \\
  b_2 &= \Delta_{3k}(\mathbf{\Delta}^2_{0k}+\Delta^2_{3k}+v^2_Fk^2+\omega^2)-i2\omega v_F\mathbf{\Delta}_{0k}\cdot\mathbf{k}, \\
  c_1 &= v_Fk^+(\Delta_{0ky}+i\Delta_{0kx})^2+v_Fk^-(\Delta^2_{3k}+\omega^2+k^2v^2_F), \\
  c_2 &= (\mathbf{\Delta}^2_{0k}+\Delta^2_{3k}+\omega^2)(\Delta_{0kx}-i\Delta_{0ky})\\
  &-v^2_F(k^-)^2(\Delta_{0kx}+i\Delta_{0ky}).
  \end{split}
\end{eqnarray}
Here for short, we do not write explicitly the discrete frequency notation $n$ in the Matsubara frequency.  $k^{\pm}=k_x\pm ik_y$, $\mathbf{\Delta}_{0k}=(\Delta_{0kx},\Delta_{0ky})$. $\Delta_{0k}$ and $\Delta_{3k}$ acquire the self-consistent solution at small $k$ as $\Delta_{0k}\propto k$ and  $\Delta_{3k}$ a constant independent of $k$, as is discussed in the main text.

Casting the vertex renormalization of Fig.6 into the form of the integral equations, we obtain
\begin{equation}\label{eq35}
\begin{split}
  &\Gamma_{\beta^{\prime}\beta}(\mathbf{k}+\mathbf{q},\mathbf{k})\tau^+=\int\frac{d^3k^{\prime}}{(2\pi)^3}V^{\alpha\alpha^{\prime}\beta\beta^{\prime}}_{\mathbf{k}-\mathbf{k}^{\prime}}\\
  &\times\overline{\tau}^+G^0_{\alpha\gamma^{\prime}}(\mathbf{k}^{\prime})\Gamma_{\gamma^{\prime}\gamma}(\mathbf{k}^{\prime}+\mathbf{q},\mathbf{k}^{\prime})
  \tilde{\tau}^{a}G^0_{\alpha^{\prime}\gamma}(\mathbf{k}^{\prime}+\mathbf{q})\overline{\tau}^-,
\end{split}
\end{equation}
where we inserted the interaction in Eq.30 of the main text. This is the equation with respect to the $(1,2)$ Nambu component of $\Gamma$, therefore $\tau^+=\tau^x+i\tau^y$ takes place on the left hand side.  The $(2,1)$ Nambu components of $\Gamma$, proportional to $\tau^-=\tau^x-i\tau^y$, satisfy the same equations as Eq.\eqref{eq35} and therefore is omitted in the following calculation.
Since $\hat{\Gamma}_{12}=-\hat{\Gamma}^{\star}_{21}$, $\hat{\Gamma}_{11}=\hat{\Gamma}_{22}$, we only need to consider the equations with respect to   $\hat{\Gamma}_{11}$ and $\hat{\Gamma}_{12}$.  For purpose of simplicity, we introduced the Pauli matrix $\tilde{\tau}^{a}$, with $a=x,y$, defined as $\tilde{\tau}^{x}=\tau^x$ and $\tilde{\tau}^{y}=i\tau^y$. From above analysis, $\tilde{\tau}^{a}$ is the direction along which $\hat{\Gamma}_{\gamma^{\prime}\gamma}$ is aligned, therefore, $a$ takes the value of $x$ and $y$ for $\gamma=\gamma^{\prime}$ and $\gamma\neq\gamma^{\prime}$ respectively.
Writing explicitly the sublattice indices, then the equations we need to solve acquire the form
\begin{equation}\label{eq36}
\begin{split}
  &\Gamma_{11}(\mathbf{k}+\mathbf{q},\mathbf{k})\tau^+=\int\frac{d^3k^{\prime}}{(2\pi)^3}V^{\alpha\alpha^{\prime}11}_{\mathbf{k}-\mathbf{k}^{\prime}}\\
 & \times\overline{\tau}^+G^0_{\alpha\gamma^{\prime}}(\mathbf{k}^{\prime})\Gamma_{\gamma^{\prime}\gamma}(\mathbf{k}^{\prime}+\mathbf{q},\mathbf{k}^{\prime})
  \tilde{\tau}^{a}G^0_{\alpha^{\prime}\gamma}(\mathbf{k}^{\prime}+\mathbf{q})\overline{\tau}^-,
\end{split}
\end{equation}
\begin{equation}\label{eq37}
\begin{split}
  &\Gamma_{12}(\mathbf{k}+\mathbf{q},\mathbf{k})\tau^+=\int\frac{d^3k^{\prime}}{(2\pi)^3}V^{\alpha\alpha^{\prime}21}_{\mathbf{k}-\mathbf{k}^{\prime}}\\
  &\times\overline{\tau}^+G^0_{\alpha\gamma^{\prime}}(\mathbf{k}^{\prime})\Gamma_{\gamma^{\prime}\gamma}(\mathbf{k}^{\prime}+\mathbf{q},\mathbf{k}^{\prime})
  \tilde{\tau}^{a}G^0_{\alpha^{\prime}\gamma}(\mathbf{k}^{\prime}+\mathbf{q})\overline{\tau}^-.
\end{split}
\end{equation}

Now we provide the technical details to solve the above coupled equations. We firstly study $\Gamma_{11}$. From Eq.(21) in the main text, one obtains $V^{1211}_{\mathbf{k}-\mathbf{k}^{\prime}}=2\pi ev_F\mathbf{A}^-_{\mathbf{k}-\mathbf{k}^{\prime}}$, $V^{2111}_{\mathbf{k}-\mathbf{k}^{\prime}}=-2\pi ev_F\mathbf{A}^+_{\mathbf{k}-\mathbf{k}^{\prime}}$, where $\mathbf{A}^{\pm}_{\mathbf{k}}=\mathbf{k}^{\pm}/|\mathbf{k}|^2$, and all other components in $V^{\alpha\alpha^{\prime}11}_{\mathbf{k}-\mathbf{k}^{\prime}}$ are zero. For each of the nonzero component, there are four terms with $(\gamma,\gamma^{\prime})=(1,1),(1,2),(2,1),(2,2)$. In the following, we firstly consider $|\mathbf{q}|\rightarrow0$ limit and then we consider a nonzero but small momentum transfer $\mathbf{q}$, because we focus on the long-wave regime of the dispersion. $\Gamma_{11}$ can be expanded into eight terms as following,
\begin{equation}\label{eq27s}
\begin{split}
 & \Gamma_{11}(\mathbf{k}+\mathbf{q},\mathbf{k})\tau^+=\int\frac{d^3k^{\prime}}{(2\pi)^3}\\
  &\times\{V^{1211}_{\mathbf{k}-\mathbf{k}^{\prime}}\overline{\tau}^+G^0_{11}(\mathbf{k}^{\prime})\Gamma_{11}(\mathbf{k}^{\prime}+\mathbf{q},\mathbf{k}^{\prime})
  \tau^{x}G^0_{21}(\mathbf{k}^{\prime}+\mathbf{q})\overline{\tau}^-\\
  &+V^{1211}_{\mathbf{k}-\mathbf{k}^{\prime}}\overline{\tau}^+G^0_{12}(\mathbf{k}^{\prime})\Gamma_{22}(\mathbf{k}^{\prime}+\mathbf{q},\mathbf{k}^{\prime})
  \tau^{x}G^0_{22}(\mathbf{k}^{\prime}+\mathbf{q})\overline{\tau}^-\\
  &+V^{2111}_{\mathbf{k}-\mathbf{k}^{\prime}}\overline{\tau}^+G^0_{21}(\mathbf{k}^{\prime})\Gamma_{11}(\mathbf{k}^{\prime}+\mathbf{q},\mathbf{k}^{\prime})
  \tau^{x}G^0_{11}(\mathbf{k}^{\prime}+\mathbf{q})\overline{\tau}^-\\
  &+V^{2111}_{\mathbf{k}-\mathbf{k}^{\prime}}\overline{\tau}^+G^0_{22}(\mathbf{k}^{\prime})\Gamma_{22}(\mathbf{k}^{\prime}+\mathbf{q},\mathbf{k}^{\prime})
  \tau^{x}G^0_{12}(\mathbf{k}^{\prime}+\mathbf{q})\overline{\tau}^-\\
  &+V^{1211}_{\mathbf{k}-\mathbf{k}^{\prime}}\overline{\tau}^+G^0_{11}(\mathbf{k}^{\prime})\Gamma_{12}(\mathbf{k}^{\prime}+\mathbf{q},\mathbf{k}^{\prime})
  i\tau^{y}G^0_{22}(\mathbf{k}^{\prime}+\mathbf{q})\overline{\tau}^-\\
  &+V^{1211}_{\mathbf{k}-\mathbf{k}^{\prime}}\overline{\tau}^+G^0_{12}(\mathbf{k}^{\prime})\Gamma_{21}(\mathbf{k}^{\prime}+\mathbf{q},\mathbf{k}^{\prime})
  i\tau^{y}G^0_{21}(\mathbf{k}^{\prime}+\mathbf{q})\overline{\tau}^-\\
  &+V^{2111}_{\mathbf{k}-\mathbf{k}^{\prime}}\overline{\tau}^+G^0_{21}(\mathbf{k}^{\prime})\Gamma_{12}(\mathbf{k}^{\prime}+\mathbf{q},\mathbf{k}^{\prime})
  i\tau^{y}G^0_{12}(\mathbf{k}^{\prime}+\mathbf{q})\overline{\tau}^-\\
  &+V^{2111}_{\mathbf{k}-\mathbf{k}^{\prime}}\overline{\tau}^+G^0_{22}(\mathbf{k}^{\prime})\Gamma_{21}(\mathbf{k}^{\prime}+\mathbf{q},\mathbf{k}^{\prime})
  i\tau^{y}G^0_{11}(\mathbf{k}^{\prime}+\mathbf{q})\overline{\tau}^-\}.
\end{split}
\end{equation}
For $|\mathbf{q}|\rightarrow0$, four of the above eight terms on the r.h.s are zero due to the rotational invariance. It is clear as we know that $a_1$, $a_2$, $b_1$, and $b_2$ are $\theta$-independent, while $c_1,c_2\propto e^{-i\theta}=\mathbf{k}^-/|\mathbf{k}|$. Moreover, $V^{1211}_{\mathbf{k}-\mathbf{k}^{\prime}}$ and $V^{2111}_{\mathbf{k}-\mathbf{k}^{\prime}}$ are also $\theta$, $\theta^{\prime}$-dependent. For the above eight terms, we will respectively encounter the integrands proportional to $\mathbf{A}^-_{\mathbf{k}-\mathbf{k}^{\prime}}\mathbf{k}^{\prime+}/|\mathbf{k}^{\prime}|$, $\mathbf{A}^-_{\mathbf{k}-\mathbf{k}^{\prime}}\mathbf{k}^{\prime-}/|\mathbf{k}^{\prime}|$, $\mathbf{A}^+_{\mathbf{k}-\mathbf{k}^{\prime}}\mathbf{k}^{\prime+}/|\mathbf{k}^{\prime}|$, $\mathbf{A}^+_{\mathbf{k}-\mathbf{k}^{\prime}}\mathbf{k}^{\prime-}/|\mathbf{k}^{\prime}|$, $\mathbf{A}^-_{\mathbf{k}-\mathbf{k}^{\prime}}\mathbf{k}^{\prime-}/|\mathbf{k}^{\prime}|$,
$\mathbf{A}^-_{\mathbf{k}-\mathbf{k}^{\prime}}\mathbf{k}^{\prime+}/|\mathbf{k}^{\prime}|$,
$\mathbf{A}^+_{\mathbf{k}-\mathbf{k}^{\prime}}\mathbf{k}^{\prime-}/|\mathbf{k}^{\prime}|$,
$\mathbf{A}^+_{\mathbf{k}-\mathbf{k}^{\prime}}\mathbf{k}^{\prime+}/|\mathbf{k}^{\prime}|$.
After interaction over $\theta^{\prime}$, only combinations between $\mathbf{A}^-_{\mathbf{k}-\mathbf{k}^{\prime}}\mathbf{k}^{\prime+}$ and $\mathbf{A}^+_{\mathbf{k}-\mathbf{k}^{\prime}}\mathbf{k}^{\prime-}$ are nonzero  as following,
\begin{equation}\label{eq28s}
\begin{split}
  \int^{2\pi}_0 d\theta^{\prime}\mathbf{A}^-_{\mathbf{k}-\mathbf{k}^{\prime}}\frac{\mathbf{k}^{\prime+}}{|\mathbf{k}^{\prime}|}&=\int^{2\pi}_0 d\theta^{\prime}\mathbf{A}^+_{\mathbf{k}-\mathbf{k}^{\prime}}\frac{\mathbf{k}^{\prime-}}{|\mathbf{k}^{\prime}|}\\
  &=-\frac{2\pi}{k^{\prime}}\Theta(|\mathbf{k}^{\prime}|-|\mathbf{k}|).
\end{split}
\end{equation}
Therefore, only the first, fourth, sixth, and seventh term are nonzero, i.e.,
\begin{equation}\label{eq29s}
\begin{split}
  &\Gamma_{11}(\mathbf{k}+\mathbf{q},\mathbf{k})\tau^+=\int\frac{d^3k^{\prime}}{(2\pi)^3} \\ &\times\{V^{1211}_{\mathbf{k}-\mathbf{k}^{\prime}}\overline{\tau}^+G^0_{11}(\mathbf{k}^{\prime})\Gamma_{11}(\mathbf{k}^{\prime}+\mathbf{q},\mathbf{k}^{\prime})
  \tau^{x}G^0_{21}(\mathbf{k}^{\prime}+\mathbf{q})\overline{\tau}^-\\
  &+V^{2111}_{\mathbf{k}-\mathbf{k}^{\prime}}\overline{\tau}^+G^0_{22}(\mathbf{k}^{\prime})\Gamma_{22}(\mathbf{k}^{\prime}+\mathbf{q},\mathbf{k}^{\prime})
  \tau^{x}G^0_{12}(\mathbf{k}^{\prime}+\mathbf{q})\overline{\tau}^-\\
  &+V^{1211}_{\mathbf{k}-\mathbf{k}^{\prime}}\overline{\tau}^+G^0_{12}(\mathbf{k}^{\prime})\Gamma_{21}(\mathbf{k}^{\prime}+\mathbf{q},\mathbf{k}^{\prime})
  i\tau^{y}G^0_{21}(\mathbf{k}^{\prime}+\mathbf{q})\overline{\tau}^-\\
  &+V^{2111}_{\mathbf{k}-\mathbf{k}^{\prime}}\overline{\tau}^+G^0_{21}(\mathbf{k}^{\prime})\Gamma_{12}(\mathbf{k}^{\prime}+\mathbf{q},\mathbf{k}^{\prime})
  i\tau^{y}G^0_{12}(\mathbf{k}^{\prime}+\mathbf{q})\overline{\tau}^-\}.
\end{split}
\end{equation}
Let us first calculate the first two terms  $[1]+[2]$ in Eq.\eqref{eq29s}. For  $|\mathbf{q}|\rightarrow0$, we arrive at,
\begin{equation}\label{eq30s}
\begin{split}
  [1]+[2]&=-\tau^+\int \frac{d^3k^{\prime}}{(2\pi)^3}\\
  &\times\{V^{1211}_{\mathbf{k}-\mathbf{k}^{\prime}}\frac{a_2(\mathbf{k}^{\prime})c^{\star}_2(\mathbf{k}^{\prime})+a_1(\mathbf{k}^{\prime})c^{\star}_1(\mathbf{k}^{\prime})}{U^2(\mathbf{k}^{\prime})}\Gamma_{11}(\mathbf{k}^{\prime})\\
  &+V^{1211\star}_{\mathbf{k}-\mathbf{k}^{\prime}}\frac{b_2(\mathbf{k}^{\prime})c_2(\mathbf{k}^{\prime})-b_1(\mathbf{k}^{\prime})c_1(\mathbf{k}^{\prime})}{U^2(\mathbf{k}^{\prime})}\Gamma_{11}(\mathbf{k}^{\prime})\},
\end{split}
\end{equation}
where we have used $\Gamma_{11}=\Gamma_{22}$. After insertion of $a_i$, $b_i$ and $c_i$ into the above terms, then we obtain
\begin{equation}\label{eq31s}
\begin{split}
  [1]+[2]&=-\tau^+\Gamma_{11}(2\pi ev_F)\int\frac{d\omega^{\prime}}{2\pi}\int^{\Lambda}_0\frac{dk^{\prime}}{2\pi}k^{\prime}g(k^{\prime},\omega^{\prime})\\
  &\times\int^{2\pi}_0\frac{d\theta^{\prime}}{2\pi}[\frac{(\mathbf{k}-\mathbf{k}^{\prime})^-}{|\mathbf{k}-\mathbf{k}^{\prime}|^2}\frac{\mathbf{k}^{\prime+}}{|\mathbf{k}^{\prime}|}+\frac{(\mathbf{k}-\mathbf{k}^{\prime})^+}{|\mathbf{k}-\mathbf{k}^{\prime}|^2}\frac{\mathbf{k}^{\prime-}}{|\mathbf{k}^{\prime}|}]\\
  &=\tau^+\Gamma_{11}(4\pi ev_F)\int\frac{d\omega^{\prime}}{2\pi}\int^{\Lambda}_k\frac{dk^{\prime}}{2\pi}g(k^{\prime},\omega^{\prime})\\
  &=\tau^+eC(e)\Gamma_{11},
\end{split}
\end{equation}
where $g(k^{\prime},\omega^{\prime})=(g_1+g_2)/U^2(\mathbf{k}^{\prime})$, and
$g_1=\Delta_{3k^{\prime}}\Delta_{0k^{\prime}}(\Delta^2_{0k^{\prime}}+\Delta^2_{3k^{\prime}}+\omega^{\prime2}+k^{\prime2}v^2_F)(\Delta^2_{0k^{\prime}}+\Delta^2_{3k^{\prime}}+\omega^{\prime2}-v^2_Fk^{\prime2})$, $g_2=2\Delta_{0k^{\prime}}\Delta_{3k^{\prime}}v^2_Fk^{\prime2}(\Delta^2_{3k^{\prime}}+\omega^{\prime2}+v^2_Fk^{\prime2}-\Delta^2_{0k^{\prime}})$
is a function of $\mathbf{k}^{\prime}$ originated from expansion of $[a_2(\mathbf{k}^{\prime})c^{\star}_2(\mathbf{k}^{\prime})+a_1(\mathbf{k}^{\prime})c^{\star}_1(\mathbf{k}^{\prime})]/U^2(\mathbf{k}^{\prime})$ and $[b_2(\mathbf{k}^{\prime})c_2(\mathbf{k}^{\prime})-b_1(\mathbf{k}^{\prime})c_1(\mathbf{k}^{\prime})]/U^2(\mathbf{k}^{\prime})$, where the odd terms of $\omega^{\prime}$ have been removed as they go to zero after integral of $\omega^{\prime}$.  We used $k\rightarrow0$  in last line of Eq.\eqref{eq31s}, and $C(e)=4\pi v_F\int\frac{d\omega^{\prime}}{2\pi}\int^{\Lambda}_k\frac{dk^{\prime}}{2\pi}g(k^{\prime},\omega^{\prime})$ is a dimensionless constant that only relies CS charge $e$ while it is $\Lambda$-independent as can be shown directly by a rescaling from $k^{\prime}_{\mu}$ to $\Lambda k^{\prime\prime}_{\mu}$. We calculate numerical the quantity $eC(e)$ as a function of $e$, a very weak $e$-dependence is found for $e>3$ where the CS superconductor can find itself a stable phase, and $eC(e)$ saturates to a small constant $eC(e)=0.292$ for large $e$.
Therefore, Eq.\eqref{eq29s} is reduced to
\begin{equation}\label{eq33s}
  [1-eC(e)]\Gamma_{11}\tau^+=[3]+[4].
\end{equation}
The terms [1]+[2] bring brought a constant correction $eC(e)\thicksim0.3$ to the coefficient in front of $\Gamma_{11}$ in the Bethe-Salpeter-type equation. Now let us calculate r.h.s on Eq.\eqref{eq33s}, after insertion of the Green's functions, at $\mathbf{q}\rightarrow0$ one obtains,
\begin{equation}\label{eq34s}
\begin{split}
  [3]+[4]&=-\tau^+(2\pi ev_F)\int\frac{d^3k^{\prime}}{(2\pi)^3}\\
  &\times[\frac{(\mathbf{k}-\mathbf{k}^{\prime})^-}{|\mathbf{k}-\mathbf{k}^{\prime}|^2}\frac{\mathbf{k}^{\prime+}}{|\mathbf{k}^{\prime}|}
  +\frac{(\mathbf{k}-\mathbf{k}^{\prime})^+}{|\mathbf{k}-\mathbf{k}^{\prime}|^2}\frac{\mathbf{k}^{\prime-}}{|\mathbf{k}^{\prime}|}]\overline{\Gamma}_{12}\\
  &\times\frac{|c_2(\mathbf{k}^{\prime})|^2-|c_1(\mathbf{k}^{\prime})|^2}{U^{2}(\mathbf{k}^{\prime})},
\end{split}
\end{equation}
where we have introduced $\Gamma_{12}=\overline{\Gamma}_{12}e^{-i\theta^{\prime}}=\overline{\Gamma}_{12}\mathbf{k}^{\prime-}/|\mathbf{k}^{\prime}|$ since we have $\Gamma_{12}\propto e^{-i\theta^{\prime}}$.  Note that although we consider $\mathbf{q}\rightarrow0$ the frequency $\nu$ is implicit in Eq.\eqref{eq34s}. Formally, $\nu\rightarrow0$, it is straightforward to find out that
\begin{equation}\label{eq37s}
  \lim_{\nu\rightarrow0}\frac{|c_2(\mathbf{k}^{\prime})|^2-|c_1(\mathbf{k}^{\prime})|^2}{U^2(\mathbf{k}^{\prime})}=\frac{(\Delta^2_{0k^{\prime}}-v^2_Fk^{\prime2})}{(\omega^2+E^2_{k^{\prime},-})(\omega^2+E^2_{k^{\prime},+})},
\end{equation}
where $E_{k,\pm}=\sqrt{\Delta^2_{3k}+(\Delta_{0k}\pm v_Fk)^2}$.
Then, we recover finite $\nu$, finite but small $|\mathbf{q}|$ and make Taylor expansion with respect to $|\mathbf{q}|$. After tedious expansion for both the numerator and denominator, it is found that all the linear terms with respect to $|\mathbf{q}|$ after expansion vanish  after the integration of $\theta^{\prime}$. The second order terms $|\mathbf{q}|^2$ are nontrivial and can modify the Bethe-Salpeter-type equations. Besides, the second order terms from the numerator and denominator are equal to each other after expansion. Therefore, Eq.\eqref{eq34s} is cast into the following form after we recover $\nu$ and a small momentum transfer $\mathbf{q}$,
\begin{equation}\label{eq38s}
\begin{split}
  [3]+[4]&=-\tau^+(4\pi ev_F)\int\frac{d^3k^{\prime}}{(2\pi)^3}\\
  &\times[\frac{(\mathbf{k}-\mathbf{k}^{\prime})^-}{|\mathbf{k}-\mathbf{k}^{\prime}|^2}\frac{\mathbf{k}^{\prime+}}{|\mathbf{k}^{\prime}|}
  +\frac{(\mathbf{k}-\mathbf{k}^{\prime})^+}{|\mathbf{k}-\mathbf{k}^{\prime}|^2}\frac{\mathbf{k}^{\prime-}}{|\mathbf{k}^{\prime}|}]\overline{\Gamma}_{12}\\
  &\times\frac{|c_2(\mathbf{k}^{\prime})|^2-|c_1(\mathbf{k}^{\prime})|^2}{U(\mathbf{k}^{\prime})U(\mathbf{k^{\prime}}+\mathbf{q})}\\
  &\simeq\tau^+(8\pi ev_F)\int\frac{d\omega^{\prime}}{2\pi}\int^{\Lambda}_k\frac{dk^{\prime}k^{\prime}}{2\pi}\overline{\Gamma}_{12}\\
  &\times\frac{(\Delta^2_{0k^{\prime}}-v^2_Fk^{\prime2})}{(\omega^2+E^2_{k^{\prime},-})[(\omega+\nu)^2+E^2_{k^{\prime},+}+\Delta^2_{0q}]},
\end{split}
\end{equation}
where $\Delta^2_{0q}=(ev_Fq/2)^2$ and we have assumed $e\gg1$ so that $\Delta^2_{0q}\gg(1+e)v^2_Fq^2$ in the last line. Since in the low-energy effective theory, the interaction between CS fermions $H_{int}$ is proportional to $e$, the large $e$ condition with $e\gg1$ requires a strong coupling between fermions and thereby a stable CS superconductor state. Here, we consider the collective modes on top of a stable CS superconductor mean-field ground state, and restrict the following discussion to $e\gg1$.

The integral of the Matsubara frequency $\omega^{\prime}$ essentially represents for a a sum of poles along the imaginary axis. We then recover the discrete notation for the frequency and then complete the sum of frequency as following,
\begin{equation}\label{eq39s}
\begin{split}
  &T\sum_n\frac{1}{[(i\omega_n)^2-E^2_{k^{\prime},-}][(i\omega_n+i\nu_2)^2-E^2_{k^{\prime},+}-\Delta^2_{0q}]}\\
  &=-\frac{1}{2E_{k^{\prime},-}}\frac{1}{(i\nu_n+\xi_{k^{\prime},+}-E_{k^{\prime},-})(i\nu_n-\xi_{k^{\prime},+}-E_{k^{\prime},-})}\\
  &-\frac{1}{2\xi_{k^{\prime},+}}\frac{1}{(i\nu_n+\xi_{k^{\prime},+}-E_{k^{\prime},-})(i\nu_n+\xi_{k^{\prime},+}+E_{k^{\prime},-})},
\end{split}
\end{equation}
where we introduced $\xi^2_{k^{\prime},+}=E^2_{k^{\prime},+}+\Delta^2_{0q}$, and we have taken the zero temperature with the Fermi-Dirac function being reduced to a step function. Then, for $e\gg 1$, $E_{k^{\prime},\pm}=\sqrt{\Delta^2_{3k^{\prime}}+\Delta^2_{0k^{\prime}}}$ so that we set $E_{k^{\prime},+}=E_{k^{\prime},-}=E_{k^{\prime}}$. Further making expansion with respect to $q$, Eq.\eqref{eq39s} is cast into
\begin{equation}\label{eq40s}
\begin{split}
  &T\sum_n\frac{1}{[(i\omega_n)^2-E^2_{k^{\prime},-}][(i\omega_n+i\nu_2)^2-E^2_{k^{\prime},+}-\Delta^2_{0q}]}\\
  &=-\frac{i\nu_n}{E_{k^{\prime}}(i\nu_n+\Delta^2_{0q}/2E_{k^{\prime}})[(i\nu_n)^2-4E^2_{k^{\prime}}(1+\Delta^2_{0q}/4E^2_{k^{\prime}})^2]}\\
  &+\frac{\Delta^2_{0q}}{4E^3_{k^{\prime}}(i\nu_n+\Delta^2_{0q}/2E_{k^{\prime}})[i\nu_n+2E_{k^{\prime}}+\Delta^2_{0q}/2E_{k^{\prime}}]}.
\end{split}
\end{equation}
It is readily known from above that with $|\mathbf{q}|\rightarrow0$, the sum of Matsubara frequency is reduced to
\begin{equation}\label{eq40bs}
\begin{split}
  &T\sum_n\frac{1}{[(i\omega_n)^2-E^2_{k^{\prime},-}][(i\omega_n+i\nu_2)^2-E^2_{k^{\prime},+}-\Delta^2_{0q}]}\\
  &=-\frac{1}{E_{k^{\prime}}}\cdot\frac{1}{(i\nu_n)^2-4E^2_{k^{\prime}}}
\end{split}
\end{equation}
Eq.\eqref{eq38s} then reads as,
\begin{equation}\label{eq41s}
  [3]+[4]=-\tau^+(8\pi ev_F)\int^{\Lambda}_k\frac{dk^{\prime}k^{\prime}}{2\pi}\overline{\Gamma}_{12}\frac{1}{E_{k^{\prime}}}\cdot\frac{\Delta^2_{0k^{\prime}}-v^2_Fk^{\prime2}}{(i\nu_n)^2-4E^2_{k^{\prime}}}
\end{equation}
Inserting the above result to Eq.\eqref{eq33s} and make analytic continuation to the retarded Green's function with $i\nu_n\rightarrow\nu+i0^+$,  one obtains,
\begin{equation}\label{eq42s}
\begin{split}
  \Gamma_{11}&=-\frac{4ev_F}{1-eC(e)}\int^{\Lambda}_kdk^{\prime}k^{\prime}\overline{\Gamma}_{12}\frac{\Delta^2_{0k^{\prime}}}{E_{k^{\prime}}}\frac{1}{\nu^2-4E^2_{k^{\prime}}}\\
  &+i\frac{4e\pi v_F}{1-eC(e)}\int^{\Lambda}_kdk^{\prime}k^{\prime}\overline{\Gamma}_{12}\frac{\Delta^2_{0k^{\prime}}}{E_{k^{\prime}}}\delta(\nu^2-4E^2_{k^{\prime}}),
\end{split}
\end{equation}
We introduce constant $I=(1-eC(e))^{-1}$, then the above equation is simplified for $\nu=2\Delta_{3k^{\prime}}$ as,
\begin{equation}\label{eq43s}
  \Gamma_{11}=ev_FI\int^{\Lambda}_kdk^{\prime}\frac{\Gamma_{12,k^{\prime}}}{E_{k^{\prime}}},
\end{equation}
where $\Gamma_{12,k^{\prime}}=\overline{\Gamma}_{12}k^{\prime}$. Interestingly, the above equation reduces exactly to one of the self-consistent equation for the CS superconductor after we introduce a rescaling of momentum with $k^{\prime}=k^{\prime\prime}/I$, such that
\begin{equation}\label{eq44s}
  \Gamma_{11}=ev_F\int^{\Lambda I}_kdk^{\prime\prime}\frac{\Gamma_{12,k^{\prime\prime}}}{E_{k^{\prime\prime}}},
\end{equation}
where $E_{k^{\prime\prime}}=\sqrt{\Delta^2_{3k^{\prime\prime}}+\Delta^2_{0k^{\prime\prime}}}$ with $\Delta^2_{3k^{\prime\prime}}=I^2\Delta^2_{3k^{\prime}}$. Since $\Delta_{3k^{\prime}}\propto\Lambda$, $\Delta^2_{3k^{\prime\prime}}$ is reduced to the original form after setting a shifted momentum cutoff $\Lambda^{\prime}=I\Lambda$.
On the other hand, the self-consistent equation of the CS superconductor gap function is reduced to the same form for $e\gg1$, as shown by Eq.(27) of the main text.

From above, we know that although the term [1],[2] in Eq.\eqref{eq30s} introduces a constant shift to the coefficient of the Bethe-Salpeter-type equation for large $e$, the constant shift can be completed absorbed by defining a new momentum cutoff if $\nu=2\Delta_{3k^{\prime}}$, i.e., with $\nu=2\Delta_{3k^{\prime}}$ one can always reduce the Bethe-Salpeter-type equation to the mean-field self-consistent equation, which is satisfied from our starting point. This directly suggests us that $\nu=2\Delta_{3k^{\prime}}$ could be the solution of the Bethe-Salpeter-type equation for zero momentum shift $\mathbf{q}\rightarrow0$ (if the other self-consistent equation corresponding to $\Gamma_{12}$ is also reduced at this condition $\nu=2\Delta_{3k^{\prime}}$), i.e., the fluctuation of superconductor order parameter should have dispersion with energy $\nu=2\Delta_{3k^{\prime}}$ at $\mathbf{q}\rightarrow0$.

Now let us keep the $|\mathbf{q}|$ terms in Eq.\eqref{eq40s}. A small $|\mathbf{q}|$ should slightly perturb the solution $\nu=2\Delta_{3k^{\prime}}$ at $\mathbf{q}\rightarrow0$. Therefore, after analytic continuation $i\nu_n\rightarrow\nu+i0^+$, one can treat $q$ as a small quantity compared to $\nu$ in Eq.\eqref{eq40s} because $|\mathbf{q}|\ll 2\Delta_{3k^{\prime}}\sim \Lambda$. Keeping the leading term in Eq.\eqref{eq40s}, one then arrives at
\begin{equation}\label{eq45s}
\begin{split}
  &T\sum_n\frac{1}{[(i\omega_n)^2-E^2_{k^{\prime},-}][(i\omega_n+i\nu_2)^2-E^2_{k^{\prime},+}-\Delta^2_{0q}]}\\
  &=-\frac{1}{E_{k^{\prime}}}\frac{1}{(i\nu_n)^2-4E^2_{k^{\prime}}-2\Delta^2_{0q}},
\end{split}
\end{equation}
such that the Eq.\eqref{eq38s} is reduced to the following form as,
\begin{equation}\label{eq46s}
\begin{split}
   [3]+[4]&=-\tau^+(8\pi ev_F)\int^{\Lambda}_k\frac{dk^{\prime}k^{\prime}}{2\pi}\\
   &\times\overline{\Gamma}_{12}\frac{1}{E_{k^{\prime}}}\cdot\frac{\Delta^2_{0k^{\prime}}-v^2_Fk^{\prime2}}{(i\nu_n)^2-4E^2_{k^{\prime}}-2\Delta^2_{0q}}
\end{split}
\end{equation}
for $i\nu_n\rightarrow\nu+i0^+$ and $\nu^2=4\Delta^2_{3k^{\prime}}+2\Delta^2_{0q}$, the above equation is again simplified to the mean-field self-consistent equation as
\begin{equation}\label{eq47s}
  \Gamma_{11}=ev_F\int^{\Lambda^{\prime}}_kdk^{\prime\prime}\frac{\Gamma_{12,k^{\prime\prime}}}{E_{k^{\prime\prime}}}.
\end{equation}
Therefore we know that $\nu^2=4\Delta^2_{3k^{\prime}}+2\Delta^2_{0q}$ could be the dispersion of the Higgs mode of the $p+ip$ CS superconductor. Before one can claim this, we have to study the other equation, Eq.\eqref{eq37}.

From the interaction vertex, it is known that the only nonzero interaction components occurring in Eq.\eqref{eq37} are $V^{1121}_{\mathbf{k}-\mathbf{k}^{\prime}}=V^{2221}_{\mathbf{k}-\mathbf{k}^{\prime}}=2\pi ev_FA^-_{\mathbf{k}-\mathbf{k}^{\prime}}$. Similar to above calculation for equation of $\Gamma_{11}$, the rotational symmetry will exclude four of eight terms, leaving us four remaining nonzero terms as following,
\begin{equation}\label{eq48s}
\begin{split}
  &\Gamma_{12}(\mathbf{k}+\mathbf{q},\mathbf{k})\tau^+=\int\frac{d^3k^{\prime}}{(2\pi)^3}\\
  &\times\{V^{1121}_{\mathbf{k}-\mathbf{k}^{\prime}}\overline{\tau}^+G^0_{12}(\mathbf{k}^{\prime})\Gamma_{21}(\mathbf{k}^{\prime}+\mathbf{q},\mathbf{k}^{\prime})
  i\tau^{y}G^0_{11}(\mathbf{k}^{\prime}+\mathbf{q})\overline{\tau}^-\\
  &+V^{2221}_{\mathbf{k}-\mathbf{k}^{\prime}}\overline{\tau}^+G^0_{21}(\mathbf{k}^{\prime})\Gamma_{12}(\mathbf{k}^{\prime}+\mathbf{q},\mathbf{k}^{\prime})
  i\tau^{y}G^0_{22}(\mathbf{k}^{\prime}+\mathbf{q})\overline{\tau}^-\\
  &+V^{1121}_{\mathbf{k}-\mathbf{k}^{\prime}}\overline{\tau}^+G^0_{11}(\mathbf{k}^{\prime})\Gamma_{11}(\mathbf{k}^{\prime}+\mathbf{q},\mathbf{k}^{\prime})
  \tau^{x}G^0_{11}(\mathbf{k}^{\prime}+\mathbf{q})\overline{\tau}^-\\
  &+V^{2221}_{\mathbf{k}-\mathbf{k}^{\prime}}\overline{\tau}^+G^0_{22}(\mathbf{k}^{\prime})\Gamma_{22}(\mathbf{k}^{\prime}+\mathbf{q},\mathbf{k}^{\prime})
  \tau^{x}G^0_{22}(\mathbf{k}^{\prime}+\mathbf{q})\overline{\tau}^-\}.
\end{split}
\end{equation}
For small $\mathbf{q}$ limit $|\mathbf{q}|\rightarrow0$, the first two terms are cast into the form,
\begin{equation}\label{eq49s}
\begin{split}
  &[1]+[2]=-e^{-i\theta}\tau^+\int \frac{d^3k^{\prime}}{(2\pi)^3}\times\\
  &\{V^{1121}_{\mathbf{k}-\mathbf{k}^{\prime}}\frac{\mathbf{k}^+}{|\mathbf{k}|}\frac{c_1(\mathbf{k}^{\prime})a_1(\mathbf{k}^{\prime})-a_2(\mathbf{k}^{\prime})c_2(\mathbf{k}^{\prime})}{U^2(\mathbf{k}^{\prime})}\Gamma_{12}e^{i\theta_{k^{\prime}}}\\
  &-V^{2221}_{\mathbf{k}-\mathbf{k}^{\prime}}\frac{\mathbf{k}^-}{|\mathbf{k}|}\frac{c^{\star}_2(\mathbf{k}^{\prime})b_2(\mathbf{k}^{\prime})+c^{\star}_1(\mathbf{k}^{\prime})b_1(\mathbf{k}^{\prime})}{U^2(\mathbf{k}^{\prime})}\Gamma_{12}e^{-i\theta_{k^{\prime}}}\},
\end{split}
\end{equation}
where we insert $1=e^{-i\theta}\mathbf{k}^+/|\mathbf{k}|$ where $e^{-i\theta}$ will be canceled later by the left hand side of Eq.\eqref{eq48s}. Making further expansion, we find that term [1] equals to term [2] and
\begin{equation}\label{eq50s}
  [1]+[2]=\tau^+e^{-i\theta}(4\pi ev_F)\int\frac{d\omega^{\prime}}{2\pi}\int^{k}_0\frac{ dk^{\prime}k^{\prime}}{2\pi k}g^{\prime}(k^{\prime},\omega^{\prime})\Gamma_{12},
\end{equation}
for $k\rightarrow0$, numerical integration shows that the term $[1]+[2]$ vanishes. If one keeps a small $k$, this will lead to higher order correction (with respect to [3] and [4] term in Eq.\eqref{eq49s}) to the Bethe-Salpeter-type equations which can be absorbed into the integral limit in the remaining two terms [3] and [4]. Then, we consider term [3] and [4] in Eq.\eqref{eq48s}, after some algebra, we obtain
\begin{equation}\label{eq51s}
\begin{split}
  &[3]+[4]=\tau^+e^{-i\theta}\int\frac{d^3k^{\prime}}{(2\pi)^3}(2\pi ev_F)\frac{(\mathbf{k}-\mathbf{k}^{\prime})^-\mathbf{k}^+}{|\mathbf{k}-\mathbf{k}^{\prime}|^2|\mathbf{k}|}\Gamma_{11}\\
  &\times\frac{\sum^2_{i=1}[a_i(\mathbf{k}^{\prime}
  )a_i(\mathbf{k}^{\prime}+\mathbf{q})+b_i(\mathbf{k}^{\prime}
  )b_i(\mathbf{k}^{\prime}+\mathbf{q})]}{U(\mathbf{k}^{\prime})U(\mathbf{k}^{\prime}+\mathbf{q})}.
\end{split}
\end{equation}
As what we did above, we keep a small but finite $|\mathbf{q}|$. A tedious calculation again shows that the expansion to linear order of $|\mathbf{q}|$ vanishes after performing the integrals, and the contributions from the denominator and numerator are the same for the second order terms. In the condition of $e\gg1$, Eq.\eqref{eq51} is further simplified as
\begin{equation}\label{eq52s}
\begin{split}
  &[3]+[4]=-\tau^+e^{-i\theta}(8\pi ev_F)\int \frac{d\omega^{\prime}}{2\pi}\int^k_0\frac{dk^{\prime}k^{\prime}}{2\pi}\\
  &\times\frac{\Gamma_{11}}{k}\frac{\Delta^2_{3k^{\prime}}-\omega(\omega+\nu)}{(\Delta^2_{3k^{\prime}}+\Delta^2_{0k^{\prime}}+\omega^2)(\Delta^2_{3k^{\prime}}+(\omega+\nu)^2+\Delta^2_{0k}+\Delta^2_{0q})}
\end{split}
\end{equation}
We then recover the discrete notation for the Matsubara frequency. After performing the sum of frequency,
%
for small $k^{\prime}$, we obtain
\begin{equation}\label{eq55s}
\begin{split}
  &[3]+[4]=-\tau^+e^{-i\theta}(4ev_F)\int^{k}_0\frac{dk^{\prime}k^{\prime}}{k}\\ &\times\Gamma_{11}\frac{\Delta^2_{0k^{\prime}}}{E_{k^{\prime}}[(i\nu_n)^2-4\Delta^2_{3k^{\prime}}-4\Delta^2_{0k^{\prime}}-2\Delta^2_{0q}]}.
\end{split}
\end{equation}
With $i\nu_n\rightarrow\nu+i0^+$, and $\nu^2=4\Delta^2_{3k^{\prime}}+2\Delta^2_{0q}$, Eq.\eqref{eq46s} is finally cast into
\begin{equation}\label{eq57s}
 \Gamma_{12}=ev_F\int^{k}_0dk^{\prime}\frac{k^{\prime}}{k}\frac{\Gamma_{11}}{E_{k^{\prime}}},
\end{equation}
which is exactly of the same form with the mean-field self-consistent equation for CS superconductor for $e\gg1$. This verifies that $\nu^2=4\Delta^2_{3k^{\prime}}+2\Delta^2_{0q}$ is the found dispersion of the Higgs mode of the $p+ip$ CS superconductor.

\section{Chern-Simons fermionization with boundaries}
The Hamiltonian Eq.(1) of the main text can be written as the sum of the bulk and the boundary sector as $H=H_{b}+H_l$. The bulk Hamiltonian reads as
\begin{equation}\label{eq51}
  H_{b}=\frac{J}{2}\sum_{\langle\mathbf{r},\mathbf{r}^{\prime}\rangle^{\prime}} [\hat{S}^+_{\mathbf{r}}\hat{S}^-_{\mathbf{r}^{\prime}}+\hat{S}^-_{\mathbf{r}}\hat{S}^+_{\mathbf{r}^{\prime}}],
\end{equation}
where the ``$\langle\mathbf{r},\mathbf{r}^{\prime}\rangle^\prime$" denotes all the nearest neighbor bonds that reside on the square lattice, which do not cross any boundaries. The boundary term $H_l$ is written as,
\begin{equation}\label{eq52}
  H_{l}=\frac{J}{4}\sum_{l} [\hat{S}^+(l)\hat{S}^-(\overline{l})+\hat{S}^-(l)\hat{S}^+(\overline{l})],
\end{equation}
where we use the notation for the coordinates of sites along the boundaries, $l=(i,1)$ or $l=(1,j)$, where $i\in[1,N_x]$ and $j\in[1,N_y]$.  $\overline{l}$ is the image of site $l$ with respect to $x=0$ or $y=0$, i.e., $\overline{l}=(i,N_y)$ or $\overline{l}=(N_x,j)$, as shown by the (red) points in Fig.S1.  The sum over $l$ runs through the boundary sites and therefore we introduce an additional factor $1/2$ in Eq.\eqref{eq52} to take care of the double counting of the couplings between $l$ and $\overline{l}$.  Inserting the CS fermionization in Eq.(2) and Eq.(3) of the main text, the boundary Hamiltonian is cast in the form

\begin{figure}[t]\label{figs1}
\includegraphics[width=0.9\linewidth]{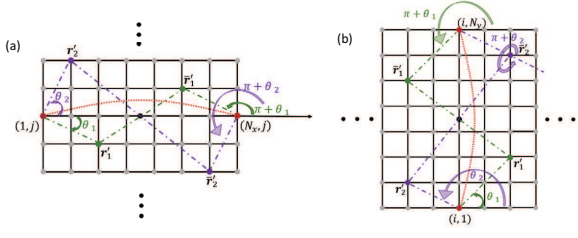}
\caption{(color online) The difference between two string operators located at boundary sites (denoted by the red points). The plot (a) and (b) indicate the calculation of the difference $\alpha(1,j)-\alpha(N_x,j)$, $\alpha(i,1)-\alpha(i,N_y)$, i.e., along $x$ and $y$ direction respectively.   For any site on the lattice, e.g.,$\mathbf{r}^{\prime}_{1}$ and $\mathbf{r}^{\prime}_{2}$, there are always inversion image site $\overline{\mathbf{r}}^{\prime}_1$ and  $\overline{\mathbf{r}}^{\prime}_2$, which contribute a difference of angles $\pi$ to the string operator. }
\end{figure}
\begin{equation}\label{eq53}
\begin{split}
  H_l&=\frac{J}{2}\sum^{N_y}_{j=1}[\hat{f}^{\dagger}(1,j)e^{ie[\alpha(1,j)-\alpha(N_x,j)]}\hat{f}(N_x,j)+h.c.]\\
  &+\frac{J}{2}\sum^{N_x}_{i=1}[\hat{f}^{\dagger}(i,1)e^{ie[\alpha(i,1)-\alpha(i,N_y)]}\hat{f}(i,N_y)+h.c.],
\end{split}
\end{equation}
where the first and second terms describe the fermion hopping crossing the boundaries along $x$ and $y$ directions respectively,  and $\alpha(\mathbf{r})=\sum_{\mathbf{r}^{\prime}\neq\mathbf{r}}\mathrm{arg}(\mathbf{r}^{\prime}-\mathbf{r})\hat{n}(\mathbf{r})$, originated from Eq.(3) of the main text, is the string operator ``located" at $\mathbf{r}$. Therefore, the boundary term depicting the spin coupling between the $l$ and $\overline{l}$ is mapped to hoppings mediated by phase operators $\mathrm{exp}[ie(\alpha(1,j)-\alpha(N_x,j))]$, $\mathrm{exp}[ie(\alpha(i,1)-\alpha(i,N_y))]$ where $\alpha(1,j)-\alpha(N_x,j)$ and $\alpha(i,1)-\alpha(i,N_y)$ are the difference of two string operators ``located" at $l$ and $\overline{l}$. It can be written explicitly as
\begin{equation}\label{eq54}
\begin{split}
  \alpha(1,j)-\alpha(N_x,j)&=\sum_{\mathbf{r}^{\prime}\neq (l,j)}\mathrm{arg}[\mathbf{r}^{\prime}-(1,j)]\hat{n}(\mathbf{r}^{\prime})\\
  &-\sum_{\mathbf{r}^{\prime}\neq (N_x,j)}\mathrm{arg}[\mathbf{r}^{\prime}-(N_x,j)]\hat{n}(\mathbf{r}^{\prime}).
\end{split}
\end{equation}
Similarly for $\alpha(i,1)-\alpha(i,N_y)$. Now we decompose the CS fermion density operator into a non-fluctuating and fluctuating parts as $\hat{n}(\mathbf{r}^{\prime})=\langle n(\mathbf{r}^{\prime})\rangle+\delta n(\mathbf{r}^{\prime})$, with the former being the expectation value with respect to the many-body ground state, i.e., the CS superconductor in our mean-field theory. If one temporarily neglects the contribution from $\langle n(\mathbf{r}^{\prime})\rangle$, the second term $\delta n(\mathbf{r}^{\prime})$ will contain fluctuation of the quantum field and can be absorbed into the fermionized bulk Hamiltonian,
\begin{equation}\label{eq55}
  H_{pbc}=t\sum_{\langle\mathbf{r},\mathbf{r}^{\prime}\rangle}[\hat{f}^{\dagger}_{\mathbf{r}}e^{ieA_{\mathbf{r},\mathbf{r}^{\prime}}}f_{\mathbf{r}^{\prime}}+h.c.].
\end{equation}
Here $\langle\mathbf{r},\mathbf{r}^{\prime}\rangle$ represents all the nearest bonds in the bulk as well as those bonds crossing the boundaries. We note in passing that due to the staggered spin configuration of the N\'{e}el state (or the staggered $\pi$-flux in the CS fermions), enforcing the periodic boundary condition means that we need to require both $N_x$ and $N_y$ being even, such that $N=N_xN_y$ being even. The total number of CS fermions $N_e$ at half-filling $\nu=1/2$ $N_e=\nu N$ is then always guaranteed to be an integer.

In addition to $H_{pbc}$, we have a remaining term, i.e., Eq.\eqref{eq54} with  $\hat{n}(\mathbf{r}^{\prime})=\langle n(\mathbf{r}^{\prime})\rangle$, resulting in
\begin{equation}\label{eq56}
\begin{split}
  \langle\alpha(1,j)-\alpha(N_x,j)\rangle&=\sum_{\mathbf{r}^{\prime}\neq (1,j)}\mathrm{arg}[\mathbf{r}^{\prime}-(1,j)]\langle\hat{n}(\mathbf{r}^{\prime})\rangle\\
  &-\sum_{\mathbf{r}^{\prime}\neq (N_x,j)}\mathrm{arg}[\mathbf{r}^{\prime}-(N_x,j)]\langle\hat{n}(\mathbf{r}^{\prime})\rangle.
\end{split}
\end{equation}
and

\begin{equation}\label{eq57}
\begin{split}
  \langle\alpha(i,1)-\alpha(i,N_y)\rangle&=\sum_{\mathbf{r}^{\prime}\neq (i,1)}\mathrm{arg}[\mathbf{r}^{\prime}-(i,1)]\langle\hat{n}(\mathbf{r}^{\prime})\rangle\\
  &-\sum_{\mathbf{r}^{\prime}\neq (i,N_y)}\mathrm{arg}[\mathbf{r}^{\prime}-(i,N_y)]\langle\hat{n}(\mathbf{r}^{\prime})\rangle.
\end{split}
\end{equation}
The planar N\'{e}el AFM order does not support out-of-plane spin expectation values such that $\langle \hat{S}^z(\mathbf{r})\rangle=0$.
In the language of CS fermions, the ground state is half-filled with $\langle\hat{n}(\mathbf{r})\rangle=1/2$, $\forall\mathbf{r}$ on the square lattice. Then, we firstly consider Eq.\eqref{eq56} related to the boundary terms along $x$-direction. In what follows, we are interested in an infinite system with periodic boundary conditions in the thermodynamic limit.
When considering the boundary condition along $x$, the boundary terms along $y$ are irrelevant, and therefore we can set $N_y\rightarrow\infty$ in the thermodynamic limit, as indicated by Fig.S1(a).
Utilizing the fact that we have
$\langle \hat{n}(-\mathbf{r})\rangle=\langle \hat{n}(\mathbf{r})\rangle$, $\forall\mathbf{r}$ in the cylinder in Fig.10(a), Eq.\eqref{eq56} is then simplified as

\begin{equation}\label{eq58}
\begin{split}
  \langle\alpha(1,j)-\alpha(N_x,j)\rangle&=\sum_{\mathbf{r}^{\prime}}~^{\prime\prime}\mathrm{arg}[\mathbf{r}^{\prime}-(1,j)]\langle\hat{n}(\mathbf{r}^{\prime})\rangle\\
  &-\sum_{\mathbf{r}^{\prime}}~^{\prime\prime}\mathrm{arg}[-\mathbf{r}^{\prime}-(N_x,j)]\langle\hat{n}(-\mathbf{r}^{\prime})\rangle\\
  &=-\pi\nu(N-2)=-\pi(N_e-1),
\end{split}
\end{equation}
where $\sum~^{\prime\prime}_{\mathbf{r}^{\prime}}$ denotes the sum over $\mathbf{r}^{\prime}$ but $\mathbf{r}^{\prime}\neq(1,j)$ and $\mathbf{r}^{\prime}\neq(N_x,j)$.
We have used $\langle\hat{n}(\mathbf{r})\rangle=\nu$, and $N_e=\langle\hat{N}_e\rangle$. We also observe the that for any $\mathbf{r}^{\prime}$ on the cylinder in Fig.S1(a), there exists an inversion image $-\mathbf{r}^{\prime}$ such that $\sum_{\mathbf{r}^{\prime}\neq (1,j),(N_x,j)}\{\mathrm{arg}[\mathbf{r}^{\prime}-(1,j)]-\mathrm{arg}[-\mathbf{r}^{\prime}-(N_x,j)]\}=-\pi$, as shown by the two example points $\mathbf{r}^{\prime}_{1,2}$ in Fig.S1(a). Similarly, the boundary condition term along $y$-direction, Eq.\eqref{eq58}, is reduced to
\begin{equation}\label{eq59}
\begin{split}
  \langle\alpha(i,1)-\alpha(i,N_y)\rangle&=\sum_{\mathbf{r}^{\prime}}~^{\prime\prime}\mathrm{arg}[\mathbf{r}^{\prime}-(i,1)]\langle\hat{n}(\mathbf{r}^{\prime})\rangle\\
  &-\sum_{\mathbf{r}^{\prime}}~^{\prime\prime}\mathrm{arg}[-\mathbf{r}^{\prime}-(i,N_y)]\langle\hat{n}(-\mathbf{r}^{\prime})\rangle\\
  &=-\pi\nu(N-2)=-\pi(N_e-1).
\end{split}
\end{equation}
After insertion of Eq.\eqref{eq58} and Eq.\eqref{eq59} into Eq.\eqref{eq53}, we obtain,
\begin{equation}\label{eq60}
\begin{split}
  H_{l}&=\frac{t}{2}\sum_{l}[e^{-ie\pi(N_e-1)}\hat{f}^{\dagger}_{l}f_{\overline{l}}+h.c.]\\
  &=\frac{t}{2}\sum_{l}[(-1)^{N_e-1}\hat{f}^{\dagger}_{l}f_{\overline{l}}+h.c.],
\end{split}
\end{equation}
where the factor $1/2$ eliminates the double counting of sites along the boundary. In the second line of Eq.\eqref{eq60}, we used $(-1)^{e(N_e-1)}=(-1)^{N_e-1}$  since $e$ is an odd integer. The above fermionized boundary Hamiltonian leads to a correction to $H_{pbc}$ in Eq.\eqref{eq55}, i.e., it contributes an additional $\mathrm{Z_2}$ phase to the CS gauge field $A_{l,\overline{l}}$ that crosses the boundary. We restore the $\mathrm{Z_2}$ phase, then $H_{pbc}$ is modified to the following total Hamiltonian
\begin{equation}\label{eq61}
\begin{split}
  H_{tot}&=t\sum_{\langle\mathbf{r},\mathbf{r}^{\prime}\rangle^{\prime}}[\hat{f}^{\dagger}_{\mathbf{r}}e^{ieA_{\mathbf{r},\mathbf{r}^{\prime}}}f_{\mathbf{r}^{\prime}}+h.c.]\\
  &+\frac{t}{2}\sum_{l}[(-1)^{N_e-1}\hat{f}^{\dagger}_{l}e^{iA_{l,\overline{l}}}f_{\overline{l}}+h.c.].
\end{split}
\end{equation}
Here the first term describes the theory of CS fermions living on a 2D lattice with open boundaries and coupled to the lattice gauge field, i.e., the bulk sector Eq.\eqref{eq51}. The second term describes the hoppings crossing the boundaries, whose hopping parameter is modified by an additional fermion parity (FP)-dependent factor $(-1)^{N_e-1}$. Therefore, we do find that the periodic boundary condition is nontrivial when one performs the CS fermionization. It generates a FP-dependent boundary condition for the CS fermions.  Without the FP-dependent boundary terms, $H_{tot}$ exactly returns back to Eq.(4) of the main text, whose ground state and the collective excitations have been carefully studied in the main text.

To make the form more concise, we rewrite Eq.\eqref{eq61} as
\begin{equation}\label{eq62}
  H_{tot}=t\sum_{\langle\mathbf{r},\mathbf{r}^{\prime}\rangle}[\hat{f}^{\dagger}_{\mathbf{r}}e^{ieA_{\mathbf{r},\mathbf{r}^{\prime}}}f_{\mathbf{r}^{\prime}}+h.c.],
\end{equation}
with an implicit boundary condition as following. Introducing the an additional row $(i,N_y+1)$ and column $(N_{x}+1,j)$ of lattice sites, the FP-dependent  boundary condition for CS fermions is found as

\begin{itemize}
  \item For $N_e$ being odd, one has $f(1,j)=f(N_{x}+1,j)$, a periodic boundary condition for CS fermions, such that in $\mathbf{k}$-space one has,
      \begin{equation}\label{eq62a}
      \begin{split}
        &\frac{1}{\sqrt{N}}\sum_{k_x,k_y}f_{k_x,k_y}e^{i(k_x+k_yj)}\\
        &=\frac{1}{\sqrt{N}}\sum_{k_x,k_y}f_{k_x,k_y}e^{i(k_x+k_yj+k_xN_x)}.
      \end{split}
      \end{equation}
 A  similar condition exists for $y$-direction, leading to $k_x=\frac{2\pi n_x}{N_x}$, $k_y=\frac{2\pi n_y}{N_y}$ with $n_{x,y}$ the integer taking the values $-N_{x,y}/2$, $-N_{x,y}/2+1$,..., $N_{x,y}/2$.
  \item For $N_e$ being even, one has $f(1,j)=-f(N_{x}+1,j)$, an anti-periodic boundary condition (APBC) for CS fermions. The APBC then generates a shift of the $\mathbf{k}$-lattice with $k_x=\frac{\pi(2n_{x}+1)}{N_x}$, $k_x=\frac{\pi(2n_y+1)}{N_y}$, with $n_{x,y}$ the integer taking the values $-N_{x,y}/2$, $-N_{x,y}/2+1$,..., $N_{x,y}/2-1$, so that $k_{x,y}$ is restricted to the BZ $[-\pi,\pi]$.
\end{itemize}

\section{Regularization of a CS superconductor}
The FP-dependent boundary conditions are essential for calculating a spin order within the CS superconductor ground state. However, before we evaluate the spin order, it will be more convenient for following calculation to firstly regularize the low-energy continuum theory of a CS superconductor. The reason for this will be apparent in the next section. We present in this section the discussion on regularization of a CS superconductor on a square lattice as an example.

If we neglect the above boundary conditions, the Hamiltonian Eq.\eqref{eq62} is the same as Eq.(4) of the main text, whose mean-field solution on the square lattice has been presented before. It gives rise to the mean-field CS superconductor on the square lattice with the Hamiltonian $H_{MF}$ in Eq.(28) (of the main text) and the order parameter Eq.(29) (of the main text). These results are obtained within a long-wavelength description of the CS superconductor, where the pairing of CS fermions takes place between $\mathbf{K}$ and $\overline{\mathbf{K}}$ in the BZ.  This is shown by the square lattice case in Fig.5 of the main text. The momentum of the CS fermions is measured from $\mathbf{K}$ and $\overline{\mathbf{K}}$, forming Cooper pairs with zero total momentum.
We also recall that in the long-wavelength limit, the mean-field solutions lead to a  constant  $\Delta_{3k}$ independent of $k$ and  $\Delta_{0k,x}=ev_Fk_{x}/2$, $\Delta_{0k,y}=ev_Fk_{y}/2$ for small $k$.

To consider the effect of  the FP-dependent boundary condition, which is essential information inherited from the lattice model, we need to first generalize the previous long-wave description of CS superconductors to a lattice CS superconductor. We, therefore, regularize the low-energy effective theory of the CS superconductor onto a square lattice. There are two ways to do it. One is to construct a mean-field theory on a lattice model, which requires more computational effort because of the effect of the gauge field on the CS fermions' high energy window. The other is to regularize the continuum Hamiltonian while keeping the low-energy effective model intact. Here, since we are interested only in the qualitative correspondence of the physical observable measured from a CS superconductor where the long-wave regime plays an essential role, we adopt the second approach without losing the relevant physics at the qualitative level.

We arrive at the CS superconductor on a square lattice by letting $k_{x/y}\rightarrow \sin k_{x/y}$ and $1\rightarrow\cos k_{x/y}$. The SC order parameter is also regularized as $\Delta_{0k,x}=(ev_F/2)\sin k_x$, $\Delta_{0k,y}=(ev_F/2)\sin k_y$, and
$\Delta_{3k}\rightarrow\Delta_3(\cos k_x\cos k_y)$ which respects the $C_{4v}$ symmetry of the square lattice. Then, we shift the  momentum by $k_x\rightarrow k_x+\pi/2$, $k_y\rightarrow k_y+\pi/2$, such that the theory now has momentums measured from the $\Gamma$ point rather than $\mathbf{K}$, $\overline{\mathbf{K}}$. This results in the lattice CS superconductor as
 $H_{latt}=\sum_{\mathbf{k}}\Psi^{\dagger}_{\mathbf{k}}\mathcal{H}(\mathbf{k})\Psi_{\mathbf{k}}$, with the basis $\Psi_{\mathbf{k}}=[\hat{f}_{\mathbf{k}A},\hat{f}_{\mathbf{k}B},\hat{f}^{\dagger}_{-\mathbf{k}A},\hat{f}^{\dagger}_{-\mathbf{k}B}]^T$,  and $\mathcal{H}(\mathbf{k})$ reads as
 \begin{equation}\label{eq65}
 \begin{split}
   \mathcal{H}(\mathbf{k})&=\tau^0\sigma^xv_F\cos k_x+\tau^z\sigma^y v_F\cos k_y+\tau^x\sigma^0\frac{ev_F}{4}\cos k_x\\
   &+\tau^y\sigma^z\frac{ev_F}{4}\cos k_y
   -\tau^y\sigma^y\Delta_3\sin k_x\sin k_y,
 \end{split}
 \end{equation}
where $\boldsymbol{\tau}$ and $\boldsymbol{\sigma}$ are Pauli matrix defined in the Nambu and sublattice space respectively. We note in passing that the regularized Hamiltonian above is in consistence with the other approach \cite{Tigrana}, i.e., the mean-field treatment of the nonlocal interaction directly on the lattice. From $H_{latt}$, we can observe the particularity of the nesting vector $\mathbf{Q}=(\pi,\pi)$ of the AFM N\'{e}el state on the square lattice (one does not need to consider other vectors, e.g., $(\pi,-\pi)$, since only one point is included in the BZ). For $\mathbf{k}=\mathbf{Q}=(\pi,\pi)$ , the inter-sublattice pairing $\Delta_3$ term vanishes. On the other hand, the intra-sublattice pairing terms $\propto\tau^{x,y}(ev_F/4)f^{\dagger}_{\mathbf{Q},a}f^{\dagger}_{-\mathbf{Q},a}$, with $a=A,B$. Since $f^{\dagger}_{\mathbf{Q},a}f^{\dagger}_{-\mathbf{Q},a}=f^{\dagger}_{\mathbf{Q},a}f^{\dagger}_{\mathbf{Q},a}=0$ due to Pauli principle, it is therefore known that the CS fermions does not form pair at $\mathbf{Q}=(\pi,\pi)$ (equivalent to $-\mathbf{Q}$). This is of the same reason as that in the 1D spinless $p$-wave superconductor, where the spinless fermions cannot form Cooper pair at $\mathbf{k}=0$. Then, the Hamiltonian at $\mathbf{k}=\mathbf{Q}$ is read off from $H_{latt}$ and Eq.\eqref{eq65} as
\begin{equation}\label{eq74}
 H_{\mathbf{Q}}=-\sqrt{2}v_Ff^{\dagger}_{\mathbf{Q},A}f_{\mathbf{Q},B}+h.c.
\end{equation}
where we made a gauge transformation that removes a global phase. After diagonalization, $H_{\mathbf{Q}}$ leads to two CS fermionic states with energy $E_{\mathbf{Q},\pm}=\pm\sqrt{2}v_F/N$, with $N$ the number of unit cells of the lattice. The state with lower energy is occupied by the CS fermions whose operator is obtained as
\begin{equation}\label{eq75}
  \tilde{\hat{f}}_{\mathbf{Q}}=\frac{1}{\sqrt{2}}(\hat{f}_{\mathbf{Q},A}-\hat{f}_{\mathbf{Q},B}).
\end{equation}
This will be a useful information for latter usage, as shown by the main text.

\section{Analytic derivation of the ground state wave function of the CS superconductors}
In this section, we provide in detail the transformation and derivation for the ground state wave function with respect to the low-energy effective Hamiltonian of a CS superconductor, using the honeycomb lattice case as an example.

Following Eq.(23) and Eq.(24) of the main text, the low-energy description of CS superconductor on honeycomb lattice reads explicitly as,
\begin{equation}\label{eq67ap}
   \mathcal{H}(\mathbf{k})=\left(
                      \begin{array}{cccc}
                                                             0 & v_Fk^- & -\Delta_{3,k} & -\frac{k^-}{k}\Delta_{0,k} \\
                                                             v_Fk^+ & 0 & \frac{k^+}{k}\Delta_{0,k} & -\Delta_{3,k} \\
                                                             -\Delta_{3,k} & \frac{k^-}{k}\Delta_{0,k} & 0 & -v_Fk^- \\
                                                             -\frac{k^+}{k}\Delta_{0,k} & -\Delta_{3,k} & -v_Fk^+ & 0 \\
                      \end{array}
                                                         \right).
\end{equation}
We can find out that the matrix $\hat{R}(\mathbf{k})$ that diagonalizes $\mathcal{H}(\mathbf{k})$, leading to a diagonal matrix  $\hat{R}^{\dagger}(\mathbf{k})\mathcal{H}(\mathbf{k})\hat{R}(\mathbf{k})$, where
 \begin{equation}\label{eq68ap}
   \hat{R}(\mathbf{k})=\left(
                      \begin{array}{cccc}
                                                             \hat{W}^{\dagger} & \sigma^z\hat{Z}^{\dagger}\sigma^z \\
                                                             \sigma^z\hat{W}^{\dagger}\sigma^z & \hat{Z}^{\dagger} &\\
                      \end{array}
                                                         \right),
\end{equation}
where $W$, $Z$ are by 2 transformation matrices acting on the sublattice space given by
 \begin{equation}\label{eq69ap}
   \hat{W}=\frac{1}{2}\left(
                      \begin{array}{cccc}
                                                             \alpha^+-\beta^+ & \frac{k^-}{k}(\alpha^++\beta^+) \\
                                                             \frac{k^+}{k}(\alpha^-+\beta^-) & \alpha^--\beta^- &\\
                      \end{array}
                                                         \right),
\end{equation}
and
 \begin{equation}\label{eq70ap}
   \hat{Z}=\frac{1}{2}\left(
                      \begin{array}{cccc}
                                                             \alpha^++\beta^+ & \frac{k^-}{k}(\alpha^+-\beta^+) \\
                                                             \frac{k^+}{k}(\alpha^--\beta^-) & \alpha^-+\beta^- &\\
                      \end{array}
                                                         \right).
\end{equation}
$\alpha^{\pm}$ and  $\beta^{\pm}$ has $\mathbf{k}$-dependence, which is not explicitly written for brevity, and they are derived from Bogoliubov transformations as,
\begin{eqnarray}
  \beta^{\pm} &=& \sqrt{\frac{1}{2}(1-\frac{\epsilon^{\pm}_{\mathbf{k}}}{E^{\pm}_{\mathbf{k}}})}, \\
  \alpha^{\pm} &=& \sqrt{\frac{1}{2}(1+\frac{\epsilon^{\pm}_{\mathbf{k}}}{E^{\pm}_{\mathbf{k}})}} ,
\end{eqnarray}
where $E^{\pm}_{\mathbf{k}}=\sqrt{\epsilon^{\pm2}_{\mathbf{k}}+\Delta^2_{3,k}}$ and $\epsilon^{\pm}_{\mathbf{k}}=v_Fk\pm\Delta_{0,k}$.
Therefore, one obtains the transformation matrix that shifts the CS fermion basis to the Bogoliubov quasi-particle basis, as  $\Gamma_{\mathbf{k}}=\hat{R}\Psi_{\mathbf{k}}$, where $\Psi_{\mathbf{k}}=[\hat{f}_{\mathbf{k},A},\hat{f}_{\mathbf{k},B},\hat{\overline{f}}^{\dagger}_{-\mathbf{k},A},\hat{\overline{f}}^{\dagger}_{-\mathbf{k},B}]^T$. For simplicity, we introduce the spinor in Nambu space $\Psi_{\mathbf{k}}=[C_{\mathbf{k}},C^{\dagger}_{-\mathbf{k}}]^T $, with $C_{\mathbf{k}}=[f_{\mathbf{k},A},f_{\mathbf{k},B}]^T\equiv[C_{\mathbf{k},1},C_{\mathbf{k},2}]^T$ and $\overline{C}^{\dagger}_{\mathbf{k}}=[\overline{f}^{\dagger}_{\mathbf{k},A},\overline{f}^{\dagger}_{\mathbf{k},B}]^T\equiv[\overline{C}^{\dagger}_{\mathbf{k},1},\overline{C}^{\dagger}_{\mathbf{k},2}]^T$, where we use the notation $C_{\mathbf{k},1/2}$ to denote for the CS fermion operators on A/B sublattice, in order to arrive at a concise form during the following derivation.  Similarly, we introduce the Bogoliubov quasi-particle spinor  $\Gamma_{\mathbf{k}}=[\gamma_{\mathbf{k}},\gamma^{\dagger}_{-\mathbf{k}}]$, with $\gamma_{\mathbf{k}}=[\gamma_{\mathbf{k},1},\gamma_{\mathbf{k},2}]^T$ and $\gamma^{\dagger}_{\mathbf{k}}=[\gamma^{\dagger}_{\mathbf{k},1},\gamma^{\dagger}_{\mathbf{k},2}]^T$.
Using these notations, we can express the Bogoliubov-particles as CS fermions as,
\begin{eqnarray}
  \gamma_{\mathbf{k},i} &=& W_{ij}C_{\mathbf{k},j}+\tilde{W}_{ij}\overline{C}^{\dagger}_{-\mathbf{k},j}, \\
  \overline{\gamma}_{-\mathbf{k},i} &=& C^{\dagger}_{\mathbf{k},j}\tilde{Z}^{\dagger}_{ji}+\overline{C}_{-\mathbf{k},j}Z^{\dagger}_{ji},
\end{eqnarray}
where have defined the off-diagonal matrix in Eq.\eqref{eq68ap} as $\hat{\tilde{W}}=\sigma^z\hat{W}\sigma^z$ and $\hat{\tilde{Z}}=\sigma^z\hat{Z}\sigma^z$. $W_{ij}$ and $Z_{ij}$'s are the components of the 2 by 2 transformation $W$ and $Z$, $i,j=1$ or $2$. The two flavors of Bogoliubov particles  $\gamma_{k,i}$ arise due to the sublattice degrees of freedom of the honeycomb lattice.  The $p+ip$-wave pairing feature is implicit in the transformation matrix $W$ and $Z$'s.

The ground state of the CS superconductor can then be written as the vacuum state of all flavors of Bogoliubov quasi-particles, therefore we require $\gamma_{\mathbf{k},1}|GS\rangle=0$ as well as $\gamma_{\mathbf{k},2}|GS\rangle=0$, $\forall\mathbf{k}$. As shown by Sec.IV of the main text, there are two Bogoliubov vacuum states corresponding to even and odd parity case respectively. Here, we derive the ground state wave function for $N_e$ being even, and the other sector can be obtained by creating an additional CS fermions. One then can write down the Bogoliubov vacuum as,
\begin{equation}\label{eq71ap}
  |GS\rangle=\prod_{\mathbf{k}}\gamma_{\mathbf{k},1}\overline{\gamma}_{-\mathbf{k},1}\gamma_{\mathbf{k},2}\overline{\gamma}_{-\mathbf{k},2}|0\rangle,
\end{equation}
where $|0\rangle$ is the vacuum state for CS fermions $C_{\mathbf{k},i}$ giving $C_{\mathbf{k},i}|0\rangle=0$. Inserting the relations between the Bogoliubov quasi-particles and the CS fermions, the ground state is cast into 16 different combinations of $C_{\mathbf{k}}$, $C^{\dagger}_{\mathbf{k}}$, however, one can simplify the equation using the Wick's theorem, i.e.,
\begin{equation}\label{eq72ap}
  \gamma_{\mathbf{k},1}\overline{\gamma}_{-\mathbf{k},1}\gamma_{\mathbf{k},2}\overline{\gamma}_{-\mathbf{k},2}=\hat{N}[\hat{P}_C[\gamma_{\mathbf{k},1}\overline{\gamma}_{-\mathbf{k},1}\gamma_{\mathbf{k},2}\overline{\gamma}_{-\mathbf{k},2}]],
\end{equation}
where $\hat{P}_C$ denotes all possible contraction of pairs of Bogoliubov particles and $\hat{N}$ denotes the normal ordering of quantum fields. Then one would encounter  several contractions which we evaluate in the following. Hereafter, we use $\hat{C}[AB]$ to represent for the contraction of two operators A and B,  defined as $\hat{C}[AB]=AB-\hat{N}[AB]$. For the contraction $\hat{C}[\gamma_{\mathbf{k},i}\gamma_{\mathbf{k},j}]$, it can be calculated as,
\begin{equation}\label{eq73ap}
\begin{split}
  &\hat{C}[\gamma_{\mathbf{k},i}\gamma_{\mathbf{k},j}]=\\
  &[W_{im}C_{\mathbf{k},m}+\tilde{W}_{im}\overline{C}^{\dagger}_{-\mathbf{k},m}]
  [W_{jn}C_{\mathbf{k},n}+\tilde{W}_{jn}\overline{C}^{\dagger}_{-\mathbf{k},n}]\\
  &-\hat{N}[(W_{im}C_{\mathbf{k},m}+\tilde{W}_{im}\overline{C}^{\dagger}_{-\mathbf{k},m})
  (W_{jn}C_{\mathbf{k},n}+\tilde{W}_{jn}\overline{C}^{\dagger}_{-\mathbf{k},n})]\\
  &=W_{im}\tilde{W}_{jn}\{C_{\mathbf{k},m},\overline{C}^{\dagger}_{-\mathbf{k},n}\}=0
\end{split}
\end{equation}
where the commutator goes to zero because of $N_e$ being even, where $\mathbf{k}$ in the reciprocal space cannot take $\mathbf{k}=0$ due to the anti-periodic boundary condition of CS fermions, as shown by Sec.IVA of the main text. Similarly, we obtain $\hat{C}[\overline{\gamma}_{-\mathbf{k},i}\overline{\gamma}_{-\mathbf{k},j}]=0$. However, we arrive at nonzero commutators for contraction between $\gamma_{\mathbf{k},i}$ and $\gamma_{-\mathbf{k},j}$ as following,
\begin{equation}\label{eq74ap}
\begin{split}
  &\hat{C}[\gamma_{\mathbf{k},i}\gamma_{-\mathbf{k},j}]=\\
  &[W_{im}C_{\mathbf{k},m}+\tilde{W}_{im}\overline{C}^{\dagger}_{-\mathbf{k},m}]
  [C^{\dagger}_{\mathbf{k},n}\tilde{Z}^{\dagger}_{nj}+\overline{C}_{-\mathbf{k},n}Z^{\dagger}_{nj}]\\
  &-\hat{N}[[W_{im}C_{\mathbf{k},m}+\tilde{W}_{im}\overline{C}^{\dagger}_{-\mathbf{k},m}]
  [C^{\dagger}_{\mathbf{k},n}\tilde{Z}^{\dagger}_{nj}+\overline{C}_{-\mathbf{k},n}Z^{\dagger}_{nj}]\\
  &=W_{im}\tilde{Z}^{\dagger}_{nj}\{C_{\mathbf{k},m},C^{\dagger}_{\mathbf{k},n}\}=W_{im}\tilde{Z}^{\dagger}_{nj}\delta_{mn}.
\end{split}
\end{equation}
and similarly one obtains $\hat{C}[\overline{\gamma}_{-\mathbf{k},i}\gamma_{\mathbf{k},j}]=-(\hat{W}\cdot\hat{\tilde{Z}}^{\dagger})_{ji}$. Then, inserting the above contractions to Eq.\eqref{eq72ap} and then evaluating the normal orderings of the remaining uncontracted CS fermions, one can obtain the ground state of the CS superconductor for even FP, which reads as,
\begin{equation}\label{eq75ap}
\begin{split}
  |GS\rangle&=-\prod_{\mathbf{k}}[W_{1n}\tilde{Z}^{\dagger}_{n2}+\tilde{W}_{1i}\tilde{Z}^{\dagger}_{n2}\overline{C}^{\dagger}_{-\mathbf{k},i}C^{\dagger}_{\mathbf{k},n}]\\
  &\times[W_{2n}\tilde{Z}^{\dagger}_{n1}+\tilde{W}_{2m}\tilde{Z}^{\dagger}_{j1}\overline{C}^{\dagger}_{-\mathbf{k},m}C^{\dagger}_{\mathbf{k},j}]|0\rangle.
\end{split}
\end{equation}
Inserting the specific elements of $\hat{W}$, $\hat{\tilde{W}}$, $\hat{Z}$ and $\hat{\tilde{Z}}$ in Eq.\eqref{eq69ap}, Eq.\eqref{eq70ap} in to Eq.\eqref{eq72ap}, the ground state is finally cast into,
\begin{equation}\label{eq76ap}
\begin{split}
  |GS\rangle&=-\prod_{\mathbf{k}}U_{\mathbf{k}}e^{G_{ij}\overline{C}^{\dagger}_{-\mathbf{k},i}C^{\dagger}_{\mathbf{k},j}}|0\rangle,
\end{split}
\end{equation}
where $U_{\mathbf{k}}=(\beta^-\alpha^++\alpha^-\beta^+)^2/4$, and the matrix $G_{ij}$, whose matrix components reflects the pairing symmetry in the sublattice space, is a 2 by 2 matrix in Nambu space as,
 \begin{equation}\label{eq77ap}
 \begin{split}
  G_{ij}=-\frac{1}{\alpha^+\beta^-+\beta^+\alpha-}
  \left(
                      \begin{array}{cccc}
                                                             g^1_{ij} & g^2_{ij} \\
                                                             g^2_{ij} &  g^1_{ij}&\\
                      \end{array}
                                                         \right).
\end{split}
\end{equation}
where $g^1_{ij}=\alpha^+\alpha^-+\beta^+\beta^-$, $g^2_{ij}=-\frac{k^+}{k}(\alpha^-+\beta^-)(\alpha^+-\beta^+)$ and $g^3_{ij}=-\frac{k^-}{k}(\alpha^++\beta^+)(\alpha^--\beta^-)$.

\section{details in calculating the spin susceptibility }
In the main text, we have reduced the calculation to evaluating the expectation in Eq.(46) of the main text. In this section, we show the details in the pertinent calculations. Due to the particularity of the momentum $\mathbf{Q}$ discussed above, we separate the $\mathbf{k}=\mathbf{Q}$ from the rest of the terms as
\begin{equation}\label{eq82}
  \sum_{\mathbf{k}}\hat{f}_{\mathbf{k},a}=\sum_{\mathbf{k}}~^{\prime}\hat{f}_{\mathbf{k},a}+\hat{f}_{\mathbf{Q},a},
\end{equation}
where the prime denotes the sum for $\mathbf{k}\neq\mathbf{Q}$. For the string operator in Eq.(46) of the main text, one then has
\begin{equation}\label{eq83}
\begin{split}
  i\alpha_0&=ie\sum_{\mathbf{r}^{\prime}\neq0}\sum_{\mathbf{p},\mathbf{q},a}\mathrm{arg}(\mathbf{r}^{\prime})\mathrm{exp}[i(\mathbf{p}-\mathbf{q})\cdot\mathbf{r}^{\prime}]\hat{f}^{\dagger}_{\mathbf{p},a}\hat{f}_{\mathbf{q},a}\\
  &= i\overline{\alpha}_0+i\alpha_{0,\mathbf{Q}},
\end{split}
\end{equation}
where we introduced the partition of the string operator $\alpha_0$ as
\begin{equation}\label{eq84}
i\overline{\alpha}_0=ie\sum_{\mathbf{r}^{\prime}\neq0}\{\sum_{\mathbf{p},\mathbf{q},a}~^{\prime}\mathrm{arg}(\mathbf{r}^{\prime})\mathrm{exp}[i(\mathbf{p}-\mathbf{q})\cdot\mathbf{r}^{\prime}]\hat{f}^{\dagger}_{\mathbf{p},a}\hat{f}_{\mathbf{q},a},
\end{equation}
and
\begin{equation}\label{eq85}
i\alpha_{0,\mathbf{Q}}=ie\sum_{\mathbf{r}^{\prime}\neq0}\{\sum_{\mathbf{p},\mathbf{q},a}~^{\prime\prime}\mathrm{arg}(\mathbf{r}^{\prime})\mathrm{exp}[i(\mathbf{p}-\mathbf{q})\cdot\mathbf{r}^{\prime}]\hat{f}^{\dagger}_{\mathbf{p},a}\hat{f}_{\mathbf{q},a}.
\end{equation}
Here the prime on the sum denotes both $\mathbf{p}\neq\mathbf{Q}$ and $\mathbf{q}\neq\mathbf{Q}$, and the double prime denotes at least one of the momenta $\mathbf{p}$, $\mathbf{q}$ equal to $\mathbf{Q}$. Further expansion of the exponential term leads to
\begin{equation}\label{eq86}
\begin{split}
\mathrm{exp}[i\overline{\alpha}_0+i\alpha_{0,\mathbf{Q}}]&\simeq \mathrm{exp}[i\overline{\alpha}_0]\mathrm{exp}[i\alpha_{0,\mathbf{Q}}]\mathrm{exp}[-[\overline{\alpha}_{0},\alpha_{0,\mathbf{Q}}]/2]\\
&\equiv\mathrm{exp}[i\overline{\alpha}_0]\mathrm{exp}[i\alpha_{\mathbf{Q}}].
\end{split}
\end{equation}
Therefore, the string operator $e^{i\alpha_{0}}$ in Eq.(46) of the main text is partitioned into $\mathrm{exp}[i\overline{\alpha}_0]$ and $\mathrm{exp}[i\alpha_{\mathbf{Q}}]$ defined by $\mathrm{exp}[i\alpha_{\mathbf{Q}}]=\mathrm{exp}[i\alpha_{0,\mathbf{Q}}]\mathrm{exp}[-[\overline{\alpha}_{0},\alpha_{0,\mathbf{Q}}]/2]$. The term $\mathrm{exp}[i\overline{\alpha}_0]\mathrm{exp}[i\alpha_{\mathbf{Q}}]$ can be further expanded with respect to the CS fermion operators. It turns out that the string operator contains either zero or a single $f^{\dagger}_{\mathbf{Q},a}$ operator, with utilizing the following property of the Bogoliubov vacuum $f_{\mathbf{Q},a}|e\rangle=0$ (since the state with momentum $\mathbf{Q}$ is not occupied in $|e\rangle$, as derived in Sec. IVC). Therefore, two nonzero terms have contribution to the off-diagonal term $_{o}\langle GS| H^{\prime}|GS\rangle_e$, leading to
\begin{equation}\label{eq87}
\begin{split}
   &_{o}\langle GS| H^{\prime}|GS\rangle_e=\\
   &-(\frac{B}{2\sqrt{2}}) \langle(\hat{f}_{\mathbf{Q},A}-\hat{f}_{\mathbf{Q},B})\hat{f}^{\dagger}_{\mathbf{Q},a}\hat{P}_0[e^{i\overline{\alpha}_0}e^{i\alpha_{\mathbf{Q}}}]\rangle_e\\
   &-(\frac{B}{2\sqrt{2}}) \langle(\hat{f}_{\mathbf{Q},A}-\hat{f}_{\mathbf{Q},B})\sum_{\mathbf{k}}~^{\prime}\hat{f}^{\dagger}_{\mathbf{k},a}\hat{P}_1[e^{i\overline{\alpha}_0}e^{i\alpha_{\mathbf{Q}}}]\rangle_e\\
\end{split}
\end{equation}
where $\hat{P}_0$ and $\hat{P}_1$ represent for the projection to the term with zero and single $f^{\dagger}_{\mathbf{Q},a}$ operator, respectively.

To proceed, let us now consider the case where the weak magnetic field $B$ is applied on a local A sublattice such that $a=A$ (the results are similar for application onto B). Then, both of the two terms in Eq.\eqref{eq87} contribute a factor $\langle 0|\hat{f}_{\mathbf{Q},A}f^{\dagger}_{\mathbf{Q},A}|0\rangle=1$, where  the unpaired CS fermion is acting on the local Hilbert space $\{|0\rangle,|1\rangle\}$ orthogonal to $|GS\rangle_e$. The rest of the terms are obviously zero, i.e., $\langle\hat{f}_{\mathbf{Q},B}\hat{f}^{\dagger}_{\mathbf{Q},A}\hat{P}_0[e^{i\overline{\alpha}_0}e^{i\alpha_{\mathbf{Q}}}]\rangle=0$ and $\langle\hat{f}_{\mathbf{Q},B}\sum^{\prime}_{\mathbf{k}}\hat{f}^{\dagger}_{\mathbf{k},A}\hat{P}_1[e^{i\overline{\alpha}_0}e^{i\alpha_{\mathbf{Q}}}]\rangle=0$ due to the FP conservation.  Therefore, the off-diagonal component $_{o}\langle GS| H^{\prime}|GS\rangle_e$ is evaluated to be a finite constant, to represent which, we introduce a finite constant $g$, namely,
\begin{equation}\label{eq88}
\begin{split}
  g&= \langle \hat{f}_{\mathbf{Q},A}\hat{f}^{\dagger}_{\mathbf{Q},A}\hat{P}_0[e^{i\overline{\alpha}_0}e^{i\alpha_{\mathbf{Q}}}]\rangle_e\\
  &+\langle\hat{f}_{\mathbf{Q},A}\sum_{\mathbf{k}}~^{\prime}\hat{f}^{\dagger}_{\mathbf{k},A}\hat{P}_1[e^{i\overline{\alpha}_0}e^{i\alpha_{\mathbf{Q}}}]\rangle_e.
\end{split}
\end{equation}
Using the above notation, the component $_{o}\langle GS| H^{\prime}|GS\rangle_e$ is cast into
\begin{equation}\label{eq87a}
  _{o}\langle GS| H^{\prime}|GS\rangle_e=-g\frac{B}{2\sqrt{2}}.
\end{equation}
Here, we note in passing that, although we apply here a specific field only onto A sublattice, any local field asymmetric with respect to A and B sublattice will generate qualitatively the same results. However, a sublattice-symmetric field, i.e., a equal perturbation being applied to the two sublattices at $\mathbf{r}_0$, can only lead to zero perturbation matrix $_{o}\langle GS| H^{\prime}|GS\rangle_e=0$, because of the cancelation from terms $(\hat{f}_{\mathbf{Q},A}-\hat{f}_{\mathbf{Q},B})$ in Eq.\eqref{eq87}. This means that the ground state degeneracy of the CS superconductor can only be lifted by a sublattice-asymmetric field as indicated by Fig.12. This suggests that the CS superconductor is much more sensitive in response to sublattice-symmetry-breaking perturbations, in agreement with the physical picture of the N\'{e}el AFM state.

Thus, Eq.(45) of the main text is then obtained as an off-diagonal matrix, $H^{\prime}=-gB/(2\sqrt{2})s^{+}-g^{\star}B/(2\sqrt{2})s^{-}$, where $\boldsymbol{s}$ is the Pauli matrix denoting the 2D FP-even/odd space. This perturbation Hamiltonian lifts the double degeneracy of the CS superconductor, giving rise to the two lifted states as shown by Fig.10 of the main text. The one with the lower energy, $\epsilon_-=-|g|B/(2\sqrt{2})$, enjoys the wave function as
\begin{equation}\label{90}
 |GS\rangle_p=\frac{1}{\sqrt{2}}(|GS\rangle_o-|GS\rangle_e),
\end{equation}
which describes the ground state of the perturbed CS superconductor $|GS\rangle_p$ by an infinitesimal magnetic field. This state is an equal weight superposition between the FP odd and even state, as a result of the zero diagonal terms in Eq.(45) of the main text.

We aim to calculate the magnetic susceptibility with an infinitesimal external field. The magnetization can then be calculated via the perturbed ground state wave function $|GS\rangle_p$.
The expectation value of the spin operator at a generic site $\mathbf{r}$ reads as $_p\langle GS|\hat{S}^x_{\mathbf{r},a}|GS\rangle_p=-\mathrm{Re}[_o\langle GS|\hat{f}^{\dagger}_{\mathbf{r},a}e^{i\alpha_{\mathbf{r}}}|GS\rangle_e]/2$.
For simplicity, we ``measure" the spin expectation at $\mathbf{r}=0$, while the magnetization of other sites can be calculated similarly. We thus obtain
\begin{equation}\label{92}
\begin{split}
  &_p\langle GS|\hat{S}^x_{\mathbf{r}=0,a}|GS\rangle_p=\\
  &-\frac{1}{2\sqrt{2}}\mathrm{Re}\{_{e}\langle GS| (\hat{f}_{\mathbf{Q},A}-\hat{f}_{\mathbf{Q},B})(\sum_{\mathbf{k}}\hat{f}^{\dagger}_{\mathbf{k},a})e^{i\alpha_{0}}|GS\rangle_e\}.
\end{split}
\end{equation}
A similar correlation function of CS fermionic fields has already been encountered in Eq.(46) of the main text. Following the above procedure, we obtain,
\begin{equation}\label{93}
\begin{split}
  &_p\langle GS|\hat{S}^x_{\mathbf{r}=0,A}|GS\rangle_p=
  -\frac{1}{2\sqrt{2}}\mathrm{Re}\{\langle \hat{f}_{\mathbf{Q},A}\hat{f}^{\dagger}_{\mathbf{Q},A}\hat{P}_0[e^{i\overline{\alpha}_0}e^{i\alpha_{\mathbf{Q}}}]\rangle_e\\
  &+\langle \hat{f}_{\mathbf{Q},A}\sum_{\mathbf{k}^{\prime}}~^{\prime}\hat{f}^{\dagger}_{\mathbf{k}^{\prime},A}\hat{P}_1[e^{i\overline{\alpha}_0}e^{i\alpha_{\mathbf{Q}}}]\rangle_e\}
  =-\frac{\mathrm{Re}[g]}{2\sqrt{2}}.
\end{split}
\end{equation}
Eq.\eqref{93} directly suggests the formation of the finite magnetization at the A sublattice located at $\mathbf{r}=0$. Similarly, for magnetization at B sublattice, it is straightforward to obtain:
\begin{equation}\label{94}
\begin{split}
  &_p\langle GS|\hat{S}^x_{\mathbf{r}=0,B}|GS\rangle_p=\frac{1}{2\sqrt{2}}\mathrm{Re}\{\langle \hat{f}_{\mathbf{Q},B}\hat{f}^{\dagger}_{\mathbf{Q},B}\hat{P}_0[e^{i\overline{\alpha}_0}e^{i\alpha_{\mathbf{Q}}}]\rangle_e\\
  &+\langle\hat{f}_{\mathbf{Q},B}\sum_{\mathbf{k}^{\prime}}~^{\prime}\hat{f}^{\dagger}_{\mathbf{k}^{\prime},B}\hat{P}_1[e^{i\overline{\alpha}_0}e^{i\alpha_{\mathbf{Q}}}]\rangle_e\}
  =\frac{\mathrm{Re}[g]}{2\sqrt{2}}.
\end{split}
\end{equation}
Interestingly, the above results clearly show that an infinitesimal sublattice-symmetric field generates finite but opposite magnetization with respect to the A and B sublattice. The magnetic susceptibility, by definition, then is obtained as,
\begin{equation}\label{eq95}
 \chi_B=-\chi_A=\lim_{B\rightarrow0}\frac{_p\langle GS|\hat{S}^x_{\mathbf{r}=0,B}|GS\rangle_p}{B}.
\end{equation}
Since $g$ is an intrinsic quantity evaluated from the ground state $|GS\rangle_e$ of the CS superconductor, which is independent of $B$,  $\chi_B$ and $\chi_A$ diverge for infinitesimal field.



\begin{thebibliography}{99}
\bibitem{footnotes1} It should be noted that the fluxes on the lattice are gauge equivalent up to $2\pi$. Moreover, for any close loop that encloses one site, the fermion number satisifes $0<\langle n_{lp}\rangle<1$, thus the physically allowed flux is restricted between 0 and $2\pi$. When using Eq.\eqref{eq7}, the flux should be regularized to the physical regime by choosing the proper gauge.
\bibitem{footnotes2} The number of sites in the closed loop should be counted in the same way as that when one counts the sites in a lattice unit cell. For example, a vertex shared by N plaquettes only contributes $1/N$ site.
\bibitem{footnote1} The $\mathbf{q}$ dependence is associated with $\langle\hat{S}^z_{\mathbf{r}}\hat{S}^z_{\mathbf{r}+\mathbf{e}_j}\rangle$ rather than $\langle\hat{S}^y_{\mathbf{r}}\hat{S}^y_{\mathbf{r}+\mathbf{e}_j}\rangle$. $\langle\hat{S}^z_{\mathbf{r}}\hat{S}^z_{\mathbf{r}+\mathbf{e}_j}\rangle$ is much smaller than $\langle\hat{S}^y_{\mathbf{r}}\hat{S}^y_{\mathbf{r}+\mathbf{e}_j}\rangle$ for the planar N\'{e}el order, making the long-wave length limit a good approximation
\end{thebibliography}
\end{document}